\documentclass[aps,pra,twocolumn,showpacs,preprintnumbers,amsmath,amssymb,superscriptaddress,longbibliography]{revtex4-2}

\usepackage{soul}
\usepackage{mathrsfs}
\usepackage{amsfonts}
\usepackage{amsmath}
\usepackage{amssymb}
\usepackage{natbib}
\usepackage{bbm}
\usepackage{hyperref}
\usepackage{xcolor}
\usepackage{graphicx}

\begin{document}
	
	
	\title{Quantum nondemolition measurement operator with spontaneous emission}

	
	\author{Ebubechukwu O. Ilo-Okeke} 
	\affiliation{New York University Shanghai, NYU-ECNU Institute of Physics at NYU Shanghai, Shanghai Frontiers Science Center of Artificial Intelligence and Deep Learning, 567 West Yangsi Road, Shanghai, 200124, China.}
	\affiliation{Department of Physics, School of Physical Sciences, Federal University of Technology, P. M. B. 1526, Owerri 460001, Nigeria.}
	%
	\author{Tim Byrnes}\email{tim.byrnes@nyu.edu}
	\affiliation{New York University Shanghai, NYU-ECNU Institute of Physics at NYU Shanghai, Shanghai Frontiers Science Center of Artificial Intelligence and Deep Learning, 567 West Yangsi Road, Shanghai, 200126, China.}
	\affiliation{State Key Laboratory of Precision Spectroscopy, School of Physical and Material Sciences, East China Normal University, Shanghai 200062, China}
	\affiliation{Center for Quantum and Topological Systems (CQTS), NYUAD Research Institute, New York University Abu Dhabi, UAE.}
	\affiliation{Department of Physics, New York University, New York, NY 10003, USA}

	\date{\today}
	
	\begin{abstract}
		We present a theory for quantum nondemolition (QND) measurements of an atomic ensemble in the presence of spontaneous emission. We derive the master equation that governs the evolution of the ground state of the atoms and the quantum state of light. Solving the master equation exactly without invoking the Holstein-Primakoff approximation and projecting out the quantum state of light, we derive a positive operator-valued measure that describes the QND measurement. We show that at high spontaneous emission conditions, the QND measurement has a unique dominant state to which the measurement collapses. We additionally investigate the behavior of the QND measurement in the limiting case of strong atom-light interactions, where we show that the positive operator valued measure becomes a projection operator. We further analyze the effect of spontaneous emission noise on atomic state preparation. We find that it limits the width of the eigenvalue spectrum available to a quantum state in a linear superposition. This effect leads to state collapse on the dominant state. We generate various non-classical states of the atom by tuning the atom-light interaction strength. We find that non-classical states such as the Schr\"odinger-cat state, whose coherence spans the entire eigenvalue spectrum of the total spin operator $J_z$ for a given spin eigenvalue $J$, lose their coherence because spontaneous emission limits the accessibility of states farther away from the dominant state.
	\end{abstract}

	\maketitle
	\section{Introduction}\label{sec:Introduction}
    Quantum nondemolition (QND) measurements~\cite{braginsky1980,grangier1998} are  an established technique for manipulating and engineering quantum mechanical systems. In a QND measurement, a quantum probe indirectly measures a system by interacting with it. Subsequent measurements reveal information about the system without causing further disturbance to it. For a measurement to be in the nondemolition regime, the measured observable of the quantum system and the probe observable must commute. In applications where repetitive measurements are performed on the same system, each measurement must preserve the system's state to qualify as a QND measurement. This approach has been implemented successfully in precision measurement applications across a wide range of physical systems, including trapped ions~\cite{blatt2008}, superconducting qubits~\cite{devoret2013}, cavity quantum electrodynamics~\cite{guerlin2007}, gravitational wave detection~\cite{eberle2010,pitkin2011}, squeezed state preparation~\cite{loudon2000,scully1997,walls2008,byrnes2021} in optics~\cite{brune1990,holland1991,ueda1992,yanagimoto2023}, and mechanical oscillators~\cite{lecocq2015}.

	In an atomic system, QND measurements are typically performed using interactions with detuned laser light~\cite{takahashi1999,kuzmich2000,higbie2005,meppelink2010,schleier-smith2010,vasilakis2015,moller2017,ilo-okeke2023}. The interactions between atoms and light are typically weak, which limits the degree of collapse of the quantum state of the atom system. This state collapse is reflected as a phase shift in the measurement of photons and can be used to generate a variety of non-classical states of atomic ensembles. Examples include squeezed states~\cite{pezze2018}, and non-Gaussian correlated states such as the N00N state, the Schrodinger cat state, and supersinglets~\cite{cabello2003Review}. These exotic quantum states are critical to various quantum protocols such as entanglement purification~\cite{bennett1996b,bennett1996c}, teleportation~\cite{bennett1996}, remote state preparation~\cite{chaudhary2021}, and clock synchronization~\cite{jozsa2000,ilo-okeke2018}.
	
\begin{figure}[ht!]
		\includegraphics[width=\columnwidth]{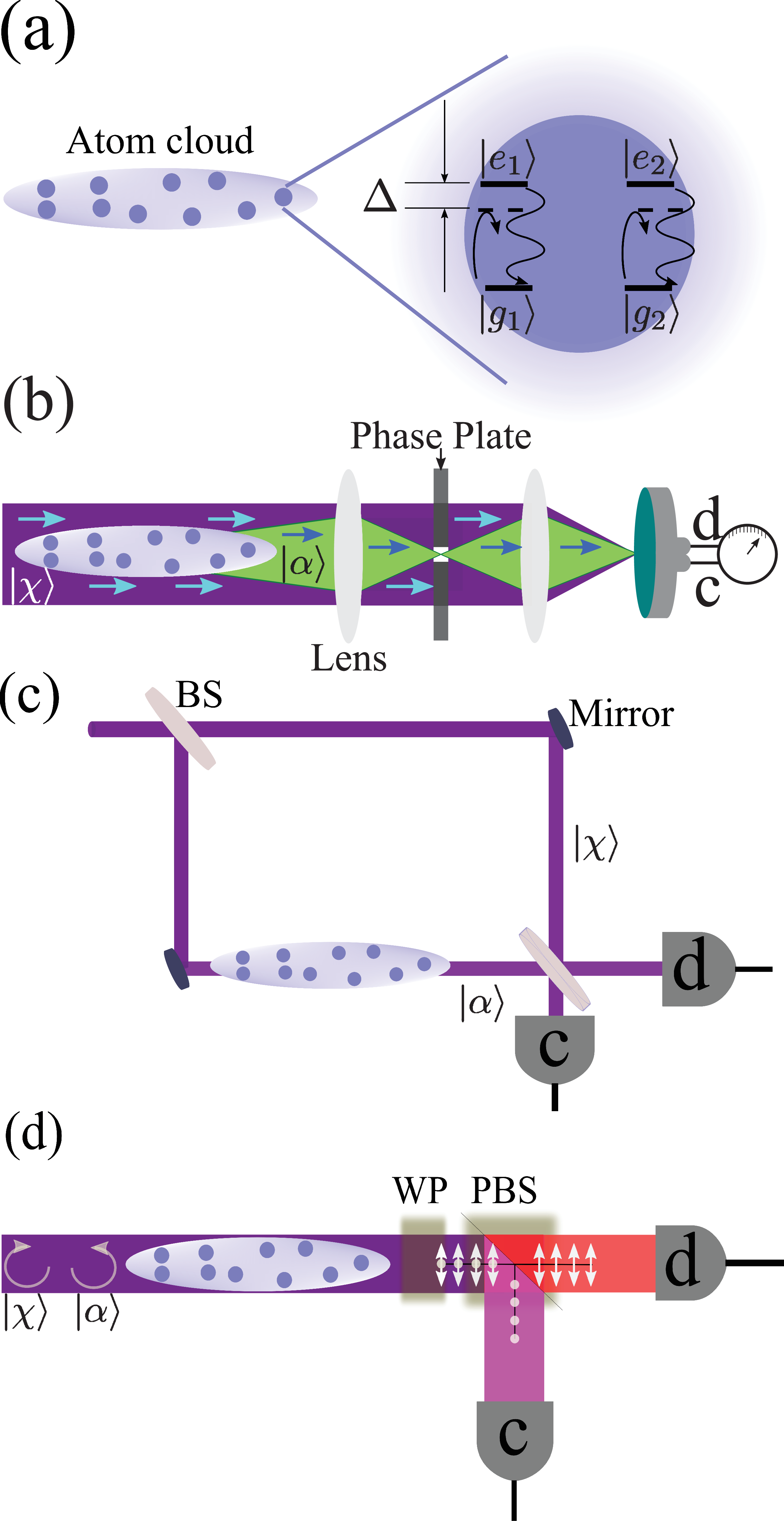}
		\caption{ QND measurement schemes considered in this paper.  Figure (a) shows the possible channels for relaxation of the atom. The wavy line represent spontaneous emission, while the straight line represents second-order off-resonance interactions. Various realizations of QND measurements using (b) phase contrast imaging, (c) Mach-Zehnder geometry, and (d) polarization measurement. }
		\label{fig1}
	\end{figure}

	While QND measurements have been successfully implemented in numerous experiments, they are affected by the spontaneous emission of the atoms~\cite{appel2009,louchet-chauvet2010,lone2015,ilo-okeke2021}. The general principle is as depicted in Fig.~\ref{fig1}, where an atom initially in its ground state interacts with a laser light detuned from the atomic resonance transition. This interaction results in a second-order off-resonance transition, as indicated in Fig.~\ref{fig1}(a). Due to the finite laser linewidth, which has a spectral frequency range, the atom has a finite probability of being resonantly excited, leading to subsequent spontaneous decay. This emission process produces photons with random phases, a range of wavelengths, and different emission directions, causing a reduction in beam intensity reaching the detector. Additionally, an excited atom can access various available states during relaxation. Thus, an excited atom can transition to various lower-lying quantum states, while acquiring residual momentum from the random photon kicks during emission. The spread of available quantum states and the atom's residual momentum contribute to the loss of coherence in its wave function~\cite{allen1975,shore1990v1,carmichael1999}.

	
	In this paper, we formulate the QND measurement using a positive operator valued measure (POVM) (\ref{eq:GeneralizedMeasurementOperator}), taking into account the effects of spontaneous emission, allowing for a convenient, yet  more realistic representation of the quantum state of the atoms from input to output. The strength of the measurement, which is dependent on the atom-light interactions, can be adjusted to be weak or strong, thanks the exact treatment not relying on the Holstein-Primakoff approximation. This tuning of measurement strength realizes different final states, such as squeezed and Schr\"odinger-cat states, from the input state. Notably, our results reveal that high spontaneous emission conditions turn the eigenvalue $m_z =0$ of the total spin operator $J_z$ into the dominant eigenstate of the POVM. This finding complements recent proposals~\cite{ilo-okeke2022,mao2022,kondappan2023} for using measurement and feedback to find the ground state of various spin systems. We note that without spontaneous emission, our approach recovers the results of Ref.~\cite{ilo-okeke2023Dowling}. We note that other works using QND measurements with photon number resolving measurements~\cite{yanagimoto2023} and atom spin squeezing~\cite{bao2020} have developed a POVM formalism. However, the POVMs were developed within the Hamiltonian formalism and thus incapable of describing the measurement outcomes of systems undergoing spontaneous emission.
	
	We have organized the remainder of the paper as follows. We begin by giving a basic introduction to the techniques and the main results of the paper in Sec.~\ref{sec:MainResults}. This is followed by Sec.~\ref{sec:SingleQubit}, where we describe the evolution of a qubit in the presence of spontaneous emission using a master equation and generalize it to many qubit systems. We formally solve the master equation in Sec.~\ref{sec:SolutionMasterEquation}. We check the validity of the solutions in Sec.~\ref{sec:AveragesSystemVariables}. We derive the POVM in the presence of spontaneous emission in Sec.~\ref{sec:MeasurementOperatorGeneral}, and derive the limiting case of the POVM becoming a projection operator in Sec.\ref{sec:NoisyProjectionMeasurement}. In Sec.~\ref{sec:Examples}, we present the effect of spontaneous emission on quantum state preparation. Finally, we provide the summary of our work in Sec.~\ref{sec:summary}

	\section{Main Result of This Paper}\label{sec:MainResults}
	In order to make the results of this paper as accessible as possible, we first present the basic idea of this work and main results of this paper. As illustrated in Fig~\ref{fig1}(a), the state of a two-level atom interacting with light undergoes a Stark shift. Such interactions shift the atom's quantum states while causing an accumulation of phase to light's state. Using the quantum master equation, we account for the spontaneous emission on the combined evolution of the atom's and light's state. 
	
	We consider the initial quantum state of light to be a coherent state. Initially, the atom is in its ground state with a unit probability. Subsequent evolution of the atom's quantum state in the presence of light has a negligible effect on the ground state probability. As such, the ground state evolves slowly due to the applied field. By adiabatically eliminating the excited state of the atom, we find the time dependence of the photon number amplitude. Additionally, we obtain the dephasing master equation governing the evolution of the ground state of an ensemble of two-level atoms and quantum state of light
	\begin{equation*}
		\frac{d\rho}{d\tau} + i\left( H\rho - \rho H^\dagger \right) = R \hat{a}J_z\rho J_z\hat{a}^\dagger,
	\end{equation*} 
	where the dimensionless time $\tau$ is $\tau = gt$, $g$ is the effective detuning, and $R$ is the effective dimensionless spontaneous atomic decay rate. The non-Hermitian Hamiltonian $H$ is 
	\begin{equation*}
		H = J_z \hat{a}^\dagger\hat{a} - i\frac{R}{2}J^2_z\hat{a}^\dagger\hat{a}.
	\end{equation*}
	The operator $\hat{a}$ is the photon annihilation operator, which acting on the vacuum state, destroys it. The operators $\hat{a}$ and $\hat{a}^\dagger$ satisfy the commutation relation $[\hat{a},\hat{a}^\dagger] = 1$. The operator $J_z$ is the total $z$ spin  operator and satisfies the commutation relations $[J_z,J_\pm] = \pm J_\pm $ and $[J_+, J_-] = 2J_z$. 
	
	We obtain the solution $\rho(\tau)$ of the master equation that gives quantum states of atom and light at any time $\tau$ as 
	\begin{equation*}
		\rho(\tau) = e^{-iH\tau}\left( e^{\int_{0}^{\tau} d\tau' \tilde{\mathcal{L}}_s()}\rho(0) \right)e^{iH^\dagger\tau},
	\end{equation*}
	where the initial state of atom and light is product of the atomic quantum state and the coherent states of light $\rho(0) = \rho_{\mathrm{atom}}\otimes\lvert\alpha,\chi\rangle\langle \alpha,\chi\rvert$, and the superoperator $\tilde{\mathcal{L}}_s ()$ is
	\begin{equation*}
		\tilde{\mathcal{L}}_s () = R\hat{a}e^{(-iJ_z - \frac{R}{2} J^2_z)\tau} J_z()J_z e^{(iJ_z - \frac{R}{2} J^2_z)\tau}\hat{a}^\dagger.
	\end{equation*}
	At the beamsplitter, light beams interfere. The unitary operator that describes the action of the beam splitter is $e^{-i\pi S_x /2}$. The operator $S_x$ is the Stokes operator for light. Counting the photons in the light beams by the detectors, $c$ and $d$, collapses the quantum state of light. The collapse of the light state described by a projection operator $\lvert n_c, n_d\rangle\langle n_c, n_d\rvert$ gives the state of the atom ensemble conditioned on counting $n_c$ and $n_d$ photons at the detectors $c$ and $d$, respectively, as
	\begin{equation*}
		\rho_{n_c,n_d} = \langle n_c,n_d\rvert e^{-i\pi S_x /2} \rho(\tau) e^{i\pi S_x /2}\lvert n_c,n_d\rangle.
	\end{equation*}
	The sequence of operations that gave rise to $\rho_{n_c,n_d}$ can be described by a generalized measurement operator
	\begin{equation*}
		\hat{\mathcal{M}}_{n_c,n_d}() = \hat{M}_{n_c,n_d} \mathcal{L}_s()\hat{M}^\dagger_{n_c,n_d}
	\end{equation*}
	where
	\begin{equation*}
		\hat{M}_{n_c,n_d} = \langle n_c,n_d \rvert e^{-i\pi S_x /2} e^{-iH\tau}\lvert\alpha,\chi\rangle,
	\end{equation*}
	similarly for $\hat{M}^\dagger_{n_c,n_d}$, and 
	\begin{equation*}
		\mathcal{L}_s() = e^{\int_{0}^{\tau}d\tau'\, R\,  \alpha e^{(-i J_z - \frac{R}{2} J^2_z)\tau'} J_z() J_ze^{(i J_z - \frac{R}{2} J^2_z)\tau'}\alpha^* }.
	\end{equation*}
	The explicit form of the operator $\hat{\mathcal{M}}_{n_c,n_d}()$ is given in Sec.~\ref{sec:MeasurementOperatorGeneral}, specifically in Eq.~(\ref{eq:GeneralizedMeasurementOperator}). The preceding equation $\hat{\mathcal{M}}_{n_c,n_d}()$ is the main result of this manuscript. The operator $\hat{\mathcal{M}}_{n_c,n_d}()$ accounts for the effects of spontaneous emission on a QND measurement of a quantum state of an atom ensemble. It takes the quantum state of an atom ensemble at the input $\rho_{\mathrm{atom}}$ and gives the output state $\rho_{n_c,n_d}$. 
	
	In the remainder of the manuscript, we give a detailed derivation of the operator $\hat{\mathcal{M}}_{n_c,n_d}()$, examine its properties, and use it to examine the effects of spontaneous emission on the correlated state preparation of atoms by calculating the the probability density distributions, and the quasiprobability distributions of quantum state of the atom ensemble. Additionally, we show that due to spontaneous emission, the measurement drives quantum state of an atomic ensemble towards an eigenvalue of $J_z$, $m_z =0$, for a given total spin $J$ value. For a spin system with more than one spin $J$ value, it drives it to a state with the minimum $J$ value for the same eigenvalue of $J_z$, $m_z=0$.

	\section{The Master Equation }\label{sec:SingleQubit}
	In this section, we consider the case of two ground states of 	an atom initially in a linear superposition, each interacting with a different excited state~\cite{higbie2005,appel2009}. We give a detailed derivation evolution of the ground states under adiabatic elimination conditions in Appendix~\ref{sec:GroundStateEvolution}. Here, by writing (\ref{eq:MasterEquation}) in terms of their relative population difference, we obtain a master equation governing the evolution of their coherences.  We then generalize the result to that of the \emph{N}-atom ensemble.

	Consider an atom with two ground states as shown in Fig.~\ref{fig1}(a). The evolution of each ground state within the adiabatic elimination is given by (\ref{eq:MasterEquation})
	\begin{equation}
		\label{eq:MasterEquationII} 
		\dot{\rho} = \sum_{k=1}^{2}-\frac{i}{\hbar} \left(\tilde{H}_{\mathrm{eff},k}\rho -\rho \tilde{H}^\dagger_{\mathrm{eff},k}\right) + \tilde{\gamma}_{\mathrm{eff},k}\hat{a}\lvert g_k\rangle\langle g_k\rvert \rho\lvert g_k\rangle \langle g_k\rvert\hat{a}^\dagger,
	\end{equation}
	where the non-Hermitian Hamiltonian is 
	\begin{equation}
		\label{eq:EffectiveHamitonianII}
		\tilde{H}_{\mathrm{eff},k} = -\hbar G_k \hat{a}^\dagger\hat{a}\lvert g_k\rangle\langle g_k \lvert - i\hbar\frac{\tilde{\gamma}_{\mathrm{eff},k}}{2}\hat{a}^\dagger\hat{a}(\lvert g_k\rangle\langle g_k \rvert)^2.
	\end{equation}
	Using the completeness relation, and atomic inversion $\hat{\sigma}_z$ operator between the two ground states 
	\begin{align}
		\label{eq:TotalPopulationOperator}
		\hat{\mathbbm{1}} & = \lvert g_1\rangle\langle g_1\rvert + \lvert g_2\rangle\langle g_2\rvert,\\
		\hat{\sigma}_z&  = \frac{\lvert g_1\rangle\langle g_1\rvert - \lvert g_2\rangle\langle g_2\rvert}{2},
	\end{align}
	and averaging over the light states for terms that contain the identity operator in (\ref{eq:MasterEquationII}), one arrives at the following effective master equation for an atom interacting with laser light,
	\begin{equation}
		\label{eq:EffectiveMasterEquation}
		\frac{d\rho}{d\tau} + i\left(H_\mathrm{eff} \rho - \rho H_\mathrm{eff}^\dagger \right) =   R\,  \hat{a} \hat{\sigma}_z\rho \hat{\sigma}_z\hat{a}^\dagger,
	\end{equation} 
	where the effective detuning $g =- (G_1 - G_2)$, gives the phase shift. The dimensionless time is $\tau = g t$. The parameter $R$ is the effective dimensionless spontaneous atomic decay rate  $R = \left(\tilde{\gamma}_{\mathrm{eff},1}	+ \tilde{\gamma}_{\mathrm{eff},2}\right)/g$,  while the effective non-Hermitian Hamiltonian is
	\begin{equation}
		\label{eq:EffectiveHamiltonian}
		H_\mathrm{eff} =  \hat{\sigma}_z\hat{a}^\dagger\hat{a} - i\frac{R}{2} \hat{\sigma}_z^2 \hat{a}^\dagger\hat{a}.
	\end{equation}
	
	\subsection{The Ensemble Case}\label{sec:sec:AtomEnsemble}
	For an ensemble consisting of $N$-atoms, the evolution of the ensemble state is the same as the product of individual atom states, $\rho_\mathrm{atom} = \rho_1\otimes\rho_2\otimes\cdots\otimes \rho_N$. Let the Pauli operator acting on the \emph{k}th atom be $\hat{\sigma}_i^{(k)}$ where $ i \in \{x,y,z\} $. We define the total spin operators $J_x = (J_+ + J_-)/2$, $J_y = (J_+ - J_-)/(2i)$ where  
	\begin{equation}
		\label{eq:TotalSpinOperators}
		\begin{split}
			J_+ = \sum_{k=1}^{N} \hat{\sigma}_+^{(k)},\qquad
			J_- = \sum_{k=1}^{N} \hat{\sigma}_-^{(k)},
		\end{split}
	\end{equation}
	$\hat{\sigma}^{(k)}_- = \lvert g_2^k\rangle\langle g_1^k\rvert$ is the 	the lowering, $\hat{\sigma}_+^{(k)} = \lvert g_1^k\rangle\langle g_2^k\rvert$ is the raising operators and the total $z$-spin operator $J_z$ is 
	\begin{equation}
		\label{eq:TotalInversionSpinOperator}
		J_z = \sum_{k =1}^{N} \hat{\sigma}_{z}^{(k)}.
	\end{equation} 
	The operators $J_z$ and $J_\pm$ satisfy the following commutation relations
	\begin{equation}
		\label{eq:commutationrelation}
		\left[J_\pm,J_z \right] = \mp J_\pm, \quad \left[ J_+,J_-\right] = 2J_z.
	\end{equation}
	The evolution of a qubit atom ensemble interacting with coherent light becomes
	\begin{equation}
		\label{eq:EffectiveMasterEquationII}
		\frac{d\rho}{d\tau} + i\left(H_\mathrm{eff} \rho - \rho H_\mathrm{eff}^\dagger \right) =   R\,  \hat{a} J_z\rho J_z\hat{a}^\dagger,
	\end{equation} 
	where the non-Hermitian Hamiltonian is
	\begin{equation}
		\label{eq:EffectiveHamiltonianII}
		H_\mathrm{eff} =  J_z\hat{a}^\dagger\hat{a} - i\frac{R}{2} J_z^2 \hat{a}^\dagger\hat{a}.
	\end{equation}
	
	The atom qubit ensemble consists of \emph{N} atoms, with the maximum spin angular momentum $J_\mathrm{max} = N /2$, the minimum spin angular momentum  $J_\mathrm{min} = 0$ for $N$ even and $J_\mathrm{min} = 1/2$ for \emph{N} odd, and $J$ takes on values $J = J_\mathrm{min}, \cdots , J_\mathrm{max}$. The eigenstates are denoted $J_z \lvert J, m_z \rangle = m_z \lvert J, m_z \rangle$, where
	$m_z = -J, -J + 1, \cdots, J$. To obtain the behavior of the density matrix in (\ref{eq:EffectiveMasterEquationII}), we use the ansatz 
	\begin{equation}
		\label{eq:ansatz}
		\rho = \sum_{m_z = -J}^{J}\sum_{m'_z=-J}^{J}\rho_{m_z,m'_z} \lvert J,m_z\rangle\langle J, m'_z\rvert \otimes \lvert \alpha_{m_z}\rangle\langle\alpha_{m'_z}\rvert,
	\end{equation}
	where $\rho_{m_z,m'_z}$ describes the matrix elements of the atomic spin, while $\alpha$ is the photon number amplitude, and  $\lvert\alpha_{m_z}\rangle = e^{-|\alpha_{m_z}|^2/2} e^{\alpha_{m_z} a^\dagger}\lvert 0 \rangle$ is the light coherent state index by the the eigenvalues of $J_z$.

	\section{Solution To The Master Equation}\label{sec:SolutionMasterEquation}
	Usually, solving the master equation is done by specifying the matrix elements. For instance, a density matrix containing $N$ elements would require solving $N$ coupled equations with a coupling matrix of dimension $N\times N$. More so, not all the density matrix elements are coupled to each other. This technique may not bode well for many applications and does not allow easy interpretation of the underlying dynamics. The matrix approach is popular because it eliminates the complexity brought about by the action of the operators which may not be tractable. Here, we solve the master equation (\ref{eq:EffectiveMasterEquationII}) to obtain the evolution of the spin qubit ensemble and light state using formal techniques for solving the first-order differential equations. It allows one to solve the density matrix equation more generally within its space, and cuts down dramatically the computational time when compared to matrix element approach. We also present the solution obtained using the matrix method approach in Appendix~\ref{sec:SolutionAtomEnsemble}.

	\subsection{Formal Solution To The Master Equation}\label{sec:FormalSolution}
	We present the detailed derivation of the formal solution in  Appendix~\ref{sec:FormalSolutionMasterEquation}, and only summarize the results here. In seeking the solution of the master equation (\ref{eq:EffectiveMasterEquationII}), we observe that its right hand side behaves as a source. However, the homogenous part of the equation is linear in $\rho$. Thus, moving into the frame rotating at a rate given by non-Hermitian Hamiltonian (\ref{eq:EffectiveHamiltonianII}) and integrating the resulting equation gives the solution
	\begin{equation}
		\label{eq:GeneralSolution}
		\rho(\tau) = e^{-iH_\mathrm{eff}\tau} \left( e^{\int_{0}^{\tau}d\tau'\,  \tilde{\mathcal{L}}_s() }\rho(0)\right)  e^{i H^\dagger_\mathrm{eff}\tau},
	\end{equation}
 	where the superoperator~\cite{breuer2002,carmichael2008} $\tilde{\mathcal{L}}_s$ is
 	\begin{equation}
 		\label{eq:TransformedSuperopertor}
 		\tilde{\mathcal{L}}_s() = R\,  \hat{a}e^{(-i J_z - \frac{R}{2} J^2_z)\tau} J_z() J_ze^{(i J_z - \frac{R}{2} J^2_z)\tau}\hat{a}^\dagger.
 	\end{equation}
	The same solution (\ref{eq:GeneralSolution}) is obtained using the ordering of   operators, see Appendix~\ref{sec:DisentanglingOperator}. The action of the exponential superoperator $e^{\int_{0}^{\tau}d\tau\,  \tilde{\mathcal{L}}_s() }$ on the initial state $\rho(0)$ is as given in (\ref{eq:SolutionTransformedState}). Because the coherent state is an eigenstate of the $\hat{a}$, the photon operators $\hat{a},\,\hat{a}^\dagger$ appearing in the superoperator do not affect the state of photons. Their effect is to contribute the statistics of the photons to the quantum state of the atoms. It is the propagators $e^{\frac{-i}{\hbar} H_\mathrm{eff}\tau}, \,e^{\frac{i}{\hbar} H^*_\mathrm{eff}\tau},$ that modify the quantum state of photons by taking its initial state at time $\tau=0$ to a different state at time $\tau$. Substituting (\ref{eq:GeneralSolution}) in (\ref{eq:EffectiveMasterEquationII}) shows that it is indeed the solution. Equation (\ref{eq:GeneralSolution}) is the focus of analysis in Sec.~\ref{sec:AveragesSystemVariables}, and it will be used to formulate the measurement operator in Sec.~\ref{sec:MeasurementOperatorGeneral}. 
	
	\section{Averages Of System Variables} \label{sec:AveragesSystemVariables}
	We begin by showing that the solution obtained using matrix elements and the formal solution are the same. The solution of the density matrix by the matrix element approach is given in Appendix \ref{sec:SolutionAtomEnsemble}.
	
	\begin{figure}[t]
		\includegraphics[width=\columnwidth]{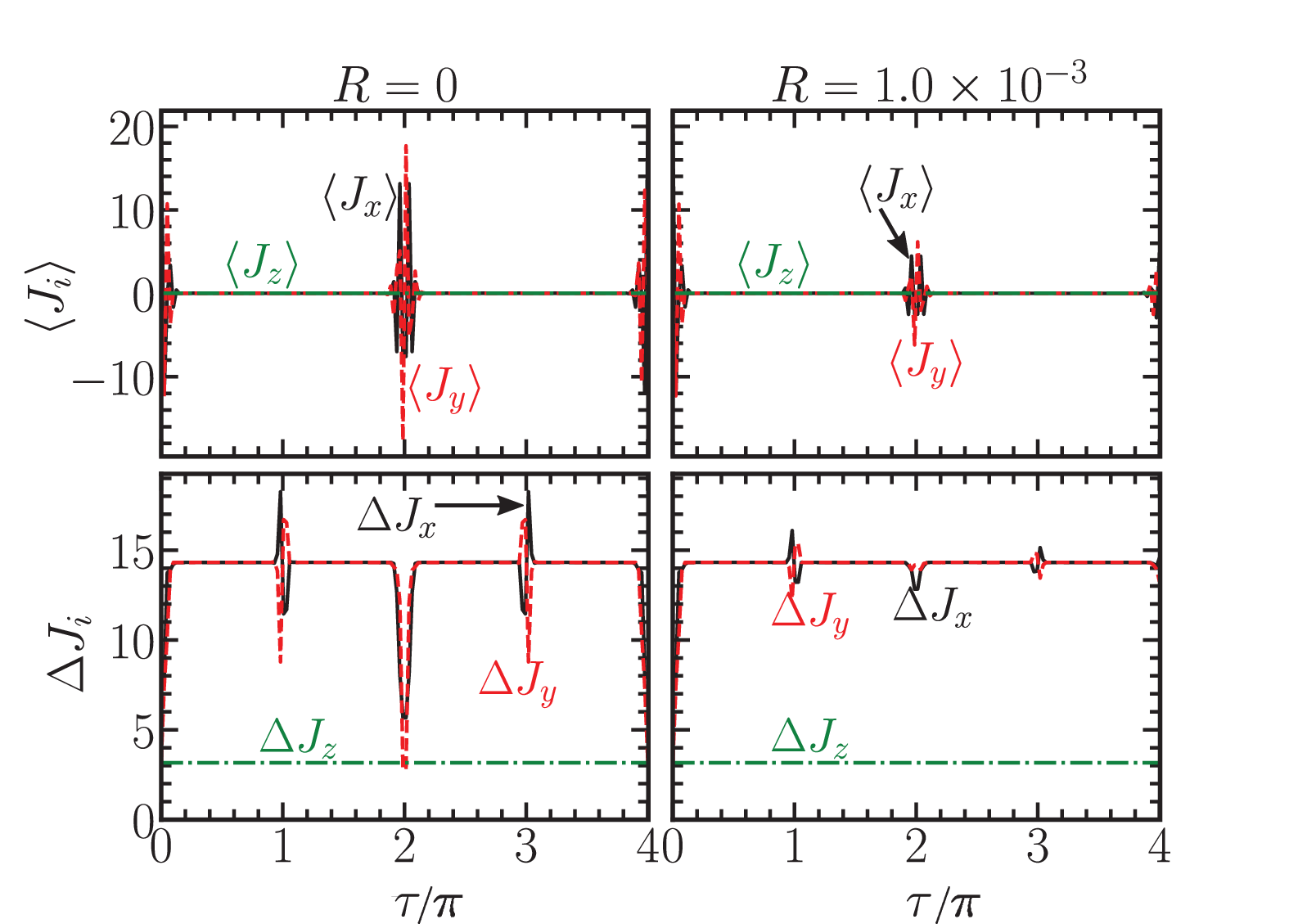}
		\caption{The spin averages with an initial spin coherent state polarized along the \emph{x}-axis. In the first row is the evolution of the mean spin variables $J_i$, $i = x,\,y, \, z$, while the second row is their standard deviation. The parameters of the figure are  $\alpha =\sqrt{50} $, $J = 20$, and the parameter $R$ is as shown in the figure. The mean spin values  and their variances are calculated using (\ref{eq:AtomReducedDensityMatrix}) in (\ref{eq:OperatorMean}) and (\ref{eq:OperatorVariance}), respectively.}
		\label{fig2}
	\end{figure}
	%
	
	\subsection{Examples of Atomic States \label{sec:sec:InitialState}}
	The theory we develop here applies to any quantum state. For illustrative purposes, we consider the spin coherent and thermal states. The spin coherent state~\cite{arecchi1972,byrnes2021} is defined as
	\begin{align}
		\label{eq:CoherentState}
		\lvert \theta, \phi\rangle & = \sum_{m_z = -J}^{J} \binom{2J}{J+m_z}^{\frac{1}{2}} \sin^{J+m_z}\left(\frac{\theta}{2}\right)\nonumber\\
		&\times \cos^{J-m_z}\left(\frac{\theta}{2}\right) e^{-i(J+m_z)\phi} \lvert J, m_z\rangle,
	\end{align}
	where $\theta$ and $\phi$ are polar and the azimuthal angle on a Bloch sphere, and   $\binom{n}{k}$ is binomial function.
	
	The other state we consider is the thermal state 
	\begin{align}
		\label{eq:ThermalState}
		\rho^\mathrm{th}_0 =  \frac{1}{\mathcal{N}}\sum_{J = J_\mathrm{min} }^{J_\mathrm{max}} \sum_{m_z = -J}^{J}\dfrac{e^{-E_0 J_z }}{P_J}  \lvert J,m_z\rangle\langle J, m_z\rvert,
	\end{align} 
	where $P_J = \sum_{m_z=-J}^{J} \langle J, m_z\rvert e^{-E_0 J_z } \lvert J,m_z\rangle$, $J_\mathrm{min} = 0\, \mathrm{or}\, 1/2$ depending on the total number of atoms $N$ is even or odd, and $\mathcal{N} = J_\mathrm{max} + 1$ for $J_\mathrm{max}$ being an integer spin and $\mathcal{N} = J_\mathrm{max} + \frac{1}{2}$ for $J_\mathrm{max}$ being a fractional spin. The dimensionless parameter $E_0$ is defined as $\hbar \omega/(k T)$, where $\hbar\omega $ is the characteristic energy of the system, $k$ is the Boltzmann constant, and $T$ is temperature.
	
	\subsection{Averages Of System Operators} \label{sec:sec:AveragesOperators}
	To place the comparison on the same footing, we use  (\ref{eq:ansatz}) which upon substituting in (\ref{eq:GeneralSolution}) gives 
	\begin{equation}
		\label{eq:formal}
		\begin{split}
			\rho(\tau)& = \sum_{m_z=-J}^{J}\sum_{m'_z=-J}^{J}\exp\bigg[ \frac{R\lvert\alpha\rvert^2 m_z m'_z}{i(m'_z - m_z) -\frac{R}{2}(m'^2_z + m_z^2)}\\
			&\times \left(e^{i\tau(m'_z - m_z) -\frac{R}{2}(m'^2_z + m_z^2)} - 1\right)\bigg]\rho_{m_z,m'_z}(0)\\
			&\lvert J, m_z\rangle\langle J,m'_z\rvert\otimes
			e^{\left[\frac{\lvert\alpha\rvert^2}{2}\left(e^{-R\tau m_z^2} + e^{-R\tau m'^2_z} -2 \right)  \right]}\\
			& \times\lvert\alpha_{m_z}\rangle\langle \alpha_{m'_z}\rvert,
		\end{split}
	\end{equation} 
	where $\rho_{m_z,m'_z} = \rho_{m_z,m'_z}(0)$ and $\alpha_{m_z}$ is 
	\begin{equation}
		\label{eq:PhotonAmplitude}
		\alpha_{m_z} = \alpha e^{-im_z\tau - \frac{R}{2}m_z^2 \tau}.
	\end{equation}
	The parameter $\alpha$ is the initial photon number amplitude. Comparing the scalar functions in (\ref{eq:formal}) to (\ref{eq:MatrixElementSolution}), we see that they are the same.  
	
	The knowledge of the density matrix allows for the averages such as the mean and standard deviation to be calculated. For instance, the mean value of any operator is calculated by performing the trace over the matrix product 
	\begin{equation}
		\label{eq:OperatorMean}
		\langle \hat{O} \rangle = \mathrm{Tr}\left(\rho(\tau)\hat{O}\right).
	\end{equation}
	Similarly, the standard deviation of the operator $\Delta \hat{O}$ is taken as the square root of the variance $(\Delta \hat{O})^2$ defined as
	\begin{equation}
		\label{eq:OperatorVariance}
		\left(\Delta \hat{O} \right)^2= \mathrm{Tr}\left(\rho(\tau)\hat{O}^2\right) -\left(\mathrm{Tr}\left(\rho(\tau)\hat{O}\right)\right)^2.
	\end{equation}
	
	\subsection{Averages Of Spin Variables} \label{sec:sec:RedeucedDensityMatrix}
	The order in which to trace over the systems does not matter. Choosing to study the effects on the spin variables, we trace over the photon states and find the reduced density matrix of the atoms $\rho_\mathrm{atom} = \mathrm{Tr}_\mathrm{light}[\rho(t)]$
	\begin{equation}
		\label{eq:AtomReducedDensityMatrix}
		\begin{split}
			\rho_{\mathrm{atom}}(t) &= \sum_{m_z=-J}^{J}\sum_{m'_z = -J}^{J} \rho_{m_z,m'_z}(0) \\
			&\times \exp\Bigg[ \left(1 + \frac{R m'_zm_z}{i(m'_z-m_z) - \frac{R}{2}(m'^2_z + m_z^2)}\right)\\
			& \times \lvert\alpha\rvert^2e^{i\tau (m'_z-m_z) -\frac{R}{2}\tau (m^2_z + m'^2_z)}  \Bigg]\lvert J,m_z\rangle\langle J,m'_z\lvert.
		\end{split}
	\end{equation}
	For the diagonal terms $m'_z=m_z$, evaluation of the exponential factor shows that the spin state is not affected by spontaneous emission. The spin averages, calculated using the reduced density matrix with the spin coherent state initially polarized along the \emph{x} axis, are presented in Fig.~\ref{fig2}. In the absence of spontaneous emission $R=0$, the atoms maintain their coherence marked by oscillations in averages of their variables, as shown in the first column of Fig.~\ref{fig2}. These oscillations would experience the collapse (loss) and revival (resurgence) as a result of coherent interaction between the atoms and light. The loss of Rabi oscillation is due to the coherent interaction, which causes each spin state to evolve at a different rate. Averaging over the different evolution rates results in their interference and leads to the loss of coherence exhibited by  the spin states. The revival of coherent oscillations is due to the discrete nature of spins. Spontaneous emission $R> 0$ erodes the revival of coherent Rabi oscillations, even in modest amounts, as shown in second column of Fig.~\ref{fig2}.

	\begin{figure}[t]
		\includegraphics[width=\columnwidth]{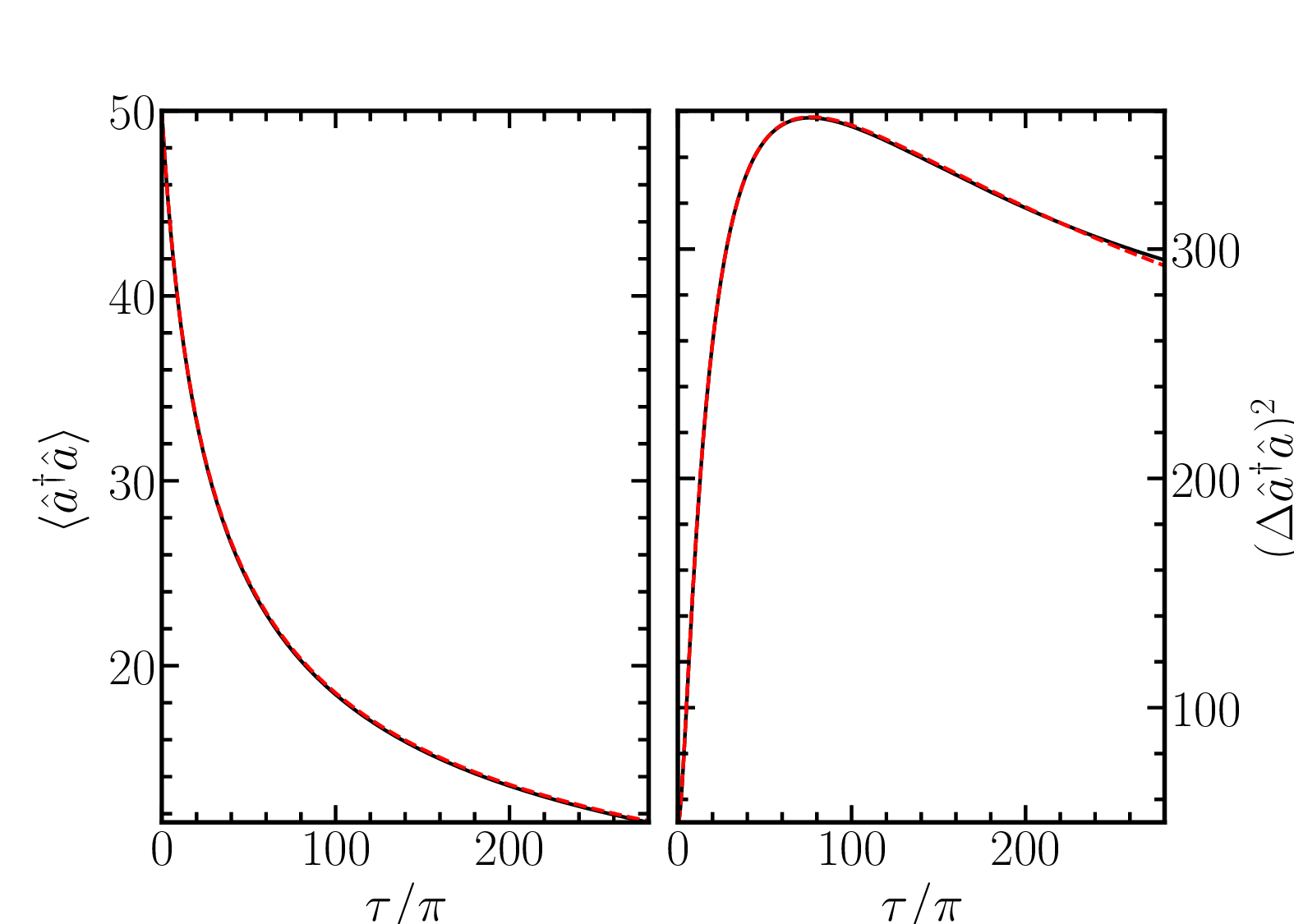}
		\caption{The photon number averages. The left subfigure shows the evolution of the mean photon number, while right subfigure is its variance. The initial photon number amplitude is $\alpha =\sqrt{50} $, the maximum spin value is $J = 20$, and the dimensionless spontaneous emission rate is $R = 0.001$. The mean photon number (solid line) and its variance (solid line) are calculated using (\ref{eq:LightReducedDensityMatrix}) in (\ref{eq:OperatorMean}) and (\ref{eq:OperatorVariance}), respectively. The dashed lines are calculated using~(\ref{eq:AveragePhotoNumber}) for the mean value and~(\ref{eq:PhotonNumberVariance}) for the variance.}  
		\label{fig3}
	\end{figure}
	
	We note that the spin variable $J_z$ commutes with the master equation (\ref{eq:EffectiveMasterEquation}). Hence the $J_z$ is a constant of motion, and its eigenstates are not affected by the spontaneous emission. It manifests in Fig.~\ref{fig2}, where the spin variable $J_z$ is not affected by spontaneous emission. Also, the fixed points of the differential equation play a role in searching for their solution~\cite{boyce2005,boas2006}. The resulting equation obtained by setting time derivative of the master equation to zero gives its the fixed points. Expanding the spin states in the eigenbasis of $J_z$ [see for instance~(\ref{eq:ansatz})], one finds that the master equation vanishes for $m'_z = m_z$. However, there is an eigenvalue for which all the terms in the master equation \emph{independently} vanish and happens for $m'_z = m_z = 0$. In the preceding section, we will discover that this eigenvalue plays a crucial role in the measurement operator, as it leads to the state $\lvert J, m_z = 0\rangle$ being the only available state under high spontaneous emission conditions. This property becomes particularly interesting when considering Refs.~\cite{ilo-okeke2022,kondappan2023,mao2022,ilo-okeke2023Dowling,chaudhary2023} proposing to use quantum nondemolition measurement and unitary feedback for quantum state preparation.

	\subsection{Averages Of Photon Number Amplitude} \label{sec:sec:RedeucedDensityLight}
	Similarly, the reduced density matrix of light obtained by tracing over the quantum state of the atoms  $\rho_\mathrm{light} = \mathrm{Tr}_\mathrm{atom}[\rho(t)]$ is 	
	\begin{equation}
		\label{eq:LightReducedDensityMatrix}
		\rho_{\mathrm{light}}(t) = \sum_{m_z=-J}^{J}\rho_{m_z,m_z}(0)\lvert\alpha_{m_z}\rangle\langle\alpha_{m_z}\rvert,
	\end{equation}
	where $\alpha_{m_z}$ is defined in (\ref{eq:PhotonAmplitude}).	Using the reduced density matrix, we calculate the evolution of the averages of photon number $\hat{a}^\dagger \hat{a}$, their mean~(\ref{eq:OperatorMean}) and variance~(\ref{eq:OperatorVariance}). Notice that the state of light is weighted by the initial probability density $\rho_{m_z,m_z}(0)$ of the atom distribution. For an atomic spin coherent state, the initial probability density $\rho_{m_z,m_z}(0)$ which is a binomial function~(\ref{eq:CoherentState}) may be put in a Gaussian form~\cite{ilo-okeke2010,ilo-okeke2016,ilo-okeke2021} in the limit of large population $N$ of atoms. The average photon number $\langle \hat{a}^\dagger\hat{a}\rangle$ calculated using the spin coherent state (\ref{eq:CoherentState}) in the large atom number limit is
	\begin{equation}
		\label{eq:AveragePhotoNumber}
		\langle \hat{a}^\dagger\hat{a}\rangle = \frac{\lvert \alpha\rvert^2}{\sqrt{1 + JR\tau\sin^2\theta}} e^{-\frac{J^2R\tau\cos^2\theta}{1+J R\tau\sin^2\theta}}.
	\end{equation}
	Similarly, the variance $\Delta\hat{a}^\dagger\hat{a}$ is
	\begin{equation}
		\label{eq:PhotonNumberVariance}
		\begin{split}
			\Delta\hat{a}^\dagger\hat{a}  = &\frac{\lvert\alpha\rvert^4}{\sqrt{1 + 2J R\tau\sin^2\theta}}e^{-\frac{2J^2R\tau\cos^2\theta}{1 + 2JR\tau\sin^2\theta}}\\ & -  \frac{\lvert \alpha\rvert^4}{1 + JR\tau\sin^2\theta} e^{-\frac{2J^2R\tau\cos^2\theta}{1+J R\tau\sin^2\theta}}\\
			& +  \frac{\lvert \alpha\rvert^2}{\sqrt{1 + JR\tau\sin^2\theta}} e^{-\frac{J^2R\tau\cos^2\theta}{1+J R\tau\sin^2\theta}}.
		\end{split}
	\end{equation}
	The first and second terms in Eq.~(\ref{eq:PhotonNumberVariance}) nearly dominate the behavior of the variance, thus determining  the essential features of the variance. Notice for $R=0$, however, one recovers the well known result, that the variance is equal to the average photon number, a consequence of the initial photon distribution being Poisson distribution.  From Eq.~(\ref{eq:AveragePhotoNumber}), it becomes clear that the intensity of the light beam that has interacted with the atoms depends on the population $N$ of the atoms and the initial statistics of the atom distribution. For instance, states for which $\theta =\pi/2$ do not suffer exponential decay rather they decay as an inverse of square root of $N$, $\langle \hat{a}^\dagger\hat{ a}\rangle = \lvert\alpha\rvert^2/\sqrt{1 + JR\tau}$. On the contrary, states for which $\theta =0\, \mathrm{or}\, \pi$ decay exponentially at a rate proportional $N^2$, $e^{-\propto N^2}$. For just one atom qubit and $\theta =0$, we recover the result of Eq.~(\ref{eq:AlphaSolution}).
	
	The results of photon number averages for $\rho_{m_z,m_z}(0)$ corresponding to that of a spin coherent state polarized along the x-axis are shown in Fig.~\ref{fig3}. For $R = 0$, the result is trivial in that the average photon number suffers no decay since the reduced density matrix of light has only components of spins in the same state $m'_z = m_z$. As a result, the photon operators $\hat{a}^\dagger,\,\hat{a}$ accumulate a phase dependent on the same spin state, see (\ref{eq:PhotonAmplitude}), such that their absolute value does not depend on any spin value and thus is a constant---the same as the initial value. However, for spontaneous emission $R > 0$, the photon operators decay in addition to the accumulated phase. Thus average photon number is affected by the spontaneous emission, as shown in Fig.~\ref{fig3}. In particular, the average photon number decays but not exponentially due to the type of initial state of the atoms, as we previously discussed. For instance, from (\ref{eq:AveragePhotoNumber}) the average photon number decays as $\langle\hat{a}^\dagger\hat{a}\rangle = \left(1 + JR\tau\right)^{-1/2}$. Thus in the short time limit and $R\ll 1$, the average photon number decays linearly, $\langle \hat{a}^\dagger\hat{a}\rangle \approx \lvert\alpha\rvert^2(1 - JR\tau/2)$, while at long time it decays as $\langle\hat{a}^\dagger\hat{a}\rangle \propto 1/\sqrt{JR\tau}$. To understand qualitatively the behavior of the variance, we look at the analytic result~(\ref{eq:PhotonNumberVariance}). For instance, the variance increases quadratically as $J^2R^2\tau^2/2$ in the region where the the average photon number is decreasing linearly for $\tau >0$. In this region the contribution from the first term is large compared to that of the second term of Eq.~(\ref{eq:PhotonNumberVariance}). However, in the region where the average photon number changes rapidly for a small change in time, the second term is large and negative leading to substantial cancellation with the first term of (\ref{eq:PhotonNumberVariance}). This cancellation brings about the concave downward curve of the variance. Hence, in general the behaviour of the variance is governed by the first two terms of (\ref{eq:PhotonNumberVariance}).

	\section{Dephasing Noise In Measurement Operator \label{sec:MeasurementOperatorGeneral}}
	In the last section, we obtained the time evolution of the density matrix $\rho(t)$ in the presence of spontaneous emission of photons by atoms in a random direction. This leads to the loss of photons manifesting as decay in the photon number amplitude. Additionally, the spontaneous emission of photons led to the dephasing of the atomic states and manifests as loss of atomic coherence. In this section, we derive the measurement operator that accounts for the effects of spontaneous emission of atoms. We characterize the operator by finding its most probable outcome, which is directly related to inference in a measurement. We also give the width of the measurement operator that influences the allowable quantum state obtained from a measurement.
	
	The measurement operator derived here applies to the scenarios shown in Fig.~\ref{fig1}(b) -- (d). For specificity, consider the setup shown in Fig.~\ref{fig1}(c). The state of light and atom in the arm of the Mach-Zehnder interferometer labeled $\lvert\alpha\rangle$ is given by (\ref{eq:GeneralSolution}). Given that the initial state of light is coherent, the phase of the light field in the state, $e^{-i\hat{H}_\mathrm{eff}\tau}\lvert\alpha\rangle$, carries the information about the atoms. At the beamsplitter, the light in the state $e^{-i\hat{H}_\mathrm{eff}\tau}\lvert\alpha\rangle$ interferes with that in the state $\lvert\chi\rangle=e^{-\frac{\lvert\chi\rvert^2}{2}} e^{\chi \hat{b}^\dagger}\lvert 0\rangle$, thereby turning the phase information into amplitude oscillations of light according to the following transformation
	\begin{equation}
		\label{eq:BeamSplitter}
		\hat{a}^\dagger = \frac{\hat{c}^\dagger + i\hat{d}^\dagger}{\sqrt{2}},\qquad \hat{b}^\dagger = \frac{i\hat{c}^\dagger + \hat{d}^\dagger}{\sqrt{2}}.
	\end{equation}
	Denoting the beamsplitter operation by a unitary operator $\hat{U}_\mathrm{BS}$, the detection of $n_c$ and $n_d$ photons at the detectors $c$ and $d$, respectively, conditions the state of the atoms to become  $\rho_{n_c,n_d}= \langle n_c,n_d\rvert \hat{U}_\mathrm{BS}\rho(t)\otimes\lvert\chi\rangle\langle\chi\rvert \hat{U}^\dagger_\mathrm{BS}\lvert n_c,n_d\rangle$,
	\begin{align}
		\label{eq:NormalizedStateBEC}
		\rho_{n_c.n_d} & =\frac{ e^{-(\lvert \chi\rvert^2 + \lvert\alpha\rvert^2)}}{n_c!n_d!} \left(\frac{\alpha e^{-i(\tau J_z -i\frac{R\tau}{2} J_z^2) } + i\chi}{\sqrt{2}}\right)^{n_c}\nonumber\\
		&\times \left(\frac{i\alpha e^{-i(\tau  J_z - i\frac{R\tau}{2} J_z^2) } + \chi}{\sqrt{2}}\right)^{n_d} \nonumber\\
		&\times e^{R \lvert\alpha\rvert^2\int_{0}^{\tau}d\tau'e^{w\tau'}J_z ()J_ze^{w*\tau'} } \rho_{\mathrm{atom}}\nonumber\\
		&\times \left(\frac{\alpha^* e^{i(\tau J_z +i\frac{R\tau}{2} J_z^2) } - i\chi^*}{\sqrt{2}}\right)^{n_c}\nonumber\\
		&\times \left(\frac{-i\alpha^* e^{i(\tau J_z +i\frac{R\tau}{2} J_z^2) } + \chi^*}{\sqrt{2}}\right)^{n_d},
	\end{align}
	where $w = -i J_z - \dfrac{R}{2} J^2_z.$
	
	Clearly the sequence can be described by a generalized measurement operator $\rho_{n_c.n_d} = \hat{\mathcal{M}}_{n_c,n_d} \rho_{\mathrm{atom}}$, where
	\begin{equation}
		\label{eq:GeneralizedMeasurementOperator}
		\hat{\mathcal{M}}_{n_c,n_d} () = \hat{M}_{n_c,n_d} \mathcal{L}_s()\hat{M}^\dagger_{n_c,n_d},
	\end{equation}
	 and the superoperator $\mathcal{L}_s ()$ is 
	\begin{equation}
		\label{eq:statepreparer}
		\mathcal{L}_s () = e^{R\lvert\alpha\rvert^2\,\int_{0}^{\tau}d\tau'\,   e^{(-i J_z - \frac{R}{2} J^2_z)\tau'} J_z() J_ze^{(i J_z - \frac{R}{2} J^2_z)\tau'} }.
	\end{equation}
 	By its action, the superoperator $\mathcal{L}_s()$ prepares from the initial state of the atoms a state that grows at a rate proportional to $R\lvert\alpha\rvert^2$. Notice that the superoperator does not have dependence on $n_c$ and $n_d$ nor does it have any of the photon operators. Hence, the superoperator does not directly contribute to the detection of photons. The operator  $\hat{M}_{n_c,n_d}$ 
	\begin{equation}
		\label{eq:measurementoperatorII}
		\begin{split}
			\hat{M}_{n_c,n_d} & = \frac{ e^{-(\frac{\lvert \chi\rvert^2 + \lvert\alpha\rvert^2}{2})}}{\sqrt{n_c!}\sqrt{n_d!}} \left(\frac{\alpha e^{-i(\tau J_z -i\frac{R \tau}{2} J_z^2) } + i\chi}{\sqrt{2}}\right)^{n_c}\\
			&\times \left(\frac{i\alpha e^{-i(\tau  J_z - i\frac{R \tau}{2} J_z^2) } + \chi}{\sqrt{2}}\right)^{n_d},
		\end{split}
	\end{equation}
	carries the information about the atoms to the detectors, where they are accessed in a measurement. For $R = 0$ one recovers the definition of $\hat{M}_{n_c,n_d}$ in~\cite{ilo-okeke2016,ilo-okeke2022,ilo-okeke2023Dowling}. Since $e^{-i\hat{H}_\mathrm{eff}\tau}e^{i\hat{H}^\dagger_\mathrm{eff}\tau}\neq \mathbbm{1}$, it is not surprising that the $\sum_{n_c,n_d}\hat{M}_{n_c,n_d}\hat{M}^\dagger_{n_c,n_d} \neq \mathbbm{1}$. However, the operator $\hat{\mathcal{M}}_{n_c,n_d} ()$ satisfies the resolution of identity,
	\begin{equation}
		\sum_{n_c=0}^{\infty}\sum_{n_d =0}^{\infty}\hat{\mathcal{M}}_{n_c,n_d}(\mathbbm{1}) =\mathbbm{1}.
	\end{equation} 
	
	The measurement operator $\hat{\mathcal{M}}_{n_c,n_d}$ can be seen to act in two steps. Given an initial state of the atoms $\rho_{\mathrm{atom}}$, the measurement in the first step prepares a different state. For example, it broadens the width of an inherently Gaussian state, such as the coherent spin state. The second step measures the state prepared in the first step via the detection of photons. The measurement thus collapses the state of light while causing the state of atoms prepared in the first step to accumulate phase $m_z\tau$ that evolves at different rates dependent on the quantum number $m_z$ and decays at a rate $R$ dependent on the square of their quantum number $m_z$. Additionally, the measurement influences the most probable outcome of the quantum state of the atoms by making it dependent on the detected photon numbers $n_c$ and $n_d$, the dimensionless time $\tau$, as well as the spontaneous emission rate $R$. We now examine each step of the measurement operator in the following.
	
	\subsection{State Preparation Step \label{sec:sec:statepreparationstep}}
	To study the behavior of the state preparation step, we use  the fact that a function of an operator $ \hat{A} $ can be decomposed into a function of its eigenstates $ \lvert a_n\rangle $ with eigenvalue $ a_n $ 
	\begin{align}
		\label{eq:OperatorEqaution}
		f(\hat{A})= \sum_{n} f(a_n) \lvert a_n\rangle\langle a_n \rvert. 
	\end{align}
	We write the superoperator $\mathcal{L}_s ()$ as 
	\begin{equation}
		\label{eq:stateprepoperator}
		\begin{split}
			\mathcal{L}_s() =&\sum_{J=0}^{J_\mathrm{max}}\sum_{J' = 0}^{J_\mathrm{max}}\sum_{m_z=-J}^{J}\sum_{m' = -J'}^{J'} e^{R\tau\lvert\alpha\rvert^2m_zm'_zf(x) }\\
			&\times\lvert J,m_z\rangle\langle J,m_z\rvert ()\lvert J',m'_z\rangle\langle J',m'_z\rvert,
		\end{split}
	\end{equation}
	where $x= i\tau [(m'_z-m_z) + i(R/2) (m^2_z + m'^2_z)]$, and $f(x)$ is defined as  
	\begin{equation}
		\label{eq:deltafunction}
		f(x) = \frac{e^{x} - 1}{x}.
	\end{equation}
	The function $f(x)$ (\ref{eq:deltafunction}) has its maximum value for $ x = 0$ (i.e. $m'_z=m_z=0$). For significant values of $R$, $0\ll R<1$, the function $f(x)$  decays quickly for large values of $m_z$, $m'_z$. Hence, the function $f(x)$ behaves as $(e^{-\frac{R\tau }{2}(m^2_z + m'^2_z)} - 1)/[-\frac{R\tau }{2}(m^2_z + m'^2_z)]$. Since the exponential term $e^{-\frac{R\tau }{2}(m^2_z + m'^2_z)}$ would be much smaller than unity at large values of $m_z$, $m'_z$, the contribution of the function $f(x)$ would scale as $[\frac{R \tau}{2}(m^2_z + m'^2_z)]^{-1}$. Hence, the amplitude of (\ref{eq:stateprepoperator}) is positive and large. As such, the superoperator $\mathcal{L}_s()$ positively amplifies the input state. For instance, for $R\ll1$ the state for which $\lvert m_z\rvert = \lvert m'_z\rvert= J$ is amplified most and scales as $e^{\lvert\alpha\rvert^2}$, while the state $m'_z = m_z = 0$ is not amplified. As such, for any input state, the preparation step increases the width of the state along the axis joining the  $m_z = m'_z= -J$ and $m_z = m'_z= J$ state (along the diagonal). The broadening of the width brings about the first step in the dephasing (scrambling) of the input state by spontaneous emission. 
	
	\subsection{Photon Detection Step \label{sec:sec:photondetectionstep}}
	To study the behavior of the operators $\hat{M}_{n_c,n_d}$, we use (\ref{eq:OperatorEqaution}) and write the detection step operators (\ref{eq:measurementoperatorII})  as
	\begin{equation}
		\label{eq:detectionstepoperator}
		\begin{split}
			\hat{M}_{n_c,n_d}  &= e^{-\frac{\lvert\alpha\rvert^2 + \lvert \chi\rvert^2}{2}} \left(\frac{\lvert \alpha\rvert^2 + \lvert\chi\rvert^2}{2} \right)^{\frac{n_c + n_d}{2}} e^{-i\frac{\pi}{2} n_d}   \\
			& \times \sum_{J=0}^{J_\mathrm{max}}\sum_{m_z=-J}^{J} e^{i(n_c + n_d)(\phi(m_z) - \theta(m_z))}  \\
			& \times e^{i(n_c \phi_{c}(m_z) + n_d\phi_{d}(m_z) )}  \\
			&\times  A(n_c,n_d,m_z)  \lvert J,m_z\rangle\langle J, m_z\rvert,
		\end{split}
	\end{equation}
	where $A(n_c,n_d,m_z)$ is defined as 
	\begin{align}
		A(n_c,n_d,m_z) & =   \frac{\left( 1 + \cos2\eta\cos2\phi(m_z) \right)^{\frac{n_c}{2}}}{\sqrt{n_c!}}\nonumber \\
		& \times \frac{\left(1 - \cos2\eta \cos2\phi(m_z) \right)^{\frac{n_d}{2}}}{\sqrt{n_d!}}.
	\end{align}
	The parameter $\eta$ is defined as  
	\begin{equation}
		\label{eq:lightamplitude}
		\tan\eta = \frac{\lvert\chi\rvert - \lvert\alpha\rvert}{\lvert\chi\rvert + \lvert\alpha\rvert},
	\end{equation} 
	whereas the phases are
	\begin{equation}
		\label{eq:phasesII}
		\begin{split}
			\theta(m_z) & = \tau(m_z - i \frac{R}{2}m^2_z),\\
			\phi(m_z)  & = \frac{\theta(m_z)}{2} + \frac{\phi_{\chi\alpha}}{2} + \frac{\pi}{4} + \frac{\phi_{p}}{2},\\
			\phi_{c} (m_z) & = \arctan\left(\tan\eta \tan\phi(m_z) \right),\\
			\phi_{d} (m_z) & = \arctan\left(\frac{\tan\phi (m_z)}{\tan\eta} \right),
		\end{split}
	\end{equation}
	and $\phi_{\chi\alpha} = \arg(\chi) - \arg(\alpha)$.
	
	For $n_c,\,n_d\gg 1$, and $\tau$ greater than zero, $A(u,v,m_z)$  is a slow-varying function of $m_z$. Using Stirling's approximation to rewrite the factorials, the complex function $A(n_c,n_d,m_z)$ may be approximated by a Gaussian
	\begin{equation}
		\begin{split}
			A(n_c,n_d,m_z) & \approx  \left(\frac{2 }{n_c + n_d}\right)^{\frac{n_c}{2}}\left(\frac{2}{n_c + n_d}\right)^{\frac{n_d}{2}}\\
			&\times \frac{e^{\frac{n_c+n_d}{2}}}{\left(4\pi^2 n_c n_d \right)^{\frac{1}{4}}} e^{-\frac{(1 - iR\xi)^2}{2\sigma^2}\left(m_z - \xi\right)^2},
		\end{split}
		\label{eq:ComplexAmplitudeApprox}
	\end{equation}
	where $\sigma$ is defined as  
	\begin{equation}
		\label{eq:Width}
		\begin{split}
			\sigma^{2} &= \bigg([(n_c + n_d)^2\cos^22\eta -  (n_c - n_d)^2]\bigg)^{-1} \\
			&\times\left(\frac{\tau^2 }{8}\frac{n_c + n_d}{n_c n_d}\right)^{-1},
		\end{split}
	\end{equation}
	and the location of the maximum $\xi$ is defined as
	\begin{equation}
		\label{eq:maximumlocation}
		\xi =  \frac{\left(1 + 4 R^2 m_0^2\right)^{1/4}\sin\left(\dfrac{\arctan(2Rm_0)}{2}\right)}{R},
	\end{equation} 
	where $m_0$ is 
	\begin{equation}
		\label{eq:peakposition}
		m_0 = \frac{1}{\tau}\left[\arcsin\left(\frac{1}{\cos2\eta} \frac{n_d - n_c}{n_d + n_c}\right) - \phi_{\chi\alpha} - \phi_{p}\right].
	\end{equation}
	In the limit of $R\ll1$, $\xi \approx m_0(1 + R^2 m_0^2)$. So, without the spontaneous emission $R=0$, $\xi = m_0$	and one recovers the results of Refs.~\cite{ilo-okeke2016,ilo-okeke2022,chaudhary2023,ilo-okeke2023Dowling}. 
	
	Substituting (\ref{eq:ComplexAmplitudeApprox}) in (\ref{eq:detectionstepoperator})  gives the photon detection operator $\hat{M}_{n_c,n_d}$ as
	\begin{align}
		\label{eq:NoisyMeasurementOperator}
		\hat{M}_{n_c,n_d} & =\frac{ e^{\frac{n_c+n_d - \lvert\chi\rvert^2 -\lvert\alpha\rvert^2 }{2}}}{\left(4\pi^2n_cn_d\right)^{1/4}}\left(\frac{\lvert\chi\rvert^2 + \lvert\alpha\rvert^2}{n_c+n_d} \right)^{\frac{n_c + n_d}{2}} e^{-i\frac{\pi}{2}n_d}\nonumber\\
		&\times \sum_{J = 0}^{J_\mathrm{max}}\sum_{m_z=-J}^{J}e^{i(n_c+n_d)\left(\frac{\phi_{\chi\alpha} + \phi_{p} -\tau m_z}{2}  +\frac{\pi}{4}\right)}\nonumber\\ 
		&\times e^{i(n_c\phi_{c}(m_z) + n_d\phi_{d}(m_z))}\nonumber\\
		&\times  e^{-\frac{1}{2\sigma_R^2}m^2_z}e^{-\frac{(1 - iR\xi)^2}{2\sigma^2}\left( m_z - \xi\right)^2}\lvert J,m_z\rangle\langle J , m_z\rvert,
	\end{align}
	where $\sigma_R^2 = \left((n_c + n_d)R\tau/2\right)^{-1}$.
	
	Because of spontaneous emission, there is a product of two Gaussian functions in the operator $\hat{M}_{n_c,n_d}$~(\ref{eq:NoisyMeasurementOperator}). The product, which is also Gaussian, 
	\begin{equation}
		\label{eq:ProductOfGausians}
		\begin{split}
			e^{-\frac{1}{2\sigma_R^2}m^2_z}e^{-\frac{(1 - iR\xi)^2}{2\sigma^2}\left( m_z - \xi\right)^2} & = e^{-\frac{(1 - iR\xi)^2\xi^2}{2(\sigma_R^2(1 - iR\xi)^2 + \sigma^2)}}\\ 
			&\times e^{-\frac{1}{2\tilde{\sigma}^2}(m_z - \tilde{m}_0)^2},
		\end{split}	
	\end{equation}
	has a width $\tilde{\sigma} $ given as 
	\begin{equation}
		\label{eq:ComplexWidthWithNoise}
		\begin{split}
			\tilde{\sigma} & = \left(\frac{n_c + n_d}{2}\tau\right)^{-1} \bigg(R+ \tau\frac{\left(1 - iR\xi\right)^2}{4n_cn_d}\\ 
			&\times  \left[ \left(n_c + n_d\right)^2\cos^22\eta - \left(n_c - n_d\right)^2 \right]\bigg)^{-1},
		\end{split}
	\end{equation} 
	and a peak located at 
	\begin{equation}
		\label{eq:PeakPositionWithNoise}
		\tilde{m}_0 = \xi \left[ 1 - \frac{\left[1 +y(1 - R^2\xi^2) \right]}{\left[1 +y (1 - R^2\xi^2) \right]^2 + 4y^2R^2\xi^2} \right],
	\end{equation}
	where the parameter $y$ is defined as
	\begin{equation}
		\label{eq:peakparameter}
		y = \frac{\tau }{4Rn_cn_d}\left[(n_c + n_d )^2\cos^22\eta - (n_c - n_d)^2\right].
	\end{equation} 
	Notice that the term in the product on the right-hand-side (RHS) of (\ref{eq:ProductOfGausians})  whose exponent is proportional to $\xi^2$ is Gaussian. It depends entirely on the photon properties like the average photon number and the photon outcomes. Thus it controls the quality of the observed light signal at the detector. For small values of $R$ and $n_c + n_d \gg\lvert n_c - n_d\rvert$, this term is nearly unity. For spontaneous emission values of order unity, $R\sim 1$, this term decays very quickly and the amplitude of the observed signal is weak depending on the value of $\xi$. Hence, a good signal in the presence of spontaneous emission would be where $n_c + n_d \gg\lvert n_c - n_d\rvert$ is satisfied. This ensures that the amplitude at the readout is as close to unity as possible since $m_0 \approx 0 $ in this limit. 
	
	On the other hand, the term on the RHS of the product (\ref{eq:ProductOfGausians}) depends on the spin quantum number $m_z$ which carries the information about the atoms. This information is used to infer the measurement of the relative photon number $n_d - n_c$~(\ref{eq:peakposition}). The spontaneous emission perturbs this information. For example, given a weak spontaneous emission rate $R\ll 1$ and $m_0 \ll 1/R$, $\xi$ is roughly the same as $m_0$ as previously discussed, while the second term of (\ref{eq:PeakPositionWithNoise}) is negligibly small. Thus, $\tilde{m}_0 \approx m_0$. However, the spectra of $m_z$ available for prediction using the photon number difference $n_d - n_c$ shrinks with increasing spontaneous emission rate $R$. This is because the width of the Gaussian $\tilde{\sigma}$~(\ref{eq:ComplexWidthWithNoise}) shrinks with an increase in the strength of spontaneous emission rate $R$. For an appreciable strength of spontaneous emission, the second term of (\ref{eq:PeakPositionWithNoise}) is no longer negligible. Instead, both the relative photon number difference $n_d - n_c$ and the total photon number $n_d + n_c$ contribute to the spectra of $m_z$ (\ref{eq:peakparameter}), thereby scrambling or masking the  outcome $m_0$. Hence, the sensitivity of the measurement to $n_d - n_c$ is being lost. Similarly, the effective width of the operator equally shrinks. For large values of spontaneous emission rate $R\sim 1$ and the  photon number difference being small compared to the total number of photons $\lvert n_d - n_c\rvert \ll (n_c + n_d)\lvert\cos2\eta\rvert$, the terms in the bracket of (\ref{eq:PeakPositionWithNoise}) nearly cancel while the width of (\ref{eq:ComplexWidthWithNoise}) approaches zero. Hence, $\tilde{m}_0$ is approximately zero, $\tilde{m}_0 \approx 0$, irrespective of photon number outcome. It is then obvious that less than a handful of the spectra of $m_z$ would be available in a measurement. In this limit, the measurement is no longer sensitive to the photon number difference $n_d - n_c$. It is important to note that the $\tilde{m}_0$ is no longer controlled only by the relative photon number difference and spontaneous emission strength $R$ (\ref{eq:maximumlocation}) but equally by the total photon number (\ref{eq:peakparameter}). Additionally, spontaneous emission limits the range of available eigenvalues of $J_z$. Hence, only a few states near $m_z = \tilde{m}_0 = 0$ are accessible due to spontaneous emission. At a very high spontaneous emission rate, the state $\lvert J,m_z  =0\rangle$ is the only stable eigenvalue of the measurement operator, since the detection probability of other eigenvalues of $J_z$ are exponentially small. 
	
	
	\section{Quantum Projection Measurement With Spontaneous Emission \label{sec:NoisyProjectionMeasurement}}
	Following the discussions in the previous section, we see that in the state preparation, where the measurement strength is weak, the spontaneous emission makes the measurement strong by shrinking the width of the amplitudes of the quantum state to a limited range. This causes the dominant quantum state to become the state $m_z =0$, as seen for example, in squeezed state preparation. Where the measurement strength is strong such as in cat state preparation, spontaneous emission causes the collapse of the superposition of the quantum state onto the state $m_z=0$. The only difference between the strengths of the measurement is that spontaneous emission speeds up the collapse of the quantum state in the latter. That the spontaneous emission seeks out only the state $m_z=0$ irrespective of the measurement strength is because the state $m_z=0$ is the only stable state of the master equation~(\ref{eq:EffectiveMasterEquation}) in the presence of spontaneous emission, as we identified in Sec.~\ref{sec:AveragesSystemVariables}. Here, we now develop the projection operator describing the collapse of the superposition of states accounting for the spontaneous emission. 
	
	\subsection{Noisy Projection Operator \label{sec:sec:NoisyProjectionOperator}}
	As discussed above, the photon detection step collapses the state of the atoms, and the relative photon number difference gives inferences about the state collapse. Using the relative photon number difference
	\begin{equation}
		\begin{split}
			u & = \frac{n_c + n_d}{2},\\
			v & = \frac{n_d - n_c}{2},
		\end{split}
	\end{equation}
	we define the photon detection operator $\hat{M}_{u,v}$
	\begin{equation}
		\label{eq:MeasurementOperatorRelativeBasis}
		\begin{split}
			\hat{M}_{u,v}& = \frac{e^{iv\left(\phi_{d,0} - \phi_{c,0} -\frac{\pi}{2}\right)}}{\left[4\pi^2\left(u^2 -v^2\right)\right]^{\frac{1}{4}}}e^{-\frac{\left(u - \frac{\lvert\alpha\rvert^2 + \lvert\chi\rvert^2}{2}\right)^2}{\lvert\alpha\rvert^2 + \lvert\chi\rvert^2}}\\
			&\times e^{iu\left[\phi_{d,0}+\phi_{c,0} + \phi_{\chi\alpha} +\phi_{p} -\tau\tilde{m}_0\right]}\\
			& \times e^{-\frac{(1 - iR\xi)^2\xi^2}{2[\sigma_R^2 (1 - iR\xi)^2 + \sigma^2]}} \sum_{J=0}^{J_\mathrm{max}}\sum_{m_z=-J}^{J} e^{-\frac{(m_z - \tilde{m}_0)^2}{2\tilde{\sigma}^2}}
			\\
			&\times e^{-i\tau (u\zeta_+ - v\zeta_- )(m_z - \tilde{m}_0)}\lvert J,m_z\rangle\langle J,m_z\rvert.
		\end{split}
	\end{equation}
	
	The phase are defined as $\phi_{c,0} = \phi_{c}(m_z=\tilde{m}_0 )$,  $\phi_{d,0} = \phi_{d}(m_z = \tilde{m}_0)$,  and 
	\begin{equation}
		\label{eq:PhaseFirstDerivative}
		\begin{split}
			\zeta_+  & = 1 - \zeta_d - \zeta_c,\\
			\zeta_- & = \zeta_d - \zeta_c,\\
			\zeta_c & = \frac{1}{2}\frac{\tan\eta}{\cos^2(\phi(\tilde{m}_0))} \frac{(1 - iR\tilde{m}_0)}{1 +\tan^2\eta\tan^2(\phi(\tilde{m}_0)) },\\
			\zeta_d & = \frac{1}{2}\frac{\tan\eta}{\cos^2(\phi(\tilde{m}_0))} \frac{(1 - iR\tilde{m}_0)}{\tan^2(\phi(\tilde{m}_0)) + \tan^2\eta}.
		\end{split}
	\end{equation}
	Notice that for $R = 0$, one recovers the results of Ref.~\cite{ilo-okeke2023Dowling}.

	Provided that the average photon numbers are large and $m_z$ is different from $\tilde{m}_0$, the term summed over in (\ref{eq:MeasurementOperatorRelativeBasis}) is exponentially small and thus negligible. For $m_z = \tilde{m}_0$, terms within the sum gives the identity matrix, and the phases vanish. Hence, we define the photon detection operator as
	\begin{equation}
		\label{eq:GeneralisedDetectionOperator}
		\begin{split}
			\hat{M}_{u,v}& = \frac{e^{iv\left(\phi_{d,0} - \phi_{c,0} -\frac{\pi}{2}\right)}}{\left[4\pi\sigma^2_c\right]^{\frac{1}{4}}}e^{-\frac{\left(u - \frac{\lvert\alpha\rvert^2 + \lvert\chi\rvert^2}{2}\right)^2}{\lvert\alpha\rvert^2 + \lvert\chi\rvert^2}}\\
			&\times e^{iu\left[\phi_{d,0}+\phi_{c,0} + \phi_{\chi\alpha} +\phi_{p} -\tau\tilde{m}_0\right]}\\
			& \times e^{-\frac{(1 - iR\xi)^2\xi^2}{2[\sigma_R^2 (1 - iR\xi)^2 + \sigma^2]}}
			\sum_{J=\lvert\tilde{m}_0\rvert}^{J_\mathrm{max}}\lvert J,\tilde{m}_0\rangle\langle J,\tilde{m}_0\rvert,
		\end{split}
	\end{equation}
	where the effective complex width $\sigma_c$ is 
	\begin{equation}
		\label{eq:EffectiveSigma}
		\sigma_c^2 = R\tau u(u^2 - v^2) +  u (u^2\cos^2\eta - v^2)\tau^2(1 - iR\xi)^2.
	\end{equation}

	Equation (\ref{eq:GeneralisedDetectionOperator}) gives the photon detection projection operator in the presence of spontaneous emission. Substituting $\hat{M}_{u,v}$ and its complex conjugate in (\ref{eq:GeneralizedMeasurementOperator}) gives the projection measurement operator for QND measurement in the presence of spontaneous emission
	\begin{equation}
		\label{eq:GeneralisedMeasurementOperator}
		\begin{split}
			\hat{\mathcal{M}}_{u,v}() =& \hat{M}_{u,v} \mathcal{L}_\mathrm{s}()\hat{M}^*_{u,v}\\
			= & \frac{e^{-2\frac{\left(u - \frac{\lvert\alpha\rvert^2 + \lvert\chi\rvert^2}{2}\right)^2}{\lvert\alpha\rvert^2 + \lvert\chi\rvert^2}}}{\left[16\pi^2\sigma_c^2\sigma_c^{*2}\right]^{\frac{1}{4}}}  e^{-\frac{(1 - iR\xi)^2\xi^2}{2[\sigma_R^2 (1 - iR\xi)^2 + \sigma^2]}}\\
			& \times e^{-\frac{(1 + iR\xi)^2\xi^2}{2[\sigma_R^2 (1 + iR\xi)^2 + \sigma^2]}}\sum_{J=\lvert\tilde{m}_0\rvert}^{J_\mathrm{max}} \lvert J, \tilde{m}_0\rangle\langle J,\tilde{m}_0\rvert \\
			&\times\mathcal{L} () \sum_{J'=\lvert\tilde{m}_0\rvert}^{J_\mathrm{max}} \lvert J', \tilde{m}_0\rangle\langle J',\tilde{m}_0\rvert.
		\end{split}
	\end{equation} 
	In measurements, it is the relative photon number difference $v$ that is used in making the prediction about the atoms. Hence, the effect of spontaneous emission is to make the relative photon number difference $v$ insensitive to detection, see (\ref{eq:peakposition}) and (\ref{eq:maximumlocation}). This comes about by shrinking the width of $m_z$ (spectra of $m_z$ from $(2J+1)$ to a limited range, in the extreme case a single value), while the amplitude of the measurement operator $\hat{\mathcal{M}}_{u,v}$ decays exponentially as $\sim e^{-\propto\xi^2}$ for values $\xi\neq 0$. Additionally, the most probable outcome $\tilde{m}_0$ is scrambled by spontaneous emission except for $\tilde{m}_0 = 0$.  All these affect the utility of using the relative photon number difference $v$ for inference about the spin state. If one is able to correctly guess the spontaneous emission rate, then the value of $\xi$ (\ref{eq:maximumlocation}) can always be known and the resulting state $\lvert J,\tilde{m}_0\rangle$ is known without ambiguity provided that the signal is detectable. However, the amplitude of such measurements is exponentially very small $\sim e^{-\propto\xi^2}$ making detection of the corresponding state very difficult unless one uses low photon number $u$, which puts the measurement in the minimally-destructive regime and will not be a projection measurement. Because the spontaneous emission shrinks the width of the eigenvalue spectrum, the eigenvalues of the total spin operator $J_z$, further away from $m_z = 0$, are suppressed and have negligible detection probability. Hence, only a handful of $m_z$ values close to $m_z =0$  are available at a high spontaneous emission rate, with non-negligible detection probability. Of these few states, only the state $m_z =\tilde{m}_0 = 0$ suffers no effects from spontaneous emission. At a very high spontaneous emission rate, the detection probability of the state $m_z =\tilde{m}_0 = 0$ becomes higher than other states (which become exponentially small), thus making $m_z =\tilde{m}_0 = 0$ the only available state. One measurement result that guarantees the outcome is for $ \xi =0$, which is achieved by post-selecting only the outcomes for which $n_d = n_c$, as this guarantees that $\tilde{m}_0 = \xi = m_0 = 0$, see (\ref{eq:peakposition}) and (\ref{eq:maximumlocation}), and (\ref{eq:PeakPositionWithNoise}). Another possibility is to let the spontaneous emission rate be as large as possible $R\sim 1$ while allowing the longer evolution time, as shown in Sec.~\ref{sec:Examples} below.
	
	\section{Examples of spontaneous emission effects on state preparation \label{sec:Examples}}
	Here we apply the measurement operator developed above to study the effect of spontaneous emission in the preparation of squeezed and Schrodinger cat states using the spin coherent state and thermal state introduced in Sec.~\ref{sec:sec:InitialState} as the initial state. Many experiments and proposals for correlated quantum state preparation have used them as initial states. For all the initial states, we assume that the ensemble of atoms consists of $N$  two-level atoms treated as effective spin-1/2 particles. A practical choice, for example, is using the hyperfine ground states.

	\subsection{ Intermediate State\label{sec:sec:StatePreparationStep}}
	Here we look at the intermediate state generated by the measurement while assuming $N$ is large and even. Using the coherent state polarised along the \emph{x} axis $\rho_0^{\mathrm{coh}} = \lvert\theta=\pi/2,\phi = 0\rangle\langle \theta = \pi/2, \phi = 0\rvert$ as an initial state, the measurement operator first prepares a state $\tilde{\rho}^{\mathrm{coh}} =\mathcal{L}_s\rho_0^{\mathrm{coh}}$ ~(\ref{eq:stateprepoperator}), 
	\begin{align}
		\label{eq:IntermediateCoherentState}
			\tilde{\rho}^{\mathrm{coh}} & =  \left(\frac{1}{2}\right)^{2J} \sum_{m_z = -J}^{J}\sum_{m'_z =-J}^{J} e^{R\tau\lvert\alpha\rvert^2m_zm'_zf(x) }\nonumber\\
			& \times \binom{2J}{J+m_z}^{\frac{1}{2}}\binom{2J}{J+m'_z}^{\frac{1}{2}} 
			\lvert J, m_z\rangle\langle J, m'_z\rvert
	\end{align}
	Similarly, for the initial state being a thermal state~(\ref{eq:ThermalState}), the state prepared by the measurement $\tilde{\rho}^\mathrm{th} = \mathcal{L}_s\rho_0^\mathrm{th}$ is 
	\begin{equation}
		\label{eq:IntermediateThermalState}
		\begin{split}
			\tilde{\rho}^\mathrm{th} &= \frac{1}{J_\mathrm{max} + 1}\sum_{J = 0}^{J_\mathrm{max}}\sum_{m_z=-J}^{J} e^{-\lvert\alpha\rvert^2\left( e^{-R\tau m_z^2} - 1\right)}\\ &\times \frac{e^{-E_0m_z}}{P_J} \lvert J, m_z\rangle\langle J,m_z\rvert
		\end{split}
	\end{equation}	
	Neither of the states prepared by the measurement is normalized for $R\neq0$ without the detection step. 
	
	\begin{figure}[t]
		\includegraphics[width=\columnwidth]{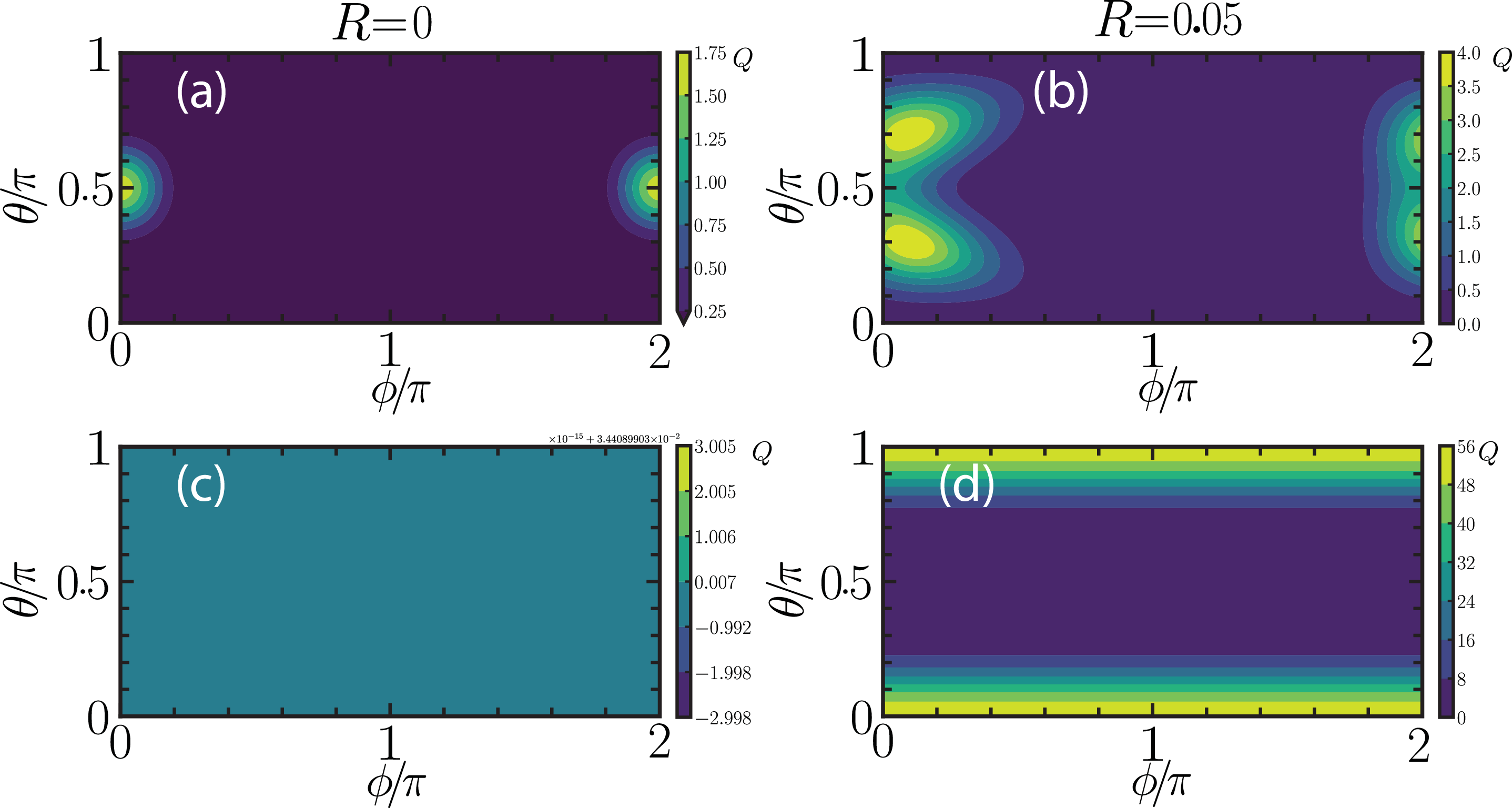}
		\caption{The $Q$ function of $\tilde{\rho}   = \mathcal{L}_s\rho_{\mathrm{atom}}$ (\ref{eq:stateprepoperator}) [see also (\ref{eq:SolutionTransformedState})] with different initial states $\rho_{\mathrm{atom}}$. In the first row is a spin coherent state~(\ref{eq:CoherentState}) initially polarized along the $x$-axis, while the second row is an thermal state~(\ref{eq:ThermalState}). The parameters of the figure are $\tau = \pi/(2J)$, $\alpha =\sqrt{20} $, $J = 10$, and $E_0 = 1\times 10^{-12}$. The parameter $R$ is as shown in the figure.}
		\label{fig4}
	\end{figure}
	
	We may visualize any state $\rho$ on a Bloch sphere using the \emph{Q} function
	\begin{equation}
		\label{eq:QfunctionJ}
		Q_{J}(\theta,\phi) = \frac{2J + 1}{4\pi}\langle\theta,\phi\lvert\rho\rvert\theta,\phi\rangle.
	\end{equation}
	For $\rho = \tilde{\rho}_0^\mathrm{coh}$, the \emph{Q} function becomes 
	\begin{align}
		\label{eq:InterQfunctionCoherentState}
			Q_J^\mathrm{coh} & = \frac{2J + 1}{4\pi}\left(\frac{1}{2}\right)^{2J}\sum_{m_z = -J}^{J} \sum_{m'_z = -J}^{J} e^{R\tau \lvert\alpha\rvert^2 m_zm'_z f(x)}\nonumber\\
			&\times \binom{2J}{J+m_z}\binom{2J}{J+m'_z}\left(\sin\frac{\theta}{2}\right)^{2J + m_z + m'_z}\nonumber\\
			& \times\left(\cos\frac{\theta}{2} \right)^{2J-m_z- m'_z} e^{i\phi(m'_z - m_z)}.
	\end{align}
	The definition of \emph{Q} (\ref{eq:QfunctionJ}) works well for spin states defined with only one total spin $J$ value such as coherent state that always occupy the maximum total spin value $J_\mathrm{max}$. For other states composed of mixture of  the total spin values $J$, the \emph{Q} function, $Q= \sum Q_J $, is
	\begin{equation}
		\label{eq:Qfunction}
		Q(\theta,\phi) = \frac{1}{J_\mathrm{max} + 1}\sum_{J = 0}^{J_\mathrm{max}} \frac{2J + 1}{4\pi}\langle\theta,\phi\lvert\rho\rvert\theta,\phi\rangle.
	\end{equation}
	The term $(J_\mathrm{max} + 1)^{-1}$ in (\ref{eq:Qfunction}) ensures its normalization since each \emph{Q}\textsubscript{\emph{J}} in the sum is individually normalized when integrated over $\theta$ and $\phi$. Using (\ref{eq:Qfunction}) we calculate the \emph{Q} function for the thermal state~(\ref{eq:IntermediateThermalState}) as 
	\begin{align}
		\label{eq:InterQfunctionThermalState}
		Q^\mathrm{th}& = \frac{1}{(J_\mathrm{max} + 1)^2} \sum_{J = 0}^{J_\mathrm{max}} \frac{2J + 1}{4\pi} \sum_{m_z=-J}^{J} \frac{e^{-E_0m_z}}{P_J}\nonumber\\
		& \times e^{-\lvert\alpha\rvert^2\left(e^{-R\tau m_z^2} - 1\right)} \binom{2J}{J + m_z} \left(\sin^2\frac{\theta}{2} \right)^{J+m_z}\nonumber\\
		&\times \left(\cos^2\frac{\theta}{2} \right)^{J-m_z}.
	\end{align} 
	
	Fig.~\ref{fig4} shows the \emph{Q} function of the different initial states. For the thermal state, the temperature is set to infinity, thus resulting in a  maximally mixed state. In the first column is the state of the atoms with the spontaneous emission rate set to zero. In this state, there is no effect from light whatsoever. This state is exactly the initial state in the absence of measurement. Since the thermal state is a maximally mixed state, there is no preferred state, and its \emph{Q} function, Fig.~\ref{fig4}(c), shows that any state in its superposition is equally probable. However, for spontaneous emission rate $R > 0$, each initial state is affected differently but still consistent as described in Sec.~\ref{sec:sec:statepreparationstep}. For instance, in an atom coherent state polarized along the \emph{x}-axis on the Bloch sphere, the superoperator  $\mathcal{L}_s()$ only broadens the width of the atomic states near the most weighted probability amplitude $m'_z =m_z = 0$ when the value of spontaneous emission rate $R$ is small. It is because as $m'_z,\,m_z$ increases, the amplitude of the atom coherent state states near $m'_z = m_z =0$ decreases exponentially faster than the amplitude of $\mathcal{L}_s()$. Thus their product increases the width of the probability amplitudes of the prepared states while suppressing the amplitudes far away from $m'_z = m_z =0$, as shown in Fig.~\ref{fig4}(b). As the value of the spontaneous emission rate $R$ increases, the trend reverses such that the amplitude of the superoperator increases faster than that of the coherent state, and their product away from zero $m'_z=m_z =0$ becomes large compared to their product near $m'_z =m_z = 0$. Hence, not only the width broadens, the position of the most probable peak equally shifts away from $m'_z = m_z = 0$. This situation is shown in Fig.~\ref{fig4}(b). 
	
	The thermal state suffers more from effects due to spontaneous emission than the spin coherent state, as shown in Fig.~\ref{fig4}. It is because each state is weighted the same, resulting in a flat probability distribution for each $m_z$ within the total spin manifold $J$, as shown in Fig~\ref{fig4}(c). Hence, for even small values of spontaneous emission $R\ll1$, the amplitude of the superoperator rises exponentially faster as $m_z$ moves away from zero than the thermal state amplitude that is flat. For each total spin manifold $J$, the amplitude of the superoperator is a maximum at $\lvert m'_z\rvert = J$. Thus the product of the thermal state amplitude and superoperator at $m_z= \pm J$ give the largest values. Hence the most probable outcome is no longer equally weighted, and the width is severely affected even for a modest spontaneous emission rate. Hence, the product of the the initial state and the superoperator peaks at $m_z = \pm J$ for each spin sector, as shown in Figs.~\ref{fig4}(d).
	
		\begin{figure*}
			\begin{minipage}[t]{\linewidth}
				\includegraphics[width=\linewidth]{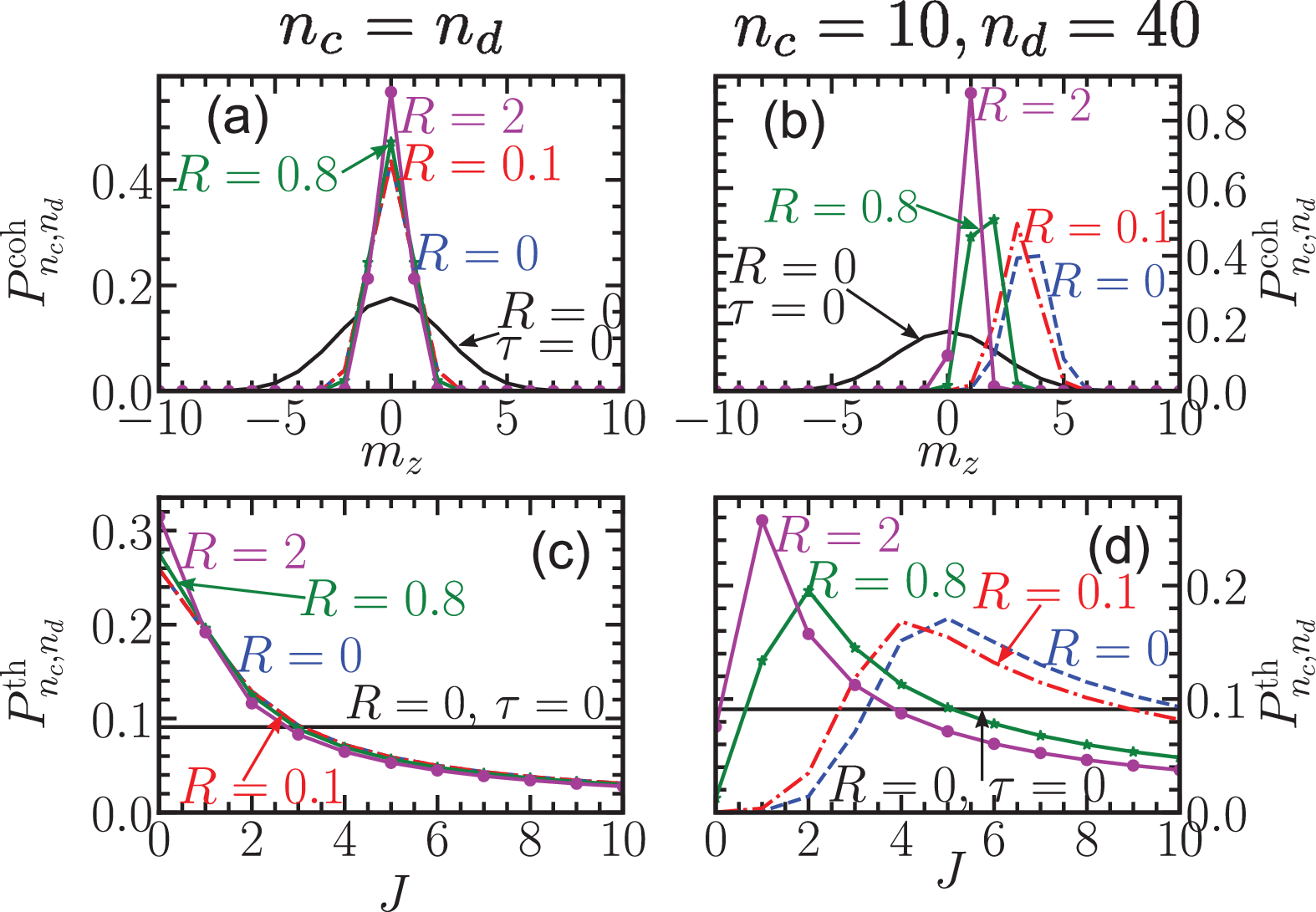}
				\caption{The conditional probability density (\ref{eq:ConditionalAtomProbability}) of the state prepared by a QND measurement for different initial states. The first row is the conditional probability density where the initial state is a coherent state polarized along \emph{x} axis as described in the text. The second row is the conditional probability density using the thermal state that is maximally mixed as the initial state. The first column takes $n_c = n_d = 20$, while in the second column $n_d = 40$ and $n_c = 10$. The dimensionless time $\tau$ for every figure is $\tau = \pi/(2J_\mathrm{max}) $ except otherwise shown in the graphs, and the dimensionless spontaneous emission rate $R$ is as shown in the figure. The parameters of the figures are $N = 20$,  $\alpha = \chi = \sqrt{20}$, and $E_0= 1\times 10^{-12}$.}
				\label{fig5}
			\end{minipage}
		\end{figure*}
	
	\subsection{Final States \label{sec:sec:FinalStates}}
	Applying the photon detection operator~(\ref{eq:detectionstepoperator}) developed in Sec.~\ref{sec:sec:photondetectionstep} to the spin coherent state (\ref{eq:IntermediateCoherentState}), we arrive at the following, 
	
	\begin{equation}
		\label{eq:ConditionalStateCoherent}
		\begin{split}
			\rho_{n_c,n_d}^{\mathrm{coh}} =& \frac{e^{-\lvert\alpha\rvert^2 - \lvert\chi\rvert^2}}{P^{\textrm{coh-L}}_{n_c,n_d}}\left(\frac{\lvert\alpha\rvert^2 + \lvert\chi\rvert^2}{2}\right)^{n_c + n_d} \left( \frac{1}{2}\right)^{2J} \\
			&\sum_{m_z = -J}^{J} \sum_{m'=-J}^{J} A(n_c,n_d,m_z) A(n_c,n_d,m'_z)\\
			&\times e^{i\frac{\tau}{2} (n_c+n_d)(m'_z - m_z)} e^{-R\tau\frac{(n_c +n_d)(m^2 + m'^2)}{4}}\\
			&\times 
			e^{in_c[\phi_{c}(m_z) - \phi_{c}(m'_z)]} 
			e^{in_d[\phi_{d}(m_z) - \phi_{d}(m'_z)]}\\
			& \times e^{R\tau \lvert \alpha\rvert^2 m_zm'_zf(x)} \binom{2J}{J+m_z}^{\frac{1}{2}} \\
			& \times\binom{2J}{J+m'_z}^{\frac{1}{2}} \lvert J,m_z\rangle\langle J,m'_z\rvert,
		\end{split}
	\end{equation}
	where $P^{\textrm{coh-L}}_{n_c,n_d}$ is the probability of obtaining $n_c$ and $n_d$ photons in a measurement using a coherent state as an initial state. Similarly, the detection operator (\ref{eq:detectionstepoperator}) acting on the thermal state (\ref{eq:IntermediateThermalState}) gives
	\begin{equation}
		\label{eq:ConditionalStateThermal}
		\begin{split}
			\rho^\mathrm{th}_{n_c,n_d} = & \frac{e^{-\lvert\alpha\rvert^2 - \lvert\chi\rvert^2}}{P^\textrm{th-L}_{n_c,n_d}}\left(\frac{\lvert\alpha\rvert^2 + \lvert\chi\rvert^2}{2}\right)^{n_c + n_d} \frac{1}{J_\mathrm{max}+1}\\ 
			&\times \sum_{J=0}^{J_\mathrm{max}} \sum_{m_z=-J}^{J} \left|A(n_d,n_d,m_z)\right|^2\\
			& \times e^{2\mathcal{R}e  [i\left( n_c \phi_c(m_z) + n_d\phi_d(m_z)\right)] } e^{-\frac{R\tau m^2_z}{2}(n_c + n_d)} \\
			& e^{-\lvert \alpha\rvert^2 \left( e^{-R\tau m_z^2} - 1 \right) } \frac{e^{-E_0 m_z}}{P_J}\lvert J, m_z\rangle\langle J,m_z\rvert,
		\end{split}
	\end{equation}
	where $\mathcal{R}e[z]$ gives the real part of a complex variable \emph{z}, and $P^{\textrm{th-L}}_{n_c,n_d}$ is the probability of obtaining $n_c$ and $n_d$ photons in a measurement with thermal state as an initial state. The photon probabilities $P^\textrm{s-L}_{n_c.n_d}$ for $\mathrm{s} = \mathrm{coh}$ or $\mathrm{th}$ is defined as 
	\begin{equation}
		\label{eq:PhotonProbability}
		P^{\textrm{s-L}}_{n_c,n_d} = \mathrm{Tr}_{J,m_z}\left(\hat{M}_{n_c,n_d}\mathcal{L}_s\rho^\mathrm{s}\hat{M}^\dagger_{n_c,n_d}\right).
	\end{equation}
	Similarly, given the atoms to be in the state \textrm{s}, the probability $P_{n_c,n_d}^\mathrm{s}$ of observing the atoms in the state $\lvert J, m_z\rangle$ conditioned on the detection of photons $n_c$ and $n_d$ is given by
	\begin{equation}
		\label{eq:ConditionalAtomProbability}
		P^{\mathrm{s}}_{n_c,n_d} =\frac{1}{P^{\textrm{s-L}}_{n_c,n_d}} \langle J, m_z\rvert \hat{M}_{n_c,n_d}\mathcal{L}_s\rho^\mathrm{s}\hat{M}^\dagger_{n_c,n_d}\lvert J,m_z\rangle.
	\end{equation}

	\subsection{Squeezed State \label{sec:sec:SqueezedState}}
	Here, we examine the effect of spontaneous emission on the preparation of a squeezed state, which is obtained for the dimensionless time $\tau$ of the order $1/J$. For definiteness, we set  time $\tau$ to be equal to $\pi\left(2J_\mathrm{max}\right)^{-1}$, and present the probability density of the atomic states~(\ref{eq:ConditionalAtomProbability}) conditioned on the detection of photons in Fig.~\ref{fig5}.

	As shown in Fig. 5, the initial probability density of the states at $\tau = 0 = R$ is unsqueezed. For the coherent state shown in the first row, the distribution is a Gaussian given by the solid line. For the maximally mixed thermal state shown in the second row, it is given by the straight line parallel to the \emph{x}-axis.  However, for $\tau = \pi(2J_{\mathrm{max}})^{-1}$ and $R=0$, it becomes squeezed. Notice that increases in $R$ lead to more squeezing of the atom probability density. For instance, in the first column where $n_c = n_d$, $m_z = m_0 = 0$ (\ref{eq:peakposition}) and the position of the peak of the probability density is unaffected. However, the width decreases with increasing $R$ as predicted in Sec.~\ref{sec:sec:photondetectionstep}. The implication is that spontaneous emission reduces the values of states that would contribute to the measurement outcome.  		 
	
	For photon number difference $n_d- n_c \neq 0$, both the peak position and width of the probability density are affected, as shown in the second column of Fig.~\ref{fig5}. The width of probability density for $\tau = \pi(2J_{\mathrm{max}})^{-1}$ and  $R=0$ is squeezed, smaller than the width at $\tau = 0$ and  $R=0$. A higher value of the spontaneous emission strength $R$ causes more reduction in the width of the probability density compared to a  smaller value of $R$. Additionally, it shifts the probability amplitude towards $m_z =0$. For the maximally mixed state shown in Fig~\ref{fig5}(d), the probability density collapses towards the lower total spin value $J$. All these agree with the predictions of Sec.~\ref{sec:sec:photondetectionstep}. 
	
	From the preceding discussion, the relative photon number difference $n_d - n_c$ plays a role in making inferences on the state of the atoms in the absence of or negligibly small spontaneous emission rate $0\leq R\ll 1$. In this limit,  the peak position $\xi$ and the width $\sigma_c$  agree with the predictions made using the relative photon number difference, $\xi = m_0 = m_z$ and $\sigma_c =\sigma$. For substantial values of the spontaneous emission rate $R <1$, the spontaneous emission renders the measurement insensitive to the relative photon number difference, except where the relative photon number difference is zero, $ n_c = n_d $. It does this by limiting the number of atomic states contributing to the inference by making the width of the probability density smaller while masking the most dominant state. Note that strong measurements reduce the width of probability density such that it is small compared to the step size of $1$ by which the eigenvalues $m_z$ of  $ J_z$ change~\cite{ilo-okeke2023Dowling}. Hence, the spontaneous emission measures the atoms strongly.

	\subsection{Schr\"odinger Cat State  \label{sec:sec:CatlState}}
	The Schr\"odinger cat state is a superposition of two coherent states realized at long interaction times $\tau$ between atoms and light.  Hence, our discussion in this section uses the spin coherent state. To achieve strong interactions, we evolve the interaction $\tau$ for a longer time by setting the interaction time to $\tau = \pi/2$. To show the quantum effects contained in the state, we calculate the Wigner distribution defined in Ref.~\cite{byrnes2021}
	\begin{equation}
		\label{eq:WignerDistribution}
		W(\theta,\phi) = \sum_{L =0}^{2J}\sum_{M=-L}^{L} \rho_{L,M} Y_{L,M}(\theta,\phi),
	\end{equation}
	where $Y_{L,M}(\theta,\phi)$ are spherical harmonics. The coefficient $\rho_{L,M}$ is defined as 
	\begin{align}
		\label{eq:RhoLM}
		\rho_{L,M} & = \sum_{m_z = -J}^{J}\sum_{m'_z=J}^{J} (-1)^{J - m_z - M}\langle J,m_z;J,-m'_z\lvert L,M \rangle\nonumber\\
		& \times \langle J m_z\lvert \rho\rvert J m'_z\rangle,
	\end{align}
	where $\langle J,m_z;J,m'_z\lvert L,M \rangle$ is the Clebsch-Gordan coefficient for combining two angular momentum eigenstates $\lvert J,m_z\rangle$ and $\lvert J,m'_z\rangle$ to the state $\lvert L,M\rangle$, and $\rho$ is the density matrix. Any quantum effect contained in the density matrix results in the negative values for the Wigner distribution.
	\begin{figure}[t]
		\begin{center}
			\includegraphics[width=\columnwidth]{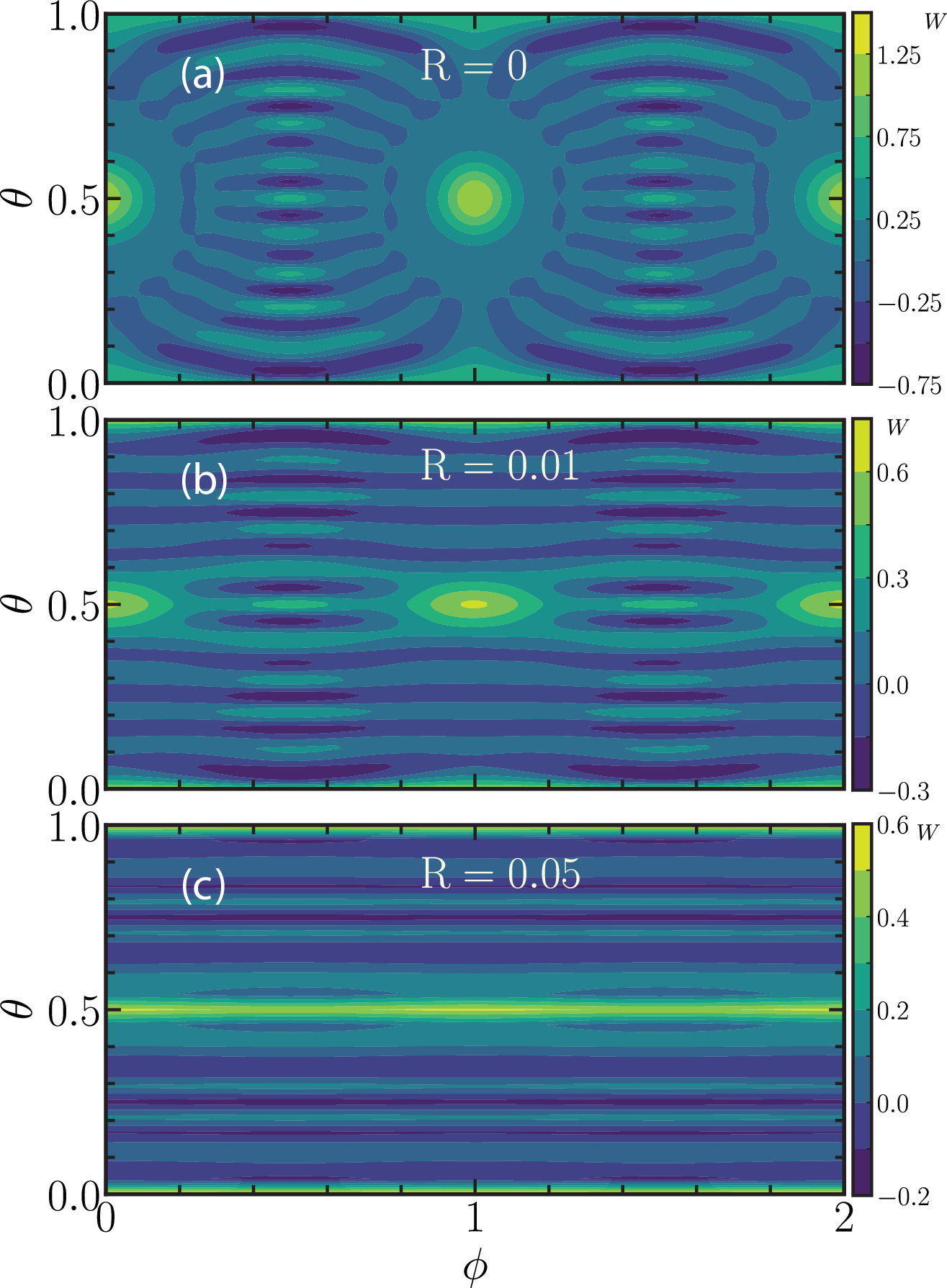}
		\end{center}
		\caption{The Wigner distribution of the state $\rho^{\mathrm{coh}}_{n_c,n_d}$ (\ref{eq:ConditionalStateCoherent}). The dimensionless time $\tau$ for the figures is $\tau = \pi/2 $, and the dimensionless spontaneous emission rate $R$ is as shown in the figure. The parameters of the figures are $N = 20$,  $\alpha = \chi = \sqrt{20}$, and $n_c = n_d = 20$.}
		\label{fig6}
	\end{figure}
	 	
	\begin{figure}[t!]
		\begin{center}
			\includegraphics[width=\columnwidth]{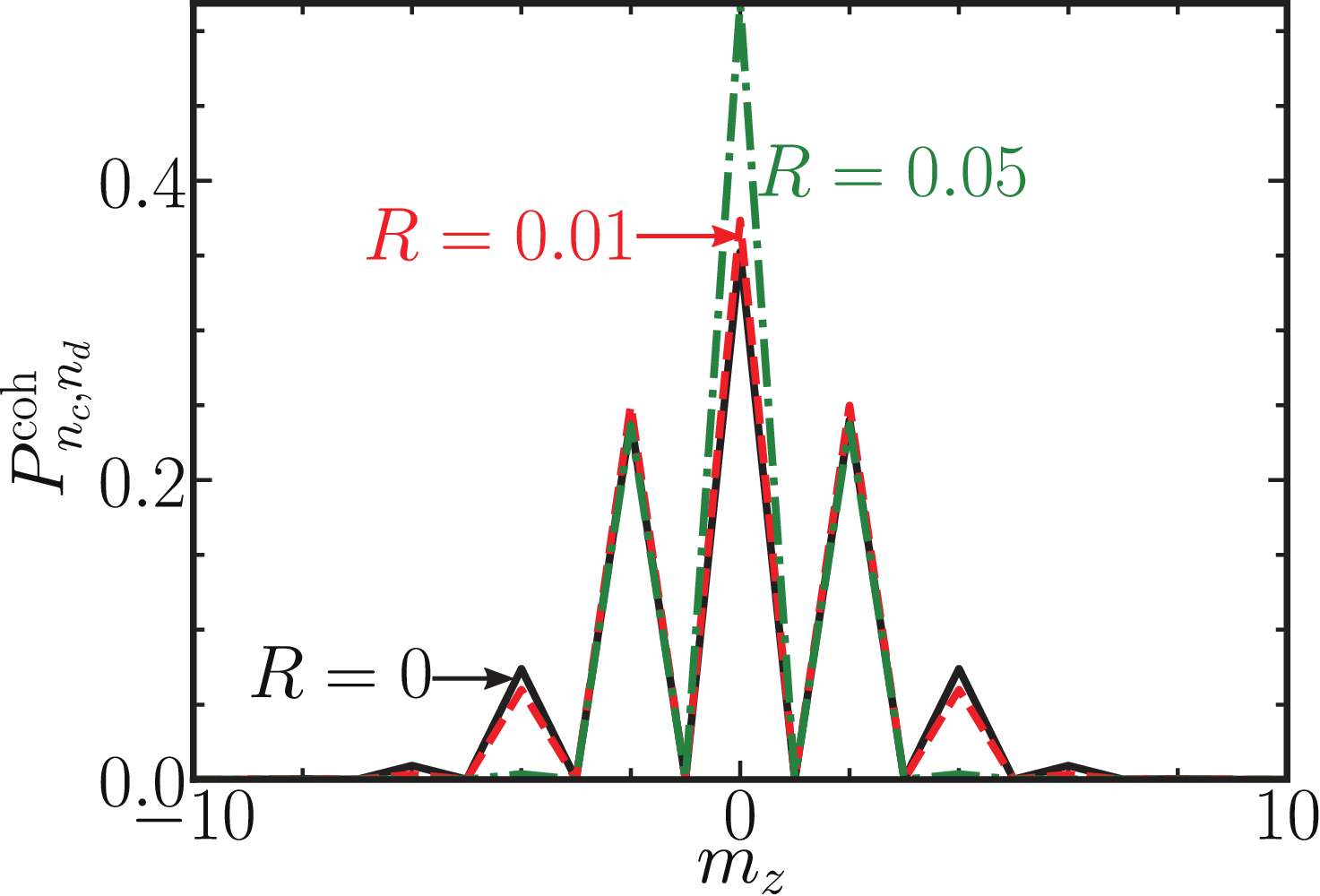}
		\end{center}
		\caption{The probability density of the state $\rho^{\mathrm{coh}}_{n_c,n_d}$ (\ref{eq:ConditionalStateCoherent}) at dimensionless time $\tau = \pi/2 $. The dimensionless spontaneous emission rate $R$ is as shown in the figure, and parameters of the figures are $N = 20$,  $\alpha = \chi = \sqrt{20}$ and $n_c = n_d = 20$.}
		\label{fig7}
	\end{figure}
	
	We present the Wigner distribution of the state in Fig.~\ref{fig6}. For $R=0$ as shown in Fig.~\ref{fig6}(a), one realizes a Schr\"odinger cat state, a macroscopic superposition of two coherent states along the equator at $\phi = 0$ and $\phi = \pi$. As the spontaneous emission rate $R$ increases, the interference between the two spin coherent states decreases. The decrease in negative values of the Wigner distribution shows that the state is becoming more classical. This feature equally manifests in Fig~\ref{fig7}, showing the conditional probability density (\ref{eq:ConditionalAtomProbability}) at the values of spontaneous emission rate $R$ given in Fig.~\ref{fig6}. At $R=0$, the probability density oscillates more rapidly than any other $R$ value. The oscillations and the width of the probability density decrease with an increase in $R$, similar to the effect observed with squeezed states in the presence of spontaneous emission. However, since the measurement is already strong, the spontaneous emission would realize only one Dicke state, which happens to be $m_z = 0$ state, $\lvert J,m_z=0\rangle$, by making the width of the distribution small. Hence, a small value of the spontaneous emission is sufficient to  destroy the quantum features realized by the measurement as seen in Figs.~\ref{fig6}(c) and ~\ref{fig7}.
	
	\section{Summary and Conclusions\label{sec:summary}}
	We derived the master equation~(\ref{eq:EffectiveMasterEquation}) for an atom interacting with light by adiabatic elimination accounting for the effects of spontaneous emission. The solution (\ref{eq:GeneralSolution}) to the master equation produces a quantum state of atoms that is dephased while giving decay in the quantum state of light associated with the optical depth in QND experiments~\cite{appel2009,sorensen1998}. Additionally, we showed that the intensity of the light beam that has interacted with the atoms depends on the initial statistics of atom distribution and may not necessarily experience exponential decay. Collapsing the quantum state of light via photon detection, we derive a positive operator valued measure~(\ref{eq:GeneralizedMeasurementOperator}) that takes any quantum state of the atoms at the input to the output. The photon number difference gives the eigenvalue of the total spin operator $J_z$ measured by the photons. However, we showed that the spontaneous emission masks the relative photon number difference, thereby making the measurement insensitive to detecting the eigenvalues of the total spin operator $J_z$.	
	
	The eigenstates of the total spin $J_z$ operator span the POVM.  Each operator term of the POVM experiences spontaneous emission in proportion to $m_z$, the eigenvalues of $J_z$. However, the operator term of the POVM corresponding to $m_z=0$ experiences no spontaneous emission. We showed that due to spontaneous emission, the measurement drives the quantum state of an atom ensemble such that it converges towards $m_z = 0$, an eigenstate of $J_z$ for a given total spin value $J$. For an ensemble with more than one spin $J$ value, its state converges to a state with the minimum $J$ value for the same eigenvalue of $J_z$, $m_z = 0$. This finding should interest  proposals~\cite{ilo-okeke2022,mao2022,kondappan2023,chaudhary2023} that use measurement and unitary feedback for quantum state preparation. 
		
	Furthermore, we tuned the atom-light interaction strength to prepare correlated states of the atoms, such as the squeezed and Schr\"odinger-cat states. We found that for a given spin eigenvalue $J$, correlations that build their coherence on $2J+1$ eigenvalues of the spin operator $J_z$ are most susceptible to spontaneous emission effects. This susceptibility is because the spontaneous emission shrinks the width of the eigenvalues available from $2J+1$ to a limited range, and a single value in the limiting case. Thus, an appreciable spontaneous emission rate turns the POVM into a projection operator whose measurement outcome is zero, $m_z =0$. Hence the Schr\"odinger-cat state is easily destroyed, even by a weak strength of spontaneous emission, compared to the squeezed state.

	\begin{acknowledgments}
		This work is supported by the National Natural Science Foundation of China (62071301); NYU-ECNU Institute of Physics at NYU Shanghai; the Joint Physics Research Institute Challenge Grant; the Science and Technology Commission of Shanghai Municipality (19XD1423000,22ZR1444600); the NYU Shanghai Boost Fund; the China Foreign Experts Program (G2021013002L); the NYU Shanghai Major-Grants Seed Fund; and the Talented Young Scientists Program (NGA-16-001) supported by the Ministry of Science and Technology of China.
	\end{acknowledgments}
	
	\appendix
	\section{Derivation of the effective ground state master equation}\label{sec:DerivationGroundStateMasterEquation}
	We consider a two-level atom as shown in Fig.~\ref{fig1}, and decompose the state $\rho_\mathrm{atom}$ of the atom in terms of its levels, $\lvert g\rangle$ and $\lvert e\rangle$, as 
	\begin{equation}
		\label{eq:AtomState}
		\rho_\mathrm{atom} = P_{ee} \lvert e\rangle\langle e\rvert  + P_{eg} \lvert e\rangle\langle g\rvert  + P_{ge} \lvert g\rangle\langle e\rvert  + P_{gg} \lvert g\rangle\langle g\rvert,
	\end{equation}
	where $P_{kk}$, $k =e,g$ are the probability of atom being in the excited and ground state, respectively, while the coefficients $P_{eg}$ and $P_{ge}$ are a measure of coherence between the ground and excited state. Substituting (\ref{eq:AtomLightState})  in (\ref{eq:DensityMatrixEvolution}), with $\rho_\mathrm{atom}$ defined in (\ref{eq:AtomState}), gives the following coupled differential equations
	\begin{align}
		\label{eq:Pee}
		\dot{P}_{ee} & = -i\left(\Omega_{ge}^*\alpha P_{ge} - \Omega_{ge}\alpha^*P_{eg} \right) - \gamma P_{ee},\\
		\label{eq:Peg}
		\dot{P}_{eg} & = -i\left(\Delta P_{eg} + \left( P_{gg} - P_{ee}\right)\Omega_{ge}^*\alpha\right) -\frac{\gamma}{2} P_{eg},\\
		\label{eq:Pge}
		\dot{P}_{ge} & =	-i\left( -\Delta P_{ge} +\Omega_{ge} \alpha^*\left( P_{ee} - P_{gg}\right) \right) -\frac{\gamma}{2}P_{ge},	\\
		\label{eq:Pgg}
		\dot{P}_{gg} & = -i\left(\Omega_{ge}\alpha^* P_{eg} -\Omega_{ge}^*\alpha P_{ge} \right) + \gamma P_{ee},\\
		\label{eq:AlphaEvol}
		\dot{\alpha} & = -i\frac{P_{ee}}{P_{ge}}\Omega_{ge},\\
		\label{eq:AlphaPEvol}
		\dot{\alpha}^* & = i\frac{P_{ee}}{P_{eg}}\Omega_{ge}^*,\\
		\label{eq:AlphaEvol2}
		\dot{\alpha} & =-i \frac{P_{eg}}{P_{gg}}\Omega_{ge},\\
		\label{eq:AlphaPEvol2}
		\dot{\alpha}^* & = i \frac{P_{ge}}{P_{gg}}\Omega_{ge}^*.
	\end{align}	 
	In deriving (\ref{eq:Pee}) -- (\ref{eq:AlphaPEvol2}), we used the fact $\lvert \alpha\rangle$ is not an eigenstate of $\hat{a}^\dagger$, and decompose $\hat{a}^\dagger\lvert \alpha\rangle $ into its orthogonal state as $\hat{a}^\dagger\lvert \alpha\rangle = \lvert \alpha_\perp\rangle + \alpha^*\lvert \alpha\rangle$. The orthogonal states $\lvert \alpha_\perp\rangle$ allow us to obtain evolution equation for $\alpha$ and $\alpha^*$. In particular, the evolution of $\alpha$ in the state $\lvert e,\alpha_\perp\rangle$  and its dual ($\alpha^*$ in the state $\langle e,\alpha_\perp\rvert$) have no corresponding part in the atom-light interactions. These states represent the free evolution part of the photons. This can be understood as the photons that come in contact with the atom when the atom is already in its excited state does not interact with it. This justifies the assumption of the phase contrast imaging~\cite{higbie2005,meppelink2010,ilo-okeke2014} that after light passes through the interaction region, two beams emerge---those that have interacted with atom and would accumulate some phase due to the interaction, and those that did not interact with the atom. 
	
	The coupled equations (\ref{eq:Pee}) -- (\ref{eq:AlphaPEvol2}) could be solved much more generally. However, we are interested in the regime where the rate of change in the probability of the excited state, as well as the coherences are negligible small $\dot{P}_{ee} = \dot{P}_{eg} = \dot{P}_{ge} = 0$. Reinterpreted, we are interested in the regime where the ground state evolves slowly in the presence of the applied field. Assuming that the atom is initial prepared in its ground state, and the detune is large, we assume for all times that $P_{gg} \gg P_{ee}$. Under these assumptions, we immediate obtain from (\ref{eq:Peg}) and (\ref{eq:Pge}) that 
	\begin{align}
		\label{eq:PegStationary}
		P_{eg} = \frac{\Omega_{ge}^*\alpha\left(-\Delta -i\dfrac{\gamma}{2}\right)}{\Delta^2  +\dfrac{\gamma^2}{4}} P_{gg},\\
		\label{eq:PgeStationary}
		P_{ge} = \frac{\Omega_{ge}\alpha^*\left(-\Delta +i\dfrac{\gamma}{2}\right)}{\Delta^2  +\dfrac{\gamma^2}{4}} P_{gg},
	\end{align}
	where (\ref{eq:PegStationary}) describes event in the Hilbert space, while (\ref{eq:PgeStationary})  describes the same event in dual Hilbert space. Upon setting the left hand side of (\ref{eq:Pee}) to zero and substituting (\ref{eq:PegStationary}) and (\ref{eq:PgeStationary}) for $P_{eg}$ and $P_{eg}$ respectively we obtain
	\begin{equation}
		\label{eq:PeeStationary}
		P_{ee} = \frac{\lvert\Omega_{ge}\alpha\rvert^2}{\left( \Delta^2 + \dfrac{\gamma^2}{4}\right)} P_{gg}.
	\end{equation}
	To justify the earlier assumption $P_{gg} \gg P_{ee}$ implies that   $\lvert\Omega_{ge}\alpha\rvert^2/\Delta^2 \ll1$ and $\gamma^2/\Delta^2 \ll 1$. 
	
	Substituting for $P_{eg}$, $P_{ge}$ and $P_{ee}$ in the evolution of $\alpha$ and its complex conjugate (\ref{eq:AlphaEvol2}) and (\ref{eq:AlphaPEvol2}) reduces to
	\begin{align}
		\label{eq:AlphaEffective}
		\dot{\alpha} & =  -i\frac{\lvert\Omega_{ge}\rvert^2\left(-\Delta -i\dfrac{\gamma}{2}\right)}{\Delta^2 + \dfrac{\gamma^2}{4}} \alpha,\\
		\label{eq:AlphaPEffective}
		\dot{\alpha}^* & = i  \frac{\lvert\Omega_{ge}\rvert^2\left(-\Delta +i\dfrac{\gamma}{2}\right)}{\Delta^2 + \dfrac{\gamma^2}{4}} \alpha^*.
	\end{align}
	It can equally be verified that similar substitution in (\ref{eq:AlphaEvol}) and (\ref{eq:AlphaPEvol}) give the same set of equations (\ref{eq:AlphaEffective}) and (\ref{eq:AlphaPEffective}) whose solution is as given in  (\ref{eq:AlphaSolution}).
	
	Substituting for $P_{eg}$, $P_{ge}$ and $P_{ee}$ in (\ref{eq:Pgg}) gives the following equation for the evolution of the ground state probability $\dot{P}_{gg}$
	\begin{equation}
		\label{eq:PggEffective}
		\begin{split}
			\dot{P}_{gg}&  = -i\frac{\lvert\Omega_{ge}\alpha\rvert^2\left(-\Delta -i\dfrac{\gamma}{2}\right)}{\Delta^2 + \dfrac{\gamma^2}{4}} P_{gg}\\ & +  i  \frac{\lvert\Omega_{ge}\alpha\rvert^2\left(-\Delta +i\dfrac{\gamma}{2}\right)}{\Delta^2 + \dfrac{\gamma^2}{4}}P_{gg}\\ & + \frac{\gamma \lvert\Omega_{ge}\alpha\rvert^2 }{\Delta^2 + \dfrac{\gamma^2}{4}} P_{gg}.
		\end{split}
	\end{equation}
	Taken these sets of equations (\ref{eq:PggEffective}), (\ref{eq:AlphaEffective}) and (\ref{eq:AlphaPEffective}), we see that a pattern emerge even though (\ref{eq:PggEffective}) is zero. There exist an effective Hamiltonian acting on the ground state of the atom and the quantum state of light of the form
	\begin{equation}
		\label{eq:EffectiveHamitonianQ}
		\tilde{H}_\mathrm{eff} = \hbar \frac{\lvert\Omega_{ge}\rvert^2\left(-\Delta -i\dfrac{\gamma}{2}\right)}{\Delta^2 + \dfrac{\gamma^2}{4}} \hat{a}^\dagger\hat{a}\lvert g\rangle\langle g \lvert.
	\end{equation}
	However, the last term appearing in (\ref{eq:PggEffective}) does not show up in the evolution of the light's complex amplitude (\ref{eq:AlphaEffective}) or (\ref{eq:AlphaPEffective}). This term is as a result of randomness introduced in the states of the atom when the atom spontaneously emits photon in some random direction. Physically, the quantum state of atoms that has undergone spontaneous emission is different from its state before photon absorption. Hence, the evolution of the ground state of the atom and the coherent state of light is 
	\begin{equation}
		\label{eq:MasterEquationQ} 
		\dot{\rho} = -\frac{i}{\hbar} \left(\tilde{H}_\mathrm{eff}\rho -\rho \tilde{H}^*_\mathrm{eff}\right) + \frac{\gamma\lvert\Omega_{ge}\rvert^2}{\Delta^2 + \frac{\gamma^2}{4}}\hat{a}\lvert g\rangle\langle g\rvert \rho\lvert g\rangle \langle g\rvert\hat{a}^\dagger.
	\end{equation}
	Equation (\ref{eq:MasterEquationQ}) is the ground state evolution master equation given in (\ref{eq:MasterEquation}) of the main text.

	\section{Effective ground state evolution in a two-level system}\label{sec:GroundStateEvolution}
	In this section, we derive the master equation for ground state evolution in the presence of spontaneous emission, within the adiabatic approximation~\cite{ilo-okeke2014}. We first derive the ground state evolution of a two-level atom interacting with coherent light. Later we extend the derivation to an atom qubit interacting with coherent light, and then generalize it to an ensemble of atom qubits.

	\subsection{The Ground State Master Equation Of A Two-Level Atom}\label{sec:sec:MasterEquationTwoLevelAtom}
	Consider a two-level atom that consists of an excited state $\lvert e\rangle $ and a ground state $\lvert g \rangle$ interacting with a monochromatic laser light beam of frequency $\omega$. With the ground state energy as the reference and within the dipole approximation, the evolution of states of light and the atom is governed by the following Hamiltonian  
	\begin{equation}
		\label{eq:Hamiltonian}
		H = \hbar \Delta \lvert e\rangle\langle e \rvert + \hbar\Omega_{ge}^*\sigma_+\hat{a} + \hbar \Omega_{ge}\hat{a}^\dagger\sigma_-,
	\end{equation}
	where $\Omega_{ge}$ is the atom-photon coupling frequency between the atomic levels, $\Delta =  \omega_e -\omega$ is the detuning of the light frequency $\omega$ from the excited state's frequency $\omega_e$, and $\sigma_+ =\lvert e\rangle\langle g\rvert$ is the atomic excitation operator, while $\sigma_- = \lvert g\rangle\langle e\rvert$ is the atomic lowering operator. The operator $\hat{a}$ is the photon annihilation operator whose action on the vacuum state $\lvert 0\rangle$ is to destroy it, $\hat{a}\lvert 0 \rangle = 0$. Additionally, the photon operators $\hat{a},\hat{a}^\dagger$ satisfies the commutation relation $[\hat{a},\hat{a}^\dagger] =1$.

	We assume the combined state of the atom and light is a product state of the form
	\begin{equation}
		\label{eq:AtomLightState}
		\rho = \rho_\mathrm{atom}\otimes \lvert \alpha\rangle\langle\alpha \rvert,
	\end{equation}
	where the light state is a coherent state, $\lvert\alpha\rangle = e^{-|\alpha|^2/2} e^{\alpha a^\dagger}\lvert 0 \rangle$, and $\alpha$ is the complex photon number amplitude of the laser light that has time dependence. The state of the atom $\rho_\mathrm{atom}$ can be decomposed in terms of its two levels, the excited and ground states. The evolution of the density matrix (\ref{eq:AtomLightState}) accounting for the spontaneous emission of the atom is
	\begin{equation}
		\label{eq:DensityMatrixEvolution}
		\dot{\rho} = -\frac{i}{\hbar }\left[H,\rho\right] +\frac{\gamma}{2}\left(2\sigma_-\rho\sigma_+ - \sigma_+\sigma_-\rho -\rho\sigma_+\sigma_- \right),
	\end{equation}
	where $\gamma$ is the atom's spontaneous decay rate.

	Initially, the atom is prepared in its ground state with a unit probability. Its subsequent evolution in the presence of light has negligible effect on the ground state probability. As such, the ground state evolves slowly due to the applied field, and the rate of change in the probability of the excited state, and the coherences of the two levels, are negligibly small. To this end, we assume $\lvert\Omega_{eg}\alpha\rvert^2/\Delta^2 \ll1$ and $\gamma^2/\Delta^2 \ll 1$, and we seek the Hamiltonian governing the evolution of the ground state and coherent state of light. In Appendix \ref{sec:DerivationGroundStateMasterEquation}, we show the detailed derivation and summarize the results below. By adiabatically eliminating the excited state evolution, we find the time dependence of the photon number amplitude
	\begin{equation}
		\label{eq:AlphaSolution}
		\alpha(t) = \alpha(0)\exp\left({i\frac{\Delta\lvert\Omega_{ge}\rvert^2 t}{\Delta^2 + \frac{\gamma^2}{4}}}\right)\exp\left( -\frac{ \gamma\lvert\Omega_{ge}\rvert^2t}{2\left(\Delta^2 + \frac{\gamma^2 }{4}\right)}\right).
	\end{equation}
	The first exponent of (\ref{eq:AlphaSolution}) shows that light accumulates a phase due to its interaction with the atom. The second exponential shows that the amplitude of light decays. This implies that not all the photons that interacted with the atom will reach the detector. Those lost are emitted in random directions away from the detectors. This decay is often described in terms of optical depth in QND experiments~\cite{appel2009,sorensen1998,sewell2012,sewell2013}. The optical depth can be controlled---making it achieve a value close to unity---by making the detuning large compared to the spontaneous emission rate. This makes the denominator of the decay term large compared to the numerator, which in turn makes the atoms more transparent to the incident beam.

	Furthermore, with large detuning from atomic resonance transition,  the coherences and excited state evolve faster than the ground state. Provided the excited state population decays at a rate faster than the time scale of the ground state evolution, the coherences and excited state probability are negligibly small. Adiabatically eliminating them causes the ground state to accumulate a global phase. Thus, the ground state of the atom evolves according to the following master equation as detailed in Appendix \ref{sec:DerivationGroundStateMasterEquation}
	\begin{equation}
		\label{eq:MasterEquation} 
		\dot{\rho} = -\frac{i}{\hbar} \left(\tilde{H}_\mathrm{eff}\rho -\rho \tilde{H}^\dagger_\mathrm{eff}\right) + \hat{\gamma}_{\mathrm{eff}}\hat{a}\lvert g\rangle\langle g\rvert \rho\lvert g\rangle \langle g\rvert\hat{a}^\dagger,
	\end{equation}
	where the effective non-Hermitian Hamiltonian is
	\begin{equation}
		\label{eq:EffectiveHamitonian}
		\tilde{H}_\mathrm{eff} = -\hbar G \hat{a}^\dagger\hat{a}\lvert g\rangle\langle g \lvert - i\hbar\frac{\tilde{\gamma}_{\mathrm{eff}}}{2}\hat{a}^\dagger\hat{a}(\lvert g\rangle\langle g \rvert)^2.
	\end{equation}
	The frequencies are defined as 
	\begin{equation}
		\label{eq:EffectiveFrequencies}
		G = \Delta\frac{\lvert\Omega_{ge}\rvert^2 }{\Delta^2 + \dfrac{\gamma^2}{4}},\qquad \tilde{\gamma}_\mathrm{eff} =  \gamma \frac{\lvert\Omega_{ge}\rvert^2}{\Delta^2 + \dfrac{\gamma^2}{4}}. 
	\end{equation}
	For a family of ground states, such as the hyperfine states of an atom, (\ref{eq:MasterEquation}) predicts that the ground state probabilities do not evolve. Instead, the ground states accumulate global phases. While this is physically unobservable,  preparing the ground states in a coherent linear superposition causes the evolution of their coherences with time because of the different phases of each ground state in the superposition. The evolution of the coherences becomes clear by writing (\ref{eq:MasterEquation}) using the relative population difference of ground states. If we are not interested in the light's evolution, one can replace the photon operators by their averages $\langle \alpha\lvert \hat{a}^\dagger\hat{a} \rvert\alpha\rangle =\lvert\alpha\rvert^2 = \langle \alpha\lvert \hat{a}^\dagger\rvert\alpha\rangle \langle \alpha\lvert\hat{a} \rvert\alpha\rangle $, and obtain the usual dephasing ground state master equation~\cite{ban2009,tempel2011}.
	We note that (\ref{eq:MasterEquation}) may be put in the Lindblad form as 
	\begin{equation}
		\label{eq:LindbladMasterEquation}
		\begin{split}
			\dot{\rho}  =& -\frac{i}{\hbar} [\hbar G \hat{a}^\dagger\hat{a}\lvert g\rangle\langle g\rvert,\rho] + \gamma_{\mathrm{eff}} \Bigg(\hat{a}\lvert g\rangle\langle g\rvert\rho\lvert g\rangle\langle g\rvert\hat{a}^\dagger \\
			&- \frac{1}{2}\Big\{\hat{a}^\dagger\hat{a}\lvert g\rangle\langle g\rvert,\rho \Big\}\Bigg).
		\end{split}
	\end{equation}

	\section{Matrix Solution of Master Equation \label{sec:SolutionAtomEnsemble}}
	Substituting the ansatz (\ref{eq:ansatz}) in (\ref{eq:EffectiveMasterEquation}), we obtain the following coupled equations,
	\begin{align}
		\label{eq:CoupledEquation_spin}
		\dot{\rho}_{m_z,m'_z} &= -\frac{R}{2}\left( m_z^2 \alpha_{m_z}\alpha^*_{m_z} + m'^2_z\alpha_{m'_z}\alpha^*_{m'_z} \right) \rho_{m_z,m'_z}\nonumber \\ & + R\alpha_{m_z}\alpha^*_{m'_z} m_z m'_z \rho_{m_z,m'_z},\\
		\label{eq:CoupledEquation_photon}
		\dot{\alpha}_{m_z} &= \left(-i m_z - \frac{R}{2}m^2_z\right)\alpha_{m_z}.
	\end{align}
	The solution of (\ref{eq:CoupledEquation_photon}) is given in (\ref{eq:PhotonAmplitude}). Similar expression is obtained for $\alpha^*_{m'_z}$ with $m_z =m'_z$. Upon substituting (\ref{eq:PhotonAmplitude}), one obtains the solution  of the spin matrix elements
	\begin{equation}
		\label{eq:MatrixElementSolution}
		\begin{split}
			\rho_{m_z,m'_z}(\tau) = 
			&\exp\bigg[\frac{\lvert \alpha\rvert^2}{2} \left(e^{-R m_z^2 \tau} + e^{-R m'^2_z\tau} -2\right)\\
			& + \frac{R\lvert\alpha\rvert^2 m_zm'_z  }{i(m'_z - m_z) -\frac{R}{2} (m_z^2+m'^2_z)}\\
			& \times \left(e^{i\tau(m'_z - m_z) -\frac{R}{2}\tau (m^2_z+m'^2_z)} - 1 \right) \bigg]\\
			&\times \rho_{m_z,m'_z}(0). 
		\end{split}
	\end{equation}

	\section{Formal Solution to the Master Equation}\label{sec:FormalSolutionMasterEquation}
	In seeking the solution of the master equation, we observe that the right hand side of the master equation (\ref{eq:EffectiveMasterEquation}) behaves as a source. Borrowing the techniques used in solving the differential equation~\cite{boyce2005,boas2006}, we treat the jump part (the right hand side) of the master equation as source. The terms on the left hand side of (\ref{eq:EffectiveMasterEquation}) governs the deterministic free evolution of the system. The solution obtained with the right hand side set to zero gives the  solution of the homogeneous differential equation. The source term gives the particular solution of the differential equation~\cite{boyce2005,boas2006}. Hence, the solution of (\ref{eq:EffectiveMasterEquation}) is a combination of the homogeneous part and the particular solution~\cite{boyce2005,boas2006}. 
	
	Finding the solution of~(\ref{eq:EffectiveMasterEquation}) begins with finding the integrating factor~\cite{boyce2005,boas2006}, which is related to the homogeneous solution by inverse transformation. The integrating factor transforms the differential equation such that the homogenous part is removed from the resulting differential equation while leaving a transformed source---the rotating frame transformation. In the case at hand, the integrating factors are  $e^{\frac{i}{\hbar}H_\mathrm{eff}\tau},\,e^{-\frac{i}{\hbar}H^*_\mathrm{eff}\tau}$. The density matrix $\rho$ is then transformed as $\tilde{\rho} = e^{\frac{i}{\hbar}H_\mathrm{eff}\tau}\rho e^{-\frac{i}{\hbar}H^*_\mathrm{eff}\tau}$, while the master equation~(\ref{eq:EffectiveMasterEquation}) reduces to
	\begin{equation}
		\label{eq:TransformedMasterEquation}
		\dot{\tilde{\rho}} = \tilde{\mathcal{L}}_s\tilde{\rho},
	\end{equation}
	where the superoperator~\cite{breuer2002,carmichael2008} $\tilde{\mathcal{L}}_s$  is  
	\begin{equation}
		\label{eq:TransformedSuperopertorApp}
		\begin{split}
			\tilde{\mathcal{L}}_s ()& = e^{\frac{i}{\hbar}H_\mathrm{eff}\tau}\hat{a}J_z()J_z\hat{a}^\dagger e^{-\frac{i}{\hbar}H^*_\mathrm{eff}\tau},\\
			\tilde{\mathcal{L}}_s() &= R\,  \hat{a}e^{\hat{w}\tau} J_z() J_ze^{\hat{w}^*\tau}\hat{a}^\dagger,
		\end{split}
	\end{equation}
	and the complex frequency $\hat{w}$ operator is
	\begin{equation}
		\label{eq:ComplexFrequencyApp}
		\hat{w} = -i J_z - \dfrac{R}{2} J^2_z.
	\end{equation} 
	In writing (\ref{eq:TransformedSuperopertorApp}), we made use of the Baker-Campbell-Hausdorff theorem, and the identity operator defined as $e^{\frac{i}{\hbar}H_\mathrm{eff}\tau} e^{-\frac{i}{\hbar}H_\mathrm{eff}\tau} = \mathbbm{1} = e^{\frac{i}{\hbar}H^*_\mathrm{eff}\tau} e^{-\frac{i}{\hbar}H^*_\mathrm{eff}\tau}$.
	
	Noting that $\tilde{\rho} (\tau=0) = \rho(0)$, (\ref{eq:TransformedMasterEquation}) can easily be integrated to give
	\begin{equation}
		\label{eq:SolutionTransformedState}
		\begin{split}
			\tilde{\rho}& = e^{\int_{0}^{\tau}d\tau'\,  \tilde{\mathcal{L}}_s() }\rho(0) \\
			&= \sum_{q = 0}^{\infty} \frac{(R \int_{0}^{\tau} d\tau')^q}{q!} (\hat{a}e^{\hat{w}\tau'} J_z)^q \rho(0) (J_ze^{\hat{w}^*\tau'}\hat{a}^\dagger)^q.
		\end{split}
	\end{equation}
	The solution $\rho(t)$ is obtained by multiplying~(\ref{eq:SolutionTransformedState}) from the left by $e^{-\frac{i}{\hbar} H_\mathrm{eff}\tau}$ and from the right by $e^{\frac{i}{\hbar} H_\mathrm{eff}^*\tau}$ to give (\ref{eq:GeneralSolution}) in the main text.

	\section{Disentangling Exponential Operator}\label{sec:DisentanglingOperator}
	Given a differential equation of the form (\ref{eq:EffectiveMasterEquation}), its solution relating the density matrix $\rho$ at any time $\tau$ to that at $\tau = 0$ is  
	\begin{equation}
		\label{eq:FormSolution}
		\rho(\tau) = \hat{U}\rho(0),
	\end{equation}
	where the operator $\hat{U}$ is defined as 
	\begin{equation}
		\label{eq:EvolutionOperator}
		\hat{U} =  e^{\tau[\mathcal{L}_H() + \mathcal{L}_s()]}.
	\end{equation}	
	The jump superoperator $\mathcal{L}_s$ is defined as  
	\begin{equation}
		\mathcal{L}_s () = R\hat{a}J_z ()J_z\hat{a}^\dagger.
	\end{equation}
	The superoperator $\mathcal{L}_H$ which controls the homogeneous evolution is defined as 
	\begin{equation}
		\mathcal{L}_H ()= \hat{w}(J_z)\hat{a}^\dagger\hat{a} () + ()\hat{w}^*(J_z)\hat{a}^\dagger\hat{a},
	\end{equation}
	where the function of the operator $\hat{w}(J_z)$ is defined in (\ref{eq:ComplexFrequencyApp}). 
	
	It is more convenient to have the operator $\hat{U}$ as a product of operators. As such, it is important to know the commutation relation between the operators in the exponent of $\hat{U}$. The superoperators $\mathcal{L}_H()$ and $\mathcal{L}_s()$ satisfy the following commutation relation
	\begin{equation}
		\label{eq:SuperoperatorCommutationRelation}
		\begin{split}
			\left[\mathcal{L}_H,\mathcal{L}_s \right]\rho &= (\mathcal{L}_H\mathcal{L}_s -\mathcal{L}_s\mathcal{L}_H)\rho\\
			& = -\left(\hat{w}(J_z)() + ()\hat{w}^*(J_z)  \right)\mathcal{L}_s\rho.
		\end{split}
	\end{equation}
	Thus any equivalent formulation of the operator $\hat{U}$ will be of the form~\cite{ilo-okeke2023}
	\begin{equation}
		\label{eq:EquivalentEvolutionOperator}
		\hat{U} = e^{p(\tau)\mathcal{L}_H}e^{r(\tau)\mathcal{L}_s}.
	\end{equation}
	To determine the forms of $p(\tau)$ and $r(\tau)$, we take the derivative of (\ref{eq:EvolutionOperator}) and (\ref{eq:EquivalentEvolutionOperator}). The derivative of (\ref{eq:EvolutionOperator}) gives 
	\begin{equation}
		\label{eq:DerivativeEvolutionOperator}
		\frac{d\hat{U}}{d\tau} = \left(\mathcal{L}_H + \mathcal{L}_s\right)\hat{U},
	\end{equation}
	while that of (\ref{eq:EquivalentEvolutionOperator}) gives
	\begin{equation}
		\label{eq:DerivativeEquivalentFormulation}
		\frac{d\hat{U}}{d\tau} = \left(\dot{p} \mathcal{L}_H + e^{p(\tau)}\dot{r} \mathcal{L}_s e^{-p(\tau)} \right)\hat{U},
	\end{equation}
	where $\dot{q} = \frac{d q}{d\tau}$.
	
	Applying the Baker-Campbell-Hausdorff relation to (\ref{eq:DerivativeEquivalentFormulation}), and equating coefficients of operators in (\ref{eq:DerivativeEvolutionOperator}) and (\ref{eq:DerivativeEquivalentFormulation}) gives the solution of the functions under the conditions $p(\tau = 0) = 0$ and $r(\tau = 0) = 0$ as
	\begin{align}
		\label{eq:CoupledEquationSolI}
		p & = \tau, \\
		\label{eq:CoupledEquationSolII}
		r & = \int^\tau_0 d\tau'\,e^{\tau'\left[ \hat{w}(J_z)() + ()\hat{w}^*(J_z)  \right]}.
	\end{align}
	Substituting (\ref{eq:CoupledEquationSolI}) and (\ref{eq:CoupledEquationSolII}) in  (\ref{eq:EquivalentEvolutionOperator}) and using (\ref{eq:EquivalentEvolutionOperator}) in (\ref{eq:FormSolution}) gives same solution as (\ref{eq:GeneralSolution}) in the main text.


\begin{thebibliography}{58}%
	\makeatletter
	\providecommand \@ifxundefined [1]{%
		\@ifx{#1\undefined}
	}%
	\providecommand \@ifnum [1]{%
		\ifnum #1\expandafter \@firstoftwo
		\else \expandafter \@secondoftwo
		\fi
	}%
	\providecommand \@ifx [1]{%
		\ifx #1\expandafter \@firstoftwo
		\else \expandafter \@secondoftwo
		\fi
	}%
	\providecommand \natexlab [1]{#1}%
	\providecommand \enquote  [1]{``#1''}%
	\providecommand \bibnamefont  [1]{#1}%
	\providecommand \bibfnamefont [1]{#1}%
	\providecommand \citenamefont [1]{#1}%
	\providecommand \href@noop [0]{\@secondoftwo}%
	\providecommand \href [0]{\begingroup \@sanitize@url \@href}%
	\providecommand \@href[1]{\@@startlink{#1}\@@href}%
	\providecommand \@@href[1]{\endgroup#1\@@endlink}%
	\providecommand \@sanitize@url [0]{\catcode `\\12\catcode `\$12\catcode
		`\&12\catcode `\#12\catcode `\^12\catcode `\_12\catcode `\%12\relax}%
	\providecommand \@@startlink[1]{}%
	\providecommand \@@endlink[0]{}%
	\providecommand \url  [0]{\begingroup\@sanitize@url \@url }%
	\providecommand \@url [1]{\endgroup\@href {#1}{\urlprefix }}%
	\providecommand \urlprefix  [0]{URL }%
	\providecommand \Eprint [0]{\href }%
	\providecommand \doibase [0]{https://doi.org/}%
	\providecommand \selectlanguage [0]{\@gobble}%
	\providecommand \bibinfo  [0]{\@secondoftwo}%
	\providecommand \bibfield  [0]{\@secondoftwo}%
	\providecommand \translation [1]{[#1]}%
	\providecommand \BibitemOpen [0]{}%
	\providecommand \bibitemStop [0]{}%
	\providecommand \bibitemNoStop [0]{.\EOS\space}%
	\providecommand \EOS [0]{\spacefactor3000\relax}%
	\providecommand \BibitemShut  [1]{\csname bibitem#1\endcsname}%
	\let\auto@bib@innerbib\@empty
	\bibitem [{\citenamefont {Braginsky}\ \emph {et~al.}(1980)\citenamefont
		{Braginsky}, \citenamefont {Vorontsov},\ and\ \citenamefont
		{Thorne}}]{braginsky1980}%
	\BibitemOpen
	\bibfield  {author} {\bibinfo {author} {\bibfnamefont {V.~B.}\ \bibnamefont
			{Braginsky}}, \bibinfo {author} {\bibfnamefont {Y.~I.}\ \bibnamefont
			{Vorontsov}},\ and\ \bibinfo {author} {\bibfnamefont {K.~S.}\ \bibnamefont
			{Thorne}},\ }\href {https://doi.org/10.1126/science.209.4456.547} {\bibfield
		{journal} {\bibinfo  {journal} {Science}\ }\textbf {\bibinfo {volume}
			{209}},\ \bibinfo {pages} {547} (\bibinfo {year} {1980})}\BibitemShut
	{NoStop}%
	\bibitem [{\citenamefont {Grangier}\ \emph {et~al.}(1998)\citenamefont
		{Grangier}, \citenamefont {Levenson},\ and\ \citenamefont
		{Poizat}}]{grangier1998}%
	\BibitemOpen
	\bibfield  {author} {\bibinfo {author} {\bibfnamefont {P.}~\bibnamefont
			{Grangier}}, \bibinfo {author} {\bibfnamefont {J.~A.}\ \bibnamefont
			{Levenson}},\ and\ \bibinfo {author} {\bibfnamefont {J.}~\bibnamefont
			{Poizat}},\ }\href@noop {} {\bibfield  {journal} {\bibinfo  {journal}
			{Nature}\ }\textbf {\bibinfo {volume} {396}},\ \bibinfo {pages} {537}
		(\bibinfo {year} {1998})}\BibitemShut {NoStop}%
	\bibitem [{\citenamefont {Blatt}\ and\ \citenamefont
		{Wineland}(2008)}]{blatt2008}%
	\BibitemOpen
	\bibfield  {author} {\bibinfo {author} {\bibfnamefont {R.}~\bibnamefont
			{Blatt}}\ and\ \bibinfo {author} {\bibfnamefont {D.}~\bibnamefont
			{Wineland}},\ }\href@noop {} {\bibfield  {journal} {\bibinfo  {journal}
			{Nature}\ }\textbf {\bibinfo {volume} {453}},\ \bibinfo {pages} {1008}
		(\bibinfo {year} {2008})}\BibitemShut {NoStop}%
	\bibitem [{\citenamefont {Devoret}\ and\ \citenamefont
		{Schoelkopf}(2013)}]{devoret2013}%
	\BibitemOpen
	\bibfield  {author} {\bibinfo {author} {\bibfnamefont {M.~H.}\ \bibnamefont
			{Devoret}}\ and\ \bibinfo {author} {\bibfnamefont {R.~J.}\ \bibnamefont
			{Schoelkopf}},\ }\href {https://doi.org/10.1126/science.1231930} {\bibfield
		{journal} {\bibinfo  {journal} {Science}\ }\textbf {\bibinfo {volume}
			{339}},\ \bibinfo {pages} {1169} (\bibinfo {year} {2013})}\BibitemShut
	{NoStop}%
	\bibitem [{\citenamefont {{C. Guerlin and J. Bernu and S. Del\'eglise and C.
				Sayrin and S. Gleyzes and S. Kuhr and M. Brune and J. -M. Raimond and S.
				Haroche}}(2007)}]{guerlin2007}%
	\BibitemOpen
	\bibfield  {author} {\bibinfo {author} {\bibnamefont {{C. Guerlin and J.
					Bernu and S. Del\'eglise and C. Sayrin and S. Gleyzes and S. Kuhr and M.
					Brune and J. -M. Raimond and S. Haroche}}},\ }\href
	{https://doi.org/10.1038/nature06057} {\bibfield  {journal} {\bibinfo
			{journal} {Nature}\ }\textbf {\bibinfo {volume} {448}},\ \bibinfo {pages}
		{889} (\bibinfo {year} {2007})}\BibitemShut {NoStop}%
	\bibitem [{\citenamefont {Eberle}\ \emph {et~al.}(2010)\citenamefont {Eberle},
		\citenamefont {Steinlechner}, \citenamefont {Bauchrowitz}, \citenamefont
		{H{\"a}ndchen}, \citenamefont {Vahlbruch}, \citenamefont {Mehmet},
		\citenamefont {M{\"u}ller-Ebhardt},\ and\ \citenamefont
		{Schnabel}}]{eberle2010}%
	\BibitemOpen
	\bibfield  {author} {\bibinfo {author} {\bibfnamefont {T.}~\bibnamefont
			{Eberle}}, \bibinfo {author} {\bibfnamefont {S.}~\bibnamefont
			{Steinlechner}}, \bibinfo {author} {\bibfnamefont {J.}~\bibnamefont
			{Bauchrowitz}}, \bibinfo {author} {\bibfnamefont {V.}~\bibnamefont
			{H{\"a}ndchen}}, \bibinfo {author} {\bibfnamefont {H.}~\bibnamefont
			{Vahlbruch}}, \bibinfo {author} {\bibfnamefont {M.}~\bibnamefont {Mehmet}},
		\bibinfo {author} {\bibfnamefont {H.}~\bibnamefont {M{\"u}ller-Ebhardt}},\
		and\ \bibinfo {author} {\bibfnamefont {R.}~\bibnamefont {Schnabel}},\ }\href
	{https://doi.org/10.1103/PhysRevLett.104.251102} {\bibfield  {journal}
		{\bibinfo  {journal} {Phys. Rev. Lett.}\ }\textbf {\bibinfo {volume} {104}},\
		\bibinfo {pages} {251102} (\bibinfo {year} {2010})}\BibitemShut {NoStop}%
	\bibitem [{\citenamefont {Pitkin}\ \emph {et~al.}(2011)\citenamefont {Pitkin},
		\citenamefont {Reid}, \citenamefont {Rowan},\ and\ \citenamefont
		{Hough}}]{pitkin2011}%
	\BibitemOpen
	\bibfield  {author} {\bibinfo {author} {\bibfnamefont {M.}~\bibnamefont
			{Pitkin}}, \bibinfo {author} {\bibfnamefont {S.}~\bibnamefont {Reid}},
		\bibinfo {author} {\bibfnamefont {S.}~\bibnamefont {Rowan}},\ and\ \bibinfo
		{author} {\bibfnamefont {J.}~\bibnamefont {Hough}},\ }\href
	{https://doi.org/10.12942/lrr-2011-5} {\bibfield  {journal} {\bibinfo
			{journal} {Living Rev. Relativ.}\ }\textbf {\bibinfo {volume} {14}},\
		\bibinfo {pages} {5} (\bibinfo {year} {2011})}\BibitemShut {NoStop}%
	\bibitem [{\citenamefont {Loudon}(2000)}]{loudon2000}%
	\BibitemOpen
	\bibfield  {author} {\bibinfo {author} {\bibfnamefont {R.}~\bibnamefont
			{Loudon}},\ }\href@noop {} {\emph {\bibinfo {title} {The Quantum Theory of
				Light}}}\ (\bibinfo  {publisher} {Oxford University Press},\ \bibinfo
	{address} {New York},\ \bibinfo {year} {2000})\BibitemShut {NoStop}%
	\bibitem [{\citenamefont {Scully}\ and\ \citenamefont
		{Zubairy}(1997)}]{scully1997}%
	\BibitemOpen
	\bibfield  {author} {\bibinfo {author} {\bibfnamefont {M.~O.}\ \bibnamefont
			{Scully}}\ and\ \bibinfo {author} {\bibfnamefont {M.~S.}\ \bibnamefont
			{Zubairy}},\ }\href@noop {} {\emph {\bibinfo {title} {Quantum Optics}}}\
	(\bibinfo  {publisher} {Cambridge University Press},\ \bibinfo {address}
	{Cambridge, UK},\ \bibinfo {year} {1997})\BibitemShut {NoStop}%
	\bibitem [{\citenamefont {Walls}\ and\ \citenamefont
		{Milburn}(2008)}]{walls2008}%
	\BibitemOpen
	\bibfield  {author} {\bibinfo {author} {\bibfnamefont {D.~F.}\ \bibnamefont
			{Walls}}\ and\ \bibinfo {author} {\bibfnamefont {G.~J.}\ \bibnamefont
			{Milburn}},\ }\href@noop {} {\emph {\bibinfo {title} {Quantum Optics}}},\
	\bibinfo {edition} {2nd}\ ed.\ (\bibinfo  {publisher} {Springer-Verlag},\
	\bibinfo {address} {Berlin},\ \bibinfo {year} {2008})\BibitemShut {NoStop}%
	\bibitem [{\citenamefont {Byrnes}\ and\ \citenamefont
		{Ilo-Okeke}(2021)}]{byrnes2021}%
	\BibitemOpen
	\bibfield  {author} {\bibinfo {author} {\bibfnamefont {T.}~\bibnamefont
			{Byrnes}}\ and\ \bibinfo {author} {\bibfnamefont {E.~O.}\ \bibnamefont
			{Ilo-Okeke}},\ }\href {https://doi.org/doi:10.1017/9781108975353} {\emph
		{\bibinfo {title} {{Quantum Atom Optics: Theory and Applications to
					Technology}}}}\ (\bibinfo  {publisher} {Cambridge University Press},\
	\bibinfo {address} {Cambridge},\ \bibinfo {year} {2021})\BibitemShut
	{NoStop}%
	\bibitem [{\citenamefont {Brune}\ \emph {et~al.}(1990)\citenamefont {Brune},
		\citenamefont {Haroche}, \citenamefont {Lefevre}, \citenamefont {Raimond},\
		and\ \citenamefont {Zagury}}]{brune1990}%
	\BibitemOpen
	\bibfield  {author} {\bibinfo {author} {\bibfnamefont {M.}~\bibnamefont
			{Brune}}, \bibinfo {author} {\bibfnamefont {S.}~\bibnamefont {Haroche}},
		\bibinfo {author} {\bibfnamefont {V.}~\bibnamefont {Lefevre}}, \bibinfo
		{author} {\bibfnamefont {J.~M.}\ \bibnamefont {Raimond}},\ and\ \bibinfo
		{author} {\bibfnamefont {N.}~\bibnamefont {Zagury}},\ }\href
	{https://doi.org/10.1103/PhysRevLett.65.976} {\bibfield  {journal} {\bibinfo
			{journal} {Phys. Rev. Lett.}\ }\textbf {\bibinfo {volume} {65}},\ \bibinfo
		{pages} {976} (\bibinfo {year} {1990})}\BibitemShut {NoStop}%
	\bibitem [{\citenamefont {Holland}\ \emph {et~al.}(1991)\citenamefont
		{Holland}, \citenamefont {Walls},\ and\ \citenamefont
		{Zoller}}]{holland1991}%
	\BibitemOpen
	\bibfield  {author} {\bibinfo {author} {\bibfnamefont {M.~J.}\ \bibnamefont
			{Holland}}, \bibinfo {author} {\bibfnamefont {D.~F.}\ \bibnamefont {Walls}},\
		and\ \bibinfo {author} {\bibfnamefont {P.}~\bibnamefont {Zoller}},\ }\href
	{https://doi.org/10.1103/PhysRevLett.67.1716} {\bibfield  {journal} {\bibinfo
			{journal} {Phys. Rev. Lett.}\ }\textbf {\bibinfo {volume} {67}},\ \bibinfo
		{pages} {1716} (\bibinfo {year} {1991})}\BibitemShut {NoStop}%
	\bibitem [{\citenamefont {Ueda}\ \emph {et~al.}(1992)\citenamefont {Ueda},
		\citenamefont {Imoto}, \citenamefont {Nagaoka},\ and\ \citenamefont
		{Ogawa}}]{ueda1992}%
	\BibitemOpen
	\bibfield  {author} {\bibinfo {author} {\bibfnamefont {M.}~\bibnamefont
			{Ueda}}, \bibinfo {author} {\bibfnamefont {N.}~\bibnamefont {Imoto}},
		\bibinfo {author} {\bibfnamefont {H.}~\bibnamefont {Nagaoka}},\ and\ \bibinfo
		{author} {\bibfnamefont {T.}~\bibnamefont {Ogawa}},\ }\href
	{https://doi.org/10.1103/PhysRevA.46.2859} {\bibfield  {journal} {\bibinfo
			{journal} {Phys. Rev. A}\ }\textbf {\bibinfo {volume} {46}},\ \bibinfo
		{pages} {2859} (\bibinfo {year} {1992})}\BibitemShut {NoStop}%
	\bibitem [{\citenamefont {Yanagimoto}\ \emph {et~al.}(2023)\citenamefont
		{Yanagimoto}, \citenamefont {Nehra}, \citenamefont {Hamerly}, \citenamefont
		{Ng}, \citenamefont {Marandi},\ and\ \citenamefont
		{Mabuchi}}]{yanagimoto2023}%
	\BibitemOpen
	\bibfield  {author} {\bibinfo {author} {\bibfnamefont {R.}~\bibnamefont
			{Yanagimoto}}, \bibinfo {author} {\bibfnamefont {R.}~\bibnamefont {Nehra}},
		\bibinfo {author} {\bibfnamefont {R.}~\bibnamefont {Hamerly}}, \bibinfo
		{author} {\bibfnamefont {E.}~\bibnamefont {Ng}}, \bibinfo {author}
		{\bibfnamefont {A.}~\bibnamefont {Marandi}},\ and\ \bibinfo {author}
		{\bibfnamefont {H.}~\bibnamefont {Mabuchi}},\ }\href
	{https://doi.org/10.1103/PRXQuantum.4.010333} {\bibfield  {journal} {\bibinfo
			{journal} {PRX Quantum}\ }\textbf {\bibinfo {volume} {4}},\ \bibinfo {pages}
		{010333} (\bibinfo {year} {2023})}\BibitemShut {NoStop}%
	\bibitem [{\citenamefont {Lecocq}\ \emph {et~al.}(2015)\citenamefont {Lecocq},
		\citenamefont {Clark}, \citenamefont {Simmonds}, \citenamefont {Aumentado},\
		and\ \citenamefont {Teufel}}]{lecocq2015}%
	\BibitemOpen
	\bibfield  {author} {\bibinfo {author} {\bibfnamefont {F.}~\bibnamefont
			{Lecocq}}, \bibinfo {author} {\bibfnamefont {J.~B.}\ \bibnamefont {Clark}},
		\bibinfo {author} {\bibfnamefont {R.~W.}\ \bibnamefont {Simmonds}}, \bibinfo
		{author} {\bibfnamefont {J.}~\bibnamefont {Aumentado}},\ and\ \bibinfo
		{author} {\bibfnamefont {J.~D.}\ \bibnamefont {Teufel}},\ }\href
	{https://doi.org/10.1103/PhysRevX.5.041037} {\bibfield  {journal} {\bibinfo
			{journal} {Phys. Rev. X}\ }\textbf {\bibinfo {volume} {5}},\ \bibinfo {pages}
		{041037} (\bibinfo {year} {2015})}\BibitemShut {NoStop}%
	\bibitem [{\citenamefont {Takahashi}\ \emph {et~al.}(1999)\citenamefont
		{Takahashi}, \citenamefont {Honda}, \citenamefont {Tanaka}, \citenamefont
		{Toyoda}, \citenamefont {Ishikawa},\ and\ \citenamefont
		{Yabuzaki}}]{takahashi1999}%
	\BibitemOpen
	\bibfield  {author} {\bibinfo {author} {\bibfnamefont {Y.}~\bibnamefont
			{Takahashi}}, \bibinfo {author} {\bibfnamefont {K.}~\bibnamefont {Honda}},
		\bibinfo {author} {\bibfnamefont {N.}~\bibnamefont {Tanaka}}, \bibinfo
		{author} {\bibfnamefont {K.}~\bibnamefont {Toyoda}}, \bibinfo {author}
		{\bibfnamefont {K.}~\bibnamefont {Ishikawa}},\ and\ \bibinfo {author}
		{\bibfnamefont {T.}~\bibnamefont {Yabuzaki}},\ }\href@noop {} {\bibfield
		{journal} {\bibinfo  {journal} {Phys. Rev. A}\ }\textbf {\bibinfo {volume}
			{60}},\ \bibinfo {pages} {4974} (\bibinfo {year} {1999})}\BibitemShut
	{NoStop}%
	\bibitem [{\citenamefont {Kuzmich}\ \emph {et~al.}(2000)\citenamefont
		{Kuzmich}, \citenamefont {Mandel},\ and\ \citenamefont
		{Bigelow}}]{kuzmich2000}%
	\BibitemOpen
	\bibfield  {author} {\bibinfo {author} {\bibfnamefont {A.}~\bibnamefont
			{Kuzmich}}, \bibinfo {author} {\bibfnamefont {L.}~\bibnamefont {Mandel}},\
		and\ \bibinfo {author} {\bibfnamefont {N.~P.}\ \bibnamefont {Bigelow}},\
	}\href {https://doi.org/10.1103/PhysRevLett.85.1594} {\bibfield  {journal}
		{\bibinfo  {journal} {Phys. Rev. Lett.}\ }\textbf {\bibinfo {volume} {85}},\
		\bibinfo {pages} {1594} (\bibinfo {year} {2000})}\BibitemShut {NoStop}%
	\bibitem [{\citenamefont {Higbie}\ \emph {et~al.}(2005)\citenamefont {Higbie},
		\citenamefont {Sadler}, \citenamefont {Inouye}, \citenamefont {Chikkatur},
		\citenamefont {Leslie}, \citenamefont {Moore}, \citenamefont {Savalli},\ and\
		\citenamefont {Stamper-Kurn}}]{higbie2005}%
	\BibitemOpen
	\bibfield  {author} {\bibinfo {author} {\bibfnamefont {J.~M.}\ \bibnamefont
			{Higbie}}, \bibinfo {author} {\bibfnamefont {L.~E.}\ \bibnamefont {Sadler}},
		\bibinfo {author} {\bibfnamefont {S.}~\bibnamefont {Inouye}}, \bibinfo
		{author} {\bibfnamefont {A.~P.}\ \bibnamefont {Chikkatur}}, \bibinfo {author}
		{\bibfnamefont {S.~R.}\ \bibnamefont {Leslie}}, \bibinfo {author}
		{\bibfnamefont {K.~L.}\ \bibnamefont {Moore}}, \bibinfo {author}
		{\bibfnamefont {V.}~\bibnamefont {Savalli}},\ and\ \bibinfo {author}
		{\bibfnamefont {D.~M.}\ \bibnamefont {Stamper-Kurn}},\ }\href@noop {}
	{\bibfield  {journal} {\bibinfo  {journal} {Phys. Rev. Lett.}\ }\textbf
		{\bibinfo {volume} {95}},\ \bibinfo {pages} {050401} (\bibinfo {year}
		{2005})}\BibitemShut {NoStop}%
	\bibitem [{\citenamefont {Meppelink}\ \emph {et~al.}(2010)\citenamefont
		{Meppelink}, \citenamefont {Rozendaal}, \citenamefont {Koller}, \citenamefont
		{Vogels},\ and\ \citenamefont {van~der Straten}}]{meppelink2010}%
	\BibitemOpen
	\bibfield  {author} {\bibinfo {author} {\bibfnamefont {R.}~\bibnamefont
			{Meppelink}}, \bibinfo {author} {\bibfnamefont {R.~A.}\ \bibnamefont
			{Rozendaal}}, \bibinfo {author} {\bibfnamefont {S.~B.}\ \bibnamefont
			{Koller}}, \bibinfo {author} {\bibfnamefont {J.~M.}\ \bibnamefont {Vogels}},\
		and\ \bibinfo {author} {\bibfnamefont {P.}~\bibnamefont {van~der Straten}},\
	}\href {https://doi.org/10.1103/PhysRevA.81.053632} {\bibfield  {journal}
		{\bibinfo  {journal} {Phys. Rev. A}\ }\textbf {\bibinfo {volume} {81}},\
		\bibinfo {pages} {053632} (\bibinfo {year} {2010})}\BibitemShut {NoStop}%
	\bibitem [{\citenamefont {Schleier-Smith}\ \emph {et~al.}(2010)\citenamefont
		{Schleier-Smith}, \citenamefont {Leroux},\ and\ \citenamefont
		{Vuleti{\'c}}}]{schleier-smith2010}%
	\BibitemOpen
	\bibfield  {author} {\bibinfo {author} {\bibfnamefont {M.~H.}\ \bibnamefont
			{Schleier-Smith}}, \bibinfo {author} {\bibfnamefont {I.~D.}\ \bibnamefont
			{Leroux}},\ and\ \bibinfo {author} {\bibfnamefont {V.}~\bibnamefont
			{Vuleti{\'c}}},\ }\href@noop {} {\bibfield  {journal} {\bibinfo  {journal}
			{Phys. Rev. Lett.}\ }\textbf {\bibinfo {volume} {104}},\ \bibinfo {pages}
		{073604} (\bibinfo {year} {2010})}\BibitemShut {NoStop}%
	\bibitem [{\citenamefont {Vasilakis}\ \emph {et~al.}(2015)\citenamefont
		{Vasilakis}, \citenamefont {Shen}, \citenamefont {Jensen}, \citenamefont
		{Balabas}, \citenamefont {Salart}, \citenamefont {Chen},\ and\ \citenamefont
		{Polzik}}]{vasilakis2015}%
	\BibitemOpen
	\bibfield  {author} {\bibinfo {author} {\bibfnamefont {G.}~\bibnamefont
			{Vasilakis}}, \bibinfo {author} {\bibfnamefont {H.}~\bibnamefont {Shen}},
		\bibinfo {author} {\bibfnamefont {K.}~\bibnamefont {Jensen}}, \bibinfo
		{author} {\bibfnamefont {M.}~\bibnamefont {Balabas}}, \bibinfo {author}
		{\bibfnamefont {D.}~\bibnamefont {Salart}}, \bibinfo {author} {\bibfnamefont
			{B.}~\bibnamefont {Chen}},\ and\ \bibinfo {author} {\bibfnamefont {E.~S.}\
			\bibnamefont {Polzik}},\ }\href@noop {} {\bibfield  {journal} {\bibinfo
			{journal} {Nature Physics}\ }\textbf {\bibinfo {volume} {11}},\ \bibinfo
		{pages} {389} (\bibinfo {year} {2015})}\BibitemShut {NoStop}%
	\bibitem [{\citenamefont {M{\o}ller}\ \emph {et~al.}(2017)\citenamefont
		{M{\o}ller}, \citenamefont {Thomas}, \citenamefont {Vasilakis}, \citenamefont
		{Zeuthen}, \citenamefont {Tsaturyan}, \citenamefont {Balabas}, \citenamefont
		{Jensen}, \citenamefont {Schliesser}, \citenamefont {Hammerer},\ and\
		\citenamefont {Polzik}}]{moller2017}%
	\BibitemOpen
	\bibfield  {author} {\bibinfo {author} {\bibfnamefont {C.~B.}\ \bibnamefont
			{M{\o}ller}}, \bibinfo {author} {\bibfnamefont {R.~A.}\ \bibnamefont
			{Thomas}}, \bibinfo {author} {\bibfnamefont {G.}~\bibnamefont {Vasilakis}},
		\bibinfo {author} {\bibfnamefont {E.}~\bibnamefont {Zeuthen}}, \bibinfo
		{author} {\bibfnamefont {Y.}~\bibnamefont {Tsaturyan}}, \bibinfo {author}
		{\bibfnamefont {M.}~\bibnamefont {Balabas}}, \bibinfo {author} {\bibfnamefont
			{K.}~\bibnamefont {Jensen}}, \bibinfo {author} {\bibfnamefont
			{A.}~\bibnamefont {Schliesser}}, \bibinfo {author} {\bibfnamefont
			{K.}~\bibnamefont {Hammerer}},\ and\ \bibinfo {author} {\bibfnamefont
			{E.~S.}\ \bibnamefont {Polzik}},\ }\href@noop {} {\bibfield  {journal}
		{\bibinfo  {journal} {Nature}\ }\textbf {\bibinfo {volume} {547}},\ \bibinfo
		{pages} {191} (\bibinfo {year} {2017})}\BibitemShut {NoStop}%
	\bibitem [{\citenamefont {Ilo-Okeke}\ \emph
		{et~al.}(2023{\natexlab{a}})\citenamefont {Ilo-Okeke}, \citenamefont
		{Kondappan}, \citenamefont {Chen}, \citenamefont {Mao}, \citenamefont
		{Ivannikov},\ and\ \citenamefont {Byrnes}}]{ilo-okeke2023}%
	\BibitemOpen
	\bibfield  {author} {\bibinfo {author} {\bibfnamefont {E.~O.}\ \bibnamefont
			{Ilo-Okeke}}, \bibinfo {author} {\bibfnamefont {M.}~\bibnamefont
			{Kondappan}}, \bibinfo {author} {\bibfnamefont {P.}~\bibnamefont {Chen}},
		\bibinfo {author} {\bibfnamefont {Y.}~\bibnamefont {Mao}}, \bibinfo {author}
		{\bibfnamefont {V.}~\bibnamefont {Ivannikov}},\ and\ \bibinfo {author}
		{\bibfnamefont {T.}~\bibnamefont {Byrnes}},\ }\href
	{https://doi.org/10.1103/PhysRevA.107.052604} {\bibfield  {journal} {\bibinfo
			{journal} {Phys. Rev. A}\ }\textbf {\bibinfo {volume} {107}},\ \bibinfo
		{pages} {052604} (\bibinfo {year} {2023}{\natexlab{a}})}\BibitemShut
	{NoStop}%
	\bibitem [{\citenamefont {Pezz{\`e}}\ \emph {et~al.}(2018)\citenamefont
		{Pezz{\`e}}, \citenamefont {Smerzi}, \citenamefont {Oberthaler},
		\citenamefont {Schmied},\ and\ \citenamefont {Treutlein}}]{pezze2018}%
	\BibitemOpen
	\bibfield  {author} {\bibinfo {author} {\bibfnamefont {L.}~\bibnamefont
			{Pezz{\`e}}}, \bibinfo {author} {\bibfnamefont {A.}~\bibnamefont {Smerzi}},
		\bibinfo {author} {\bibfnamefont {M.~K.}\ \bibnamefont {Oberthaler}},
		\bibinfo {author} {\bibfnamefont {R.}~\bibnamefont {Schmied}},\ and\ \bibinfo
		{author} {\bibfnamefont {P.}~\bibnamefont {Treutlein}},\ }\href
	{https://doi.org/10.1103/RevModPhys.90.035005} {\bibfield  {journal}
		{\bibinfo  {journal} {Rev. Mod. Phys.}\ }\textbf {\bibinfo {volume} {90}},\
		\bibinfo {pages} {035005} (\bibinfo {year} {2018})}\BibitemShut {NoStop}%
	\bibitem [{\citenamefont {Cabello}(2003)}]{cabello2003Review}%
	\BibitemOpen
	\bibfield  {author} {\bibinfo {author} {\bibfnamefont {A.}~\bibnamefont
			{Cabello}},\ }\href {https://doi.org/10.1080/09500340308234551} {\bibfield
		{journal} {\bibinfo  {journal} {Journal of Modern Optics}\ }\textbf {\bibinfo
			{volume} {50}},\ \bibinfo {pages} {1049} (\bibinfo {year}
		{2003})}\BibitemShut {NoStop}%
	\bibitem [{\citenamefont {Bennett}\ \emph
		{et~al.}(1996{\natexlab{a}})\citenamefont {Bennett}, \citenamefont
		{{DiVincenzo}}, \citenamefont {Smolin},\ and\ \citenamefont
		{Wootters}}]{bennett1996b}%
	\BibitemOpen
	\bibfield  {author} {\bibinfo {author} {\bibfnamefont {C.~H.}\ \bibnamefont
			{Bennett}}, \bibinfo {author} {\bibfnamefont {D.~P.}\ \bibnamefont
			{{DiVincenzo}}}, \bibinfo {author} {\bibfnamefont {J.~A.}\ \bibnamefont
			{Smolin}},\ and\ \bibinfo {author} {\bibfnamefont {W.~K.}\ \bibnamefont
			{Wootters}},\ }\href@noop {} {\bibfield  {journal} {\bibinfo  {journal}
			{Phys. Rev. A}\ }\textbf {\bibinfo {volume} {54}},\ \bibinfo {pages} {3824}
		(\bibinfo {year} {1996}{\natexlab{a}})}\BibitemShut {NoStop}%
	\bibitem [{\citenamefont {Bennett}\ \emph
		{et~al.}(1996{\natexlab{b}})\citenamefont {Bennett}, \citenamefont
		{Bernstein}, \citenamefont {Popescu},\ and\ \citenamefont
		{Schumacher}}]{bennett1996c}%
	\BibitemOpen
	\bibfield  {author} {\bibinfo {author} {\bibfnamefont {C.~H.}\ \bibnamefont
			{Bennett}}, \bibinfo {author} {\bibfnamefont {H.~J.}\ \bibnamefont
			{Bernstein}}, \bibinfo {author} {\bibfnamefont {S.}~\bibnamefont {Popescu}},\
		and\ \bibinfo {author} {\bibfnamefont {B.}~\bibnamefont {Schumacher}},\
	}\href@noop {} {\bibfield  {journal} {\bibinfo  {journal} {Phys. Rev. A}\
		}\textbf {\bibinfo {volume} {53}},\ \bibinfo {pages} {2046} (\bibinfo {year}
		{1996}{\natexlab{b}})}\BibitemShut {NoStop}%
	\bibitem [{\citenamefont {Bennett}\ \emph
		{et~al.}(1996{\natexlab{c}})\citenamefont {Bennett}, \citenamefont
		{Brassard}, \citenamefont {Popescu}, \citenamefont {Schumacher},
		\citenamefont {Smolin},\ and\ \citenamefont {Wootters}}]{bennett1996}%
	\BibitemOpen
	\bibfield  {author} {\bibinfo {author} {\bibfnamefont {C.~H.}\ \bibnamefont
			{Bennett}}, \bibinfo {author} {\bibfnamefont {G.}~\bibnamefont {Brassard}},
		\bibinfo {author} {\bibfnamefont {S.}~\bibnamefont {Popescu}}, \bibinfo
		{author} {\bibfnamefont {B.}~\bibnamefont {Schumacher}}, \bibinfo {author}
		{\bibfnamefont {J.~A.}\ \bibnamefont {Smolin}},\ and\ \bibinfo {author}
		{\bibfnamefont {W.~K.}\ \bibnamefont {Wootters}},\ }\href@noop {} {\bibfield
		{journal} {\bibinfo  {journal} {Phys. Rev. Lett.}\ }\textbf {\bibinfo
			{volume} {76}},\ \bibinfo {pages} {722} (\bibinfo {year}
		{1996}{\natexlab{c}})}\BibitemShut {NoStop}%
	\bibitem [{\citenamefont {Chaudhary}\ \emph {et~al.}(2021)\citenamefont
		{Chaudhary}, \citenamefont {Fadel}, \citenamefont {Ilo-Okeke}, \citenamefont
		{Pyrkov}, \citenamefont {Ivannikov},\ and\ \citenamefont
		{Byrnes}}]{chaudhary2021}%
	\BibitemOpen
	\bibfield  {author} {\bibinfo {author} {\bibfnamefont {M.}~\bibnamefont
			{Chaudhary}}, \bibinfo {author} {\bibfnamefont {M.}~\bibnamefont {Fadel}},
		\bibinfo {author} {\bibfnamefont {E.~O.}\ \bibnamefont {Ilo-Okeke}}, \bibinfo
		{author} {\bibfnamefont {A.~N.}\ \bibnamefont {Pyrkov}}, \bibinfo {author}
		{\bibfnamefont {V.}~\bibnamefont {Ivannikov}},\ and\ \bibinfo {author}
		{\bibfnamefont {T.}~\bibnamefont {Byrnes}},\ }\href
	{https://doi.org/10.1103/PhysRevA.103.062417} {\bibfield  {journal} {\bibinfo
			{journal} {Phys. Rev. A}\ }\textbf {\bibinfo {volume} {103}},\ \bibinfo
		{pages} {062417} (\bibinfo {year} {2021})}\BibitemShut {NoStop}%
	\bibitem [{\citenamefont {Jozsa}\ \emph {et~al.}(2000)\citenamefont {Jozsa},
		\citenamefont {Abrams}, \citenamefont {Dowling},\ and\ \citenamefont
		{Williams}}]{jozsa2000}%
	\BibitemOpen
	\bibfield  {author} {\bibinfo {author} {\bibfnamefont {R.}~\bibnamefont
			{Jozsa}}, \bibinfo {author} {\bibfnamefont {D.~S.}\ \bibnamefont {Abrams}},
		\bibinfo {author} {\bibfnamefont {J.~P.}\ \bibnamefont {Dowling}},\ and\
		\bibinfo {author} {\bibfnamefont {C.~P.}\ \bibnamefont {Williams}},\
	}\href@noop {} {\bibfield  {journal} {\bibinfo  {journal} {Phys. Rev. Lett.}\
		}\textbf {\bibinfo {volume} {85}},\ \bibinfo {pages} {2010} (\bibinfo {year}
		{2000})}\BibitemShut {NoStop}%
	\bibitem [{\citenamefont {Ilo-Okeke}\ \emph {et~al.}(2018)\citenamefont
		{Ilo-Okeke}, \citenamefont {Tessler}, \citenamefont {Dowling},\ and\
		\citenamefont {Byrnes}}]{ilo-okeke2018}%
	\BibitemOpen
	\bibfield  {author} {\bibinfo {author} {\bibfnamefont {E.~O.}\ \bibnamefont
			{Ilo-Okeke}}, \bibinfo {author} {\bibfnamefont {L.}~\bibnamefont {Tessler}},
		\bibinfo {author} {\bibfnamefont {J.~P.}\ \bibnamefont {Dowling}},\ and\
		\bibinfo {author} {\bibfnamefont {T.}~\bibnamefont {Byrnes}},\ }\href@noop {}
	{\bibfield  {journal} {\bibinfo  {journal} {npj Quantum Inf}\ }\textbf
		{\bibinfo {volume} {4}},\ \bibinfo {pages} {40} (\bibinfo {year}
		{2018})}\BibitemShut {NoStop}%
	\bibitem [{\citenamefont {Appel}\ \emph {et~al.}(2009)\citenamefont {Appel},
		\citenamefont {Windpassinger}, \citenamefont {Oblak}, \citenamefont {Hoff},
		\citenamefont {Kj{\ae}rgaard},\ and\ \citenamefont {Polzik}}]{appel2009}%
	\BibitemOpen
	\bibfield  {author} {\bibinfo {author} {\bibfnamefont {J.}~\bibnamefont
			{Appel}}, \bibinfo {author} {\bibfnamefont {P.~J.}\ \bibnamefont
			{Windpassinger}}, \bibinfo {author} {\bibfnamefont {D.}~\bibnamefont
			{Oblak}}, \bibinfo {author} {\bibfnamefont {U.~B.}\ \bibnamefont {Hoff}},
		\bibinfo {author} {\bibfnamefont {N.}~\bibnamefont {Kj{\ae}rgaard}},\ and\
		\bibinfo {author} {\bibfnamefont {E.~S.}\ \bibnamefont {Polzik}},\
	}\href@noop {} {\bibfield  {journal} {\bibinfo  {journal} {Proc. Natl. Acad.
				Sci. USA}\ }\textbf {\bibinfo {volume} {106}},\ \bibinfo {pages} {10960}
		(\bibinfo {year} {2009})}\BibitemShut {NoStop}%
	\bibitem [{\citenamefont {Louchet-Chauvet}\ \emph {et~al.}(2010)\citenamefont
		{Louchet-Chauvet}, \citenamefont {Appel}, \citenamefont {Renema},
		\citenamefont {Oblak}, \citenamefont {Kjaergaard},\ and\ \citenamefont
		{Polzik}}]{louchet-chauvet2010}%
	\BibitemOpen
	\bibfield  {author} {\bibinfo {author} {\bibfnamefont {A.}~\bibnamefont
			{Louchet-Chauvet}}, \bibinfo {author} {\bibfnamefont {J.}~\bibnamefont
			{Appel}}, \bibinfo {author} {\bibfnamefont {J.~J.}\ \bibnamefont {Renema}},
		\bibinfo {author} {\bibfnamefont {D.}~\bibnamefont {Oblak}}, \bibinfo
		{author} {\bibfnamefont {N.}~\bibnamefont {Kjaergaard}},\ and\ \bibinfo
		{author} {\bibfnamefont {E.~S.}\ \bibnamefont {Polzik}},\ }\href
	{https://doi.org/10.1088/1367-2630/12/6/065032} {\bibfield  {journal}
		{\bibinfo  {journal} {New Journal of Physics}\ }\textbf {\bibinfo {volume}
			{12}},\ \bibinfo {pages} {065032} (\bibinfo {year} {2010})}\BibitemShut
	{NoStop}%
	\bibitem [{\citenamefont {Lone}\ and\ \citenamefont {Byrnes}(2015)}]{lone2015}%
	\BibitemOpen
	\bibfield  {author} {\bibinfo {author} {\bibfnamefont {M.~Q.}\ \bibnamefont
			{Lone}}\ and\ \bibinfo {author} {\bibfnamefont {T.}~\bibnamefont {Byrnes}},\
	}\href@noop {} {\bibfield  {journal} {\bibinfo  {journal} {Phys. Rev. A}\
		}\textbf {\bibinfo {volume} {92}},\ \bibinfo {pages} {011401(R)} (\bibinfo
		{year} {2015})}\BibitemShut {NoStop}%
	\bibitem [{\citenamefont {Ilo-Okeke}\ \emph {et~al.}(2021)\citenamefont
		{Ilo-Okeke}, \citenamefont {Sunami}, \citenamefont {Foot},\ and\
		\citenamefont {Byrnes}}]{ilo-okeke2021}%
	\BibitemOpen
	\bibfield  {author} {\bibinfo {author} {\bibfnamefont {E.~O.}\ \bibnamefont
			{Ilo-Okeke}}, \bibinfo {author} {\bibfnamefont {S.}~\bibnamefont {Sunami}},
		\bibinfo {author} {\bibfnamefont {C.~J.}\ \bibnamefont {Foot}},\ and\
		\bibinfo {author} {\bibfnamefont {T.}~\bibnamefont {Byrnes}},\ }\href
	{https://doi.org/10.1103/PhysRevA.104.053324} {\bibfield  {journal} {\bibinfo
			{journal} {Phys. Rev. A}\ }\textbf {\bibinfo {volume} {104}},\ \bibinfo
		{pages} {053324} (\bibinfo {year} {2021})}\BibitemShut {NoStop}%
	\bibitem [{\citenamefont {Allen}\ and\ \citenamefont
		{Eberly}(1975)}]{allen1975}%
	\BibitemOpen
	\bibfield  {author} {\bibinfo {author} {\bibfnamefont {L.}~\bibnamefont
			{Allen}}\ and\ \bibinfo {author} {\bibfnamefont {J.~H.}\ \bibnamefont
			{Eberly}},\ }\href@noop {} {\emph {\bibinfo {title} {Optical {R}esonance and
				{T}wo-Level {A}toms}}}\ (\bibinfo  {publisher} {(John Wiley},\ \bibinfo
	{address} {New York},\ \bibinfo {year} {1975)})\BibitemShut {NoStop}%
	\bibitem [{\citenamefont {Shore}(1990)}]{shore1990v1}%
	\BibitemOpen
	\bibfield  {author} {\bibinfo {author} {\bibfnamefont {B.~W.}\ \bibnamefont
			{Shore}},\ }\href@noop {} {\emph {\bibinfo {title} {{The theory of coherent
					atomic excitation }}}},\ Vol.~\bibinfo {volume} {1}\ (\bibinfo  {publisher}
	{Wiley-Interscience},\ \bibinfo {address} {New York},\ \bibinfo {year}
	{1990})\BibitemShut {NoStop}%
	\bibitem [{\citenamefont {Carmichael}(1999)}]{carmichael1999}%
	\BibitemOpen
	\bibfield  {author} {\bibinfo {author} {\bibfnamefont {H.~J.}\ \bibnamefont
			{Carmichael}},\ }\href@noop {} {\emph {\bibinfo {title} {{Statistical Methods
					in Quantum Optics 1}}}}\ (\bibinfo  {publisher} {Springer-Verlag},\ \bibinfo
	{address} {Berlin},\ \bibinfo {year} {1999})\BibitemShut {NoStop}%
	\bibitem [{\citenamefont {Ilo-Okeke}\ \emph {et~al.}(2022)\citenamefont
		{Ilo-Okeke}, \citenamefont {Ji}, \citenamefont {Chen}, \citenamefont {Mao},
		\citenamefont {Kondappan}, \citenamefont {Ivannikov}, \citenamefont {Xiao},\
		and\ \citenamefont {Byrnes}}]{ilo-okeke2022}%
	\BibitemOpen
	\bibfield  {author} {\bibinfo {author} {\bibfnamefont {E.~O.}\ \bibnamefont
			{Ilo-Okeke}}, \bibinfo {author} {\bibfnamefont {Y.}~\bibnamefont {Ji}},
		\bibinfo {author} {\bibfnamefont {P.}~\bibnamefont {Chen}}, \bibinfo {author}
		{\bibfnamefont {Y.}~\bibnamefont {Mao}}, \bibinfo {author} {\bibfnamefont
			{M.}~\bibnamefont {Kondappan}}, \bibinfo {author} {\bibfnamefont
			{V.}~\bibnamefont {Ivannikov}}, \bibinfo {author} {\bibfnamefont
			{Y.}~\bibnamefont {Xiao}},\ and\ \bibinfo {author} {\bibfnamefont
			{T.}~\bibnamefont {Byrnes}},\ }\href
	{https://doi.org/10.1103/PhysRevA.106.033314} {\bibfield  {journal} {\bibinfo
			{journal} {Phys. Rev. A}\ }\textbf {\bibinfo {volume} {106}},\ \bibinfo
		{pages} {033314} (\bibinfo {year} {2022})}\BibitemShut {NoStop}%
	\bibitem [{\citenamefont {Mao}\ \emph {et~al.}(2022)\citenamefont {Mao},
		\citenamefont {Chaudhary}, \citenamefont {Kondappan}, \citenamefont {Shi},
		\citenamefont {Ilo-Okeke}, \citenamefont {Ivannikov},\ and\ \citenamefont
		{Byrnes}}]{mao2022}%
	\BibitemOpen
	\bibfield  {author} {\bibinfo {author} {\bibfnamefont {Y.}~\bibnamefont
			{Mao}}, \bibinfo {author} {\bibfnamefont {M.}~\bibnamefont {Chaudhary}},
		\bibinfo {author} {\bibfnamefont {M.}~\bibnamefont {Kondappan}}, \bibinfo
		{author} {\bibfnamefont {J.}~\bibnamefont {Shi}}, \bibinfo {author}
		{\bibfnamefont {E.~O.}\ \bibnamefont {Ilo-Okeke}}, \bibinfo {author}
		{\bibfnamefont {V.}~\bibnamefont {Ivannikov}},\ and\ \bibinfo {author}
		{\bibfnamefont {T.}~\bibnamefont {Byrnes}},\ }\href@noop {} {} (\bibinfo
	{year} {2022}),\ \bibinfo {note} {arXiv:2202.09100[quant-ph]}\BibitemShut
	{NoStop}%
	\bibitem [{\citenamefont {Kondappan}\ \emph {et~al.}(2023)\citenamefont
		{Kondappan}, \citenamefont {Chaudhary}, \citenamefont {Ilo-Okeke},
		\citenamefont {Ivannikov},\ and\ \citenamefont {Byrnes}}]{kondappan2023}%
	\BibitemOpen
	\bibfield  {author} {\bibinfo {author} {\bibfnamefont {M.}~\bibnamefont
			{Kondappan}}, \bibinfo {author} {\bibfnamefont {M.}~\bibnamefont
			{Chaudhary}}, \bibinfo {author} {\bibfnamefont {E.~O.}\ \bibnamefont
			{Ilo-Okeke}}, \bibinfo {author} {\bibfnamefont {V.}~\bibnamefont
			{Ivannikov}},\ and\ \bibinfo {author} {\bibfnamefont {T.}~\bibnamefont
			{Byrnes}},\ }\href {https://doi.org/10.1103/PhysRevA.107.042616} {\bibfield
		{journal} {\bibinfo  {journal} {Phys. Rev. A}\ }\textbf {\bibinfo {volume}
			{107}},\ \bibinfo {pages} {042616} (\bibinfo {year} {2023})}\BibitemShut
	{NoStop}%
	\bibitem [{\citenamefont {Ilo-Okeke}\ \emph
		{et~al.}(2023{\natexlab{b}})\citenamefont {Ilo-Okeke}, \citenamefont {Chen},
		\citenamefont {Li}, \citenamefont {Anusionwu}, \citenamefont {Ivannikov},\
		and\ \citenamefont {Byrnes}}]{ilo-okeke2023Dowling}%
	\BibitemOpen
	\bibfield  {author} {\bibinfo {author} {\bibfnamefont {E.~O.}\ \bibnamefont
			{Ilo-Okeke}}, \bibinfo {author} {\bibfnamefont {P.}~\bibnamefont {Chen}},
		\bibinfo {author} {\bibfnamefont {S.}~\bibnamefont {Li}}, \bibinfo {author}
		{\bibfnamefont {B.~C.}\ \bibnamefont {Anusionwu}}, \bibinfo {author}
		{\bibfnamefont {V.}~\bibnamefont {Ivannikov}},\ and\ \bibinfo {author}
		{\bibfnamefont {T.}~\bibnamefont {Byrnes}},\ }\href
	{https://doi.org/doi.org/10.1116/5.0141921} {\bibfield  {journal} {\bibinfo
			{journal} {AVS Quantum Sci.}\ }\textbf {\bibinfo {volume} {5}},\ \bibinfo
		{pages} {025004} (\bibinfo {year} {2023}{\natexlab{b}})}\BibitemShut
	{NoStop}%
	\bibitem [{\citenamefont {Bao}\ \emph {et~al.}(2020)\citenamefont {Bao},
		\citenamefont {Duan}, \citenamefont {Jin}, \citenamefont {Lu}, \citenamefont
		{Li}, \citenamefont {Qu}, \citenamefont {Wang}, \citenamefont {Novikova},
		\citenamefont {Mikhailov}, \citenamefont {Zhao}, \citenamefont {M{\o}lmer},
		\citenamefont {Shen},\ and\ \citenamefont {Xiao}}]{bao2020}%
	\BibitemOpen
	\bibfield  {author} {\bibinfo {author} {\bibfnamefont {H.}~\bibnamefont
			{Bao}}, \bibinfo {author} {\bibfnamefont {J.}~\bibnamefont {Duan}}, \bibinfo
		{author} {\bibfnamefont {S.}~\bibnamefont {Jin}}, \bibinfo {author}
		{\bibfnamefont {X.}~\bibnamefont {Lu}}, \bibinfo {author} {\bibfnamefont
			{P.}~\bibnamefont {Li}}, \bibinfo {author} {\bibfnamefont {W.}~\bibnamefont
			{Qu}}, \bibinfo {author} {\bibfnamefont {M.}~\bibnamefont {Wang}}, \bibinfo
		{author} {\bibfnamefont {I.}~\bibnamefont {Novikova}}, \bibinfo {author}
		{\bibfnamefont {E.~E.}\ \bibnamefont {Mikhailov}}, \bibinfo {author}
		{\bibfnamefont {K.~F.}\ \bibnamefont {Zhao}}, \bibinfo {author}
		{\bibfnamefont {K.}~\bibnamefont {M{\o}lmer}}, \bibinfo {author}
		{\bibfnamefont {H.}~\bibnamefont {Shen}},\ and\ \bibinfo {author}
		{\bibfnamefont {Y.}~\bibnamefont {Xiao}},\ }\href
	{https://doi.org/https://doi.org/10.1038/s41586-020-2243-7} {\bibfield
		{journal} {\bibinfo  {journal} {Nature}\ }\textbf {\bibinfo {volume} {581}},\
		\bibinfo {pages} {159} (\bibinfo {year} {2020})}\BibitemShut {NoStop}%
	\bibitem [{\citenamefont {Ilo-Okeke}\ and\ \citenamefont
		{Byrnes}(2014)}]{ilo-okeke2014}%
	\BibitemOpen
	\bibfield  {author} {\bibinfo {author} {\bibfnamefont {E.~O.}\ \bibnamefont
			{Ilo-Okeke}}\ and\ \bibinfo {author} {\bibfnamefont {T.}~\bibnamefont
			{Byrnes}},\ }\href@noop {} {\bibfield  {journal} {\bibinfo  {journal} {Phys.
				Rev. Lett.}\ }\textbf {\bibinfo {volume} {112}},\ \bibinfo {pages} {233602}
		(\bibinfo {year} {2014})}\BibitemShut {NoStop}%
	\bibitem [{\citenamefont {S{\o}rensen}\ \emph {et~al.}(1998)\citenamefont
		{S{\o}rensen}, \citenamefont {Hald},\ and\ \citenamefont
		{Polzik}}]{sorensen1998}%
	\BibitemOpen
	\bibfield  {author} {\bibinfo {author} {\bibfnamefont {J.~L.}\ \bibnamefont
			{S{\o}rensen}}, \bibinfo {author} {\bibfnamefont {J.}~\bibnamefont {Hald}},\
		and\ \bibinfo {author} {\bibfnamefont {E.~S.}\ \bibnamefont {Polzik}},\
	}\href {https://doi.org/10.1103/PhysRevLett.80.3487} {\bibfield  {journal}
		{\bibinfo  {journal} {Phys. Rev. Lett.}\ }\textbf {\bibinfo {volume} {80}},\
		\bibinfo {pages} {3487} (\bibinfo {year} {1998})}\BibitemShut {NoStop}%
	\bibitem [{\citenamefont {Sewell}\ \emph {et~al.}(2012)\citenamefont {Sewell},
		\citenamefont {Koschorreck}, \citenamefont {Napolitano}, \citenamefont
		{Dubost}, \citenamefont {Behbood},\ and\ \citenamefont
		{Mitchell}}]{sewell2012}%
	\BibitemOpen
	\bibfield  {author} {\bibinfo {author} {\bibfnamefont {R.~J.}\ \bibnamefont
			{Sewell}}, \bibinfo {author} {\bibfnamefont {M.}~\bibnamefont {Koschorreck}},
		\bibinfo {author} {\bibfnamefont {M.}~\bibnamefont {Napolitano}}, \bibinfo
		{author} {\bibfnamefont {B.}~\bibnamefont {Dubost}}, \bibinfo {author}
		{\bibfnamefont {N.}~\bibnamefont {Behbood}},\ and\ \bibinfo {author}
		{\bibfnamefont {M.~W.}\ \bibnamefont {Mitchell}},\ }\href@noop {} {\bibfield
		{journal} {\bibinfo  {journal} {Phys. Rev. Lett.}\ }\textbf {\bibinfo
			{volume} {109}},\ \bibinfo {pages} {253605} (\bibinfo {year}
		{2012})}\BibitemShut {NoStop}%
	\bibitem [{\citenamefont {Sewell}\ \emph {et~al.}(2013)\citenamefont {Sewell},
		\citenamefont {Napolitano}, \citenamefont {Behbood}, \citenamefont
		{Colangelo},\ and\ \citenamefont {Mitchell}}]{sewell2013}%
	\BibitemOpen
	\bibfield  {author} {\bibinfo {author} {\bibfnamefont {R.}~\bibnamefont
			{Sewell}}, \bibinfo {author} {\bibfnamefont {M.}~\bibnamefont {Napolitano}},
		\bibinfo {author} {\bibfnamefont {N.}~\bibnamefont {Behbood}}, \bibinfo
		{author} {\bibfnamefont {G.}~\bibnamefont {Colangelo}},\ and\ \bibinfo
		{author} {\bibfnamefont {M.~W.}\ \bibnamefont {Mitchell}},\ }\href
	{https://doi.org/https://doi.org/10.1038/nphoton.2013.100} {\bibfield
		{journal} {\bibinfo  {journal} {Nature Photonics}\ }\textbf {\bibinfo
			{volume} {7}},\ \bibinfo {pages} {517} (\bibinfo {year} {2013})}\BibitemShut
	{NoStop}%
	\bibitem [{\citenamefont {Ban}(2009)}]{ban2009}%
	\BibitemOpen
	\bibfield  {author} {\bibinfo {author} {\bibfnamefont {M.}~\bibnamefont
			{Ban}},\ }\href {https://doi.org/10.1103/PhysRevA.80.064103} {\bibfield
		{journal} {\bibinfo  {journal} {Phys. Rev. A}\ }\textbf {\bibinfo {volume}
			{80}},\ \bibinfo {pages} {064103} (\bibinfo {year} {2009})}\BibitemShut
	{NoStop}%
	\bibitem [{\citenamefont {Tempel}\ and\ \citenamefont
		{Aspuru-Guzik}(2011)}]{tempel2011}%
	\BibitemOpen
	\bibfield  {author} {\bibinfo {author} {\bibfnamefont {D.~G.}\ \bibnamefont
			{Tempel}}\ and\ \bibinfo {author} {\bibfnamefont {A.}~\bibnamefont
			{Aspuru-Guzik}},\ }\href
	{https://doi.org/https://doi.org/10.1016/j.chemphys.2011.03.014} {\bibfield
		{journal} {\bibinfo  {journal} {Chemical Physics}\ }\textbf {\bibinfo
			{volume} {391}},\ \bibinfo {pages} {130} (\bibinfo {year}
		{2011})}\BibitemShut {NoStop}%
	\bibitem [{\citenamefont {{H. -P. Breuer and F.
				Petruccione}}(2002)}]{breuer2002}%
	\BibitemOpen
	\bibfield  {author} {\bibinfo {author} {\bibnamefont {{H. -P. Breuer and F.
					Petruccione}}},\ }\href@noop {} {\emph {\bibinfo {title} {{The Theory of Open
					Quantum Systems}}}}\ (\bibinfo  {publisher} {Oxford University Press},\
	\bibinfo {address} {New York},\ \bibinfo {year} {2002})\BibitemShut {NoStop}%
	\bibitem [{\citenamefont {Carmichael}(2008)}]{carmichael2008}%
	\BibitemOpen
	\bibfield  {author} {\bibinfo {author} {\bibfnamefont {H.~J.}\ \bibnamefont
			{Carmichael}},\ }\href@noop {} {\emph {\bibinfo {title} {{Statistical Methods
					in Quantum Optics 2}}}}\ (\bibinfo  {publisher} {Springer-Verlag},\ \bibinfo
	{address} {Berlin},\ \bibinfo {year} {2008})\BibitemShut {NoStop}%
	\bibitem [{\citenamefont {Arecchi}\ \emph {et~al.}(1972)\citenamefont
		{Arecchi}, \citenamefont {Courtens}, \citenamefont {Gilmore},\ and\
		\citenamefont {Thomas}}]{arecchi1972}%
	\BibitemOpen
	\bibfield  {author} {\bibinfo {author} {\bibfnamefont {F.~T.}\ \bibnamefont
			{Arecchi}}, \bibinfo {author} {\bibfnamefont {E.}~\bibnamefont {Courtens}},
		\bibinfo {author} {\bibfnamefont {R.}~\bibnamefont {Gilmore}},\ and\ \bibinfo
		{author} {\bibfnamefont {H.}~\bibnamefont {Thomas}},\ }\href@noop {}
	{\bibfield  {journal} {\bibinfo  {journal} {Phys. Rev. A}\ }\textbf {\bibinfo
			{volume} {6}},\ \bibinfo {pages} {2211} (\bibinfo {year} {1972})}\BibitemShut
	{NoStop}%
	\bibitem [{\citenamefont {{W. E. Boyce and R. C. DiPrima}}(2005)}]{boyce2005}%
	\BibitemOpen
	\bibfield  {author} {\bibinfo {author} {\bibnamefont {{W. E. Boyce and R. C.
					DiPrima}}},\ }\href@noop {} {\emph {\bibinfo {title} {{Elementary
					Differential Equations and Boundary Value Problems}}}},\ \bibinfo {edition}
	{eight}\ ed.\ (\bibinfo  {publisher} {John Wiley {$\&$} Sons},\ \bibinfo
	{address} {New Jersey},\ \bibinfo {year} {2005})\BibitemShut {NoStop}%
	\bibitem [{\citenamefont {{M. L. Boas}}(2006)}]{boas2006}%
	\BibitemOpen
	\bibfield  {author} {\bibinfo {author} {\bibnamefont {{M. L. Boas}}},\
	}\href@noop {} {\emph {\bibinfo {title} {{Mathematical Methods In The
					Physical Sciences}}}},\ \bibinfo {edition} {3rd}\ ed.\ (\bibinfo  {publisher}
	{John Wiley {$\&$} Sons},\ \bibinfo {address} {New Jersey},\ \bibinfo {year}
	{2006})\BibitemShut {NoStop}%
	\bibitem [{\citenamefont {Chaudhary}\ \emph {et~al.}(2023)\citenamefont
		{Chaudhary}, \citenamefont {Ilo-Okeke}, \citenamefont {Ivannikov},\ and\
		\citenamefont {Byrnes}}]{chaudhary2023}%
	\BibitemOpen
	\bibfield  {author} {\bibinfo {author} {\bibfnamefont {M.}~\bibnamefont
			{Chaudhary}}, \bibinfo {author} {\bibfnamefont {E.~O.}\ \bibnamefont
			{Ilo-Okeke}}, \bibinfo {author} {\bibfnamefont {V.}~\bibnamefont
			{Ivannikov}},\ and\ \bibinfo {author} {\bibfnamefont {T.}~\bibnamefont
			{Byrnes}},\ }\href@noop {} {} (\bibinfo {year} {2023}),\ \bibinfo {note}
	{arXiv:2302.07526[quant-ph]}\BibitemShut {NoStop}%
	\bibitem [{\citenamefont {Ilo-Okeke}\ and\ \citenamefont
		{Zozulya}(2010)}]{ilo-okeke2010}%
	\BibitemOpen
	\bibfield  {author} {\bibinfo {author} {\bibfnamefont {E.~O.}\ \bibnamefont
			{Ilo-Okeke}}\ and\ \bibinfo {author} {\bibfnamefont {A.~A.}\ \bibnamefont
			{Zozulya}},\ }\href@noop {} {\bibfield  {journal} {\bibinfo  {journal} {Phys.
				Rev. A}\ }\textbf {\bibinfo {volume} {82}},\ \bibinfo {pages} {053603}
		(\bibinfo {year} {2010})}\BibitemShut {NoStop}%
	\bibitem [{\citenamefont {Ilo-Okeke}\ and\ \citenamefont
		{Byrnes}(2016)}]{ilo-okeke2016}%
	\BibitemOpen
	\bibfield  {author} {\bibinfo {author} {\bibfnamefont {E.~O.}\ \bibnamefont
			{Ilo-Okeke}}\ and\ \bibinfo {author} {\bibfnamefont {T.}~\bibnamefont
			{Byrnes}},\ }\href@noop {} {\bibfield  {journal} {\bibinfo  {journal} {Phys.
				Rev. A}\ }\textbf {\bibinfo {volume} {94}},\ \bibinfo {pages} {013617}
		(\bibinfo {year} {2016})}\BibitemShut {NoStop}%
\end{thebibliography}
%
%
	%
	%
\end{document}